# Spatial Meteorological, Socio-Economic, and Political Risks in Probabilistic Electricity Demand Forecasting

Monika Zimmermann[a,*], Florian Ziel[a]

[a] *Chair of Environmental Economics, esp. Economics of Renewable Energy*
*University of Duisburg-Essen*
*Germany*

**Abstract**

Unlike traditional commodities, electricity is difficult to store and, thus, delivered via a cross-country interconnected grid that requires a continuous balance between supply and demand. This characteristic necessitates high-resolution mid-term electricity demand forecasts to inform financial and operational decisions. The primary challenge in creating these forecasts lies in disentangling the uncertain, spatially varying, and cross-country-dependent impacts on load. Additionally, the non-stationarity of socio-economic and political load effects adds complexity. To address these challenges, we present a novel forecasting method for mid-term hourly electricity demand that is probabilistic and multivariate across 24 European countries. This approach combines multivariate simulated socio-economic, political and temperature trajectories that account for climate change within Generalized Additive Models (GAMs). Built as interpretable additive model in smooth effects, it provides practitioners with a tool to track the hourly load impact of each input at gigawatt precision. We evaluate our method using over 9 years of data (2015-2024), demonstrating significantly improved CRPS compared to standard benchmarks. Our results indicate that a single Pan-European socio-economic and political trend explains much of the non-stationarity in load across Europe. Risk scenario analyses highlight the vulnerability of countries dependent on electric heating during extreme weather events, underscoring the need for probabilistic, cross-country dependent forecasting as electric heating becomes more prevalent across Europe.

## 1. Introduction

***Motivation.*** Electricity assets stand apart from traditional financial or commodity assets due to a unique physical characteristic: electricity is difficult to store. This necessitates a continuous balance between supply and demand fundamentally shaping financial operations within electricity markets. Pricing, bidding, trading, and hedging strategies must all anticipate this balance, relying on accurate forecasts of supply and demand. These forecasts are required at an hourly, or even finer, resolution, as electricity markets operate on such timescales to account for intra-day fluctuations in demand and supply.

Beyond financial operations, the physical characteristics of electricity influence how industry stakeholders operate. Since electricity is difficult to store, it cannot be transported by conventional means. Instead, it is delivered via an interconnected European electricity grid. To maintain grid stability, prevent large-scale outages, and avoid damage to grid infrastructure or end-user devices, a constant voltage must be preserved throughout its power lines. While energy storage systems provide ad-hoc flexibility, their cost-effective operation depends on informed decisions about when to charge and discharge. As a result,

*Corresponding author
*Email addresses:* Monika.Zimmermann@uni-due.de (Monika Zimmermann), Florian.Ziel@uni-due.de (Florian Ziel)

to maintain constant voltage, operators primarily adjust the electricity supply through domestic power production and electricity imports or exports to match forecasted demand.

While extensive research has been conducted on short-term (hours to a few days) electricity demand forecasting, the reviews by Verwiebe et al. (2021) and Davis et al. (2016) highlight a notable scarcity of studies addressing hourly mid-term (weeks to one year) forecasts. This represents a significant gap, as both academic models and financial or industry operators rely on such forecasts, as argued by González Grandón et al. (2024); Behm et al. (2020); Agrawal et al. (2018); Hong et al. (2014). For instance, the reviews of Ghelasi & Ziel (2024); Ziel (2018) highlight that mid-term hourly electricity demand forecasts are commonly used as input variables for electricity price forecasting models. Similarly, financial operators need fine resolution mid-term electricity demand forecasts to handle electricity futures, forwards and power purchase agreements, see e.g. the overview by Jedrzejewski et al. (2022) and studies Taheri et al. (2025); Kandpal et al. (2024). Energy storage operators use these forecasts to optimize arbitrage opportunities from seasonal load fluctuations, manage battery degradation, and determine optimal storage sizing, as summarized by Merrick et al. (2024); Sharma et al. (2021). Additionally, power plant and grid operators require mid-term forecasts to plan production and schedule infrastructure maintenance, see Prajapat et al. (2017).

***Probabilistic and Multivariate Demand Forecasting.*** To address the aforementioned gap, this work provides both academics and practitioners with a model to forecast hourly electricity demand, referred to as load in the following, for horizons of up to one year. The primary challenge in this forecasting task lies in disentangling the various deterministic and uncertain effects on load, as well as accounting for their varying spatial levels and cross-country dependencies. Following Pierrot & Goude (2011), the factors influencing load can be categorized into four main groups:

Firstly, load is shaped by deterministic calendar patterns attributed to country-specific behaviors and working routines leading to higher electricity consumption during winter, weekdays, and daytime, and lower consumption during summer, weekends, nights, and holidays. Secondly, load is affected by uncertain meteorological factors, such as air temperature, humidity, cloud cover, wind speed, and climate change, which impact electric heating, cooling, and lighting. While weather-related factors primarily affect load at the national level, reflecting their regional formation and a country-specific prevalence of electric heating or cooling, underlying meteorological conditions, such as persistent pressure systems over Europe and climate change, have a transnational impact on load. Thirdly, uncertain socio-economic and political factors influence load through their non-stationary trends. Given the strong interconnection of the electricity grid and economies in Europe along with the increasing legislative power of the European Parliament, these load effects are transnational. Load effects due to recent crises, such as COVID-19 and the Russian invasion of Ukraine, alongside rising electricity demand from heat pumps and electric vehicles driven by decarbonization goals, illustrate these impacts. Lastly, remaining uncertain autoregressive terms, primarily national in scope, shape load.

With highly uncertain weather, climate, socio-economic, and political conditions in the mid-term future, along with their transnational impacts on load Do et al. (2024); Romano et al. (2024); Tzortzis et al. (2023), mid-term load modeling needs to account for the inherent cross-country dependent risks.

To incorporate these risks, our proposed forecasting model is probabilistic and multivariate across 24 European countries.

***Existing Research***. Existing methods that are applied in probabilistic load forecasting are wide-ranging, from classical quantile, density and trajectory ensembles forecasting, see e.g. Haben et al. (2023) for a concise overview, to sophisticated machine learning models based on e.g. different types of neural networks, gaussian process models or quantile regression forests, see e.g. Baviera & Manzoni (2024); Wang et al. (2024); Yang et al. (2024); Baviera & Messuti (2023); Li et al. (2023); Zhang et al. (2023); Brusaferri et al. (2022). While some studies have incorporated weather uncertainty into probabilistic load forecasting, see Bowala et al. (2024); Ludwig et al. (2023); Dordonnat et al. (2016), to the best of our knowledge, no existing model assesses a cross-country dependent impact of meteorological risk alongside socio-economic and political risks on electricity load. Recent advancements in the load forecasting literature encompass hybrid models combining diverse forecasting methods, as summarized by Petropoulos et al. (2022) and Hong et al. (2016). Examples include studies by De Vilmarest et al. (2024); Lu et al. (2023); Dudek (2022) that utilized among others combinations with Generalized Additive Models (GAMs).

Building upon early GAM-based methods in load forecasting, see e.g. Pierrot & Goude (2011); Fan & Hyndman (2012); Goude et al. (2014), the effectiveness of GAMs in probabilistic load forecasting is now well-established, as summarized by Antoniadis et al. (2024). For instance, winning methods in the IEEE DataPort Competition by De Vilmarest & Goude (2022) employed GAMs within model ensembles. Similarly, GAMs consistently ranked highly in previous forecasting competitions, securing top positions in GEFCom 2014 Hong et al. (2016), see Gaillard et al. (2016); Dordonnat et al. (2016).

By applying linear model structures to non-linear functions along with discretization techniques, GAMs remain interpretable and efficient in estimation while capturing non-linear relationships Lepore et al. (2022); Wood (2017). Motivated by this combination of attributes and the success of GAMs in both forecasting competitions and recent research on probabilistic load forecasting, see e.g. Gilbert et al. (2023); Browell & Fasiolo (2021), we adopted this framework as the foundation for our proposed model.

***Contributions***. Specifically, our hourly mid-term forecasting approach combines interpretable GAMs in smoothed temperatures incorporating a climate trend, non-stationary socio-economic and political state variables and calendar effects, with autoregressive post-processing. It creates probabilistic forecasts by ensembling load trajectories with multivariate simulated temperature, socio-economic and political state, and autoregressive forecasts as inputs. With our approach, we contribute five key innovations to the field of probabilistic mid-term load forecasting:

(i) **High-Resolution Interpretable Model:** We develop a GAM-based probabilistic forecasting model with a comprehensive set of high-resolution stochastic and deterministic load drivers as inputs. We apply the model to track the hourly load impact of each input at gigawatt precision across risk scenarios decomposing the uncertainty in mid-term load forecasting. This interpretability makes the model a valuable tool for risk assessments and to support the decision-making processes of financial and industry operators.

(ii) **Nuanced Modelling of Transnational Dependencies:** We employ a nuanced multivariate model that accounts for country-specific load characteristics, such as seasonalities, holidays and temperature sensitivity, alongside pan-European dependencies due to underlying meteorological, socio-economic and political conditions.

(iii) **Unit-Root Socio-Economic and Political Effects:** To capture the gradual changes in mid-term load levels driven by socio-economic and political factors, our model integrates multivariate state variables that represent aggregated socio-economic and political conditions. These variables are modeled with persistent trend behavior, specifically a unit root, and are compared to modeling approaches assuming stationarity.

(iv) **Comprehensive Robustness Check:** The robustness of our multivariate model is demonstrated through an extensive evaluation study incorporating 24 European countries[1], illustrated by Figure 1, for more than 9 years (2015-2024) including volatile periods due to the COVID-19 pandemic and the energy crisis caused by the Russian invasion of Ukraine.

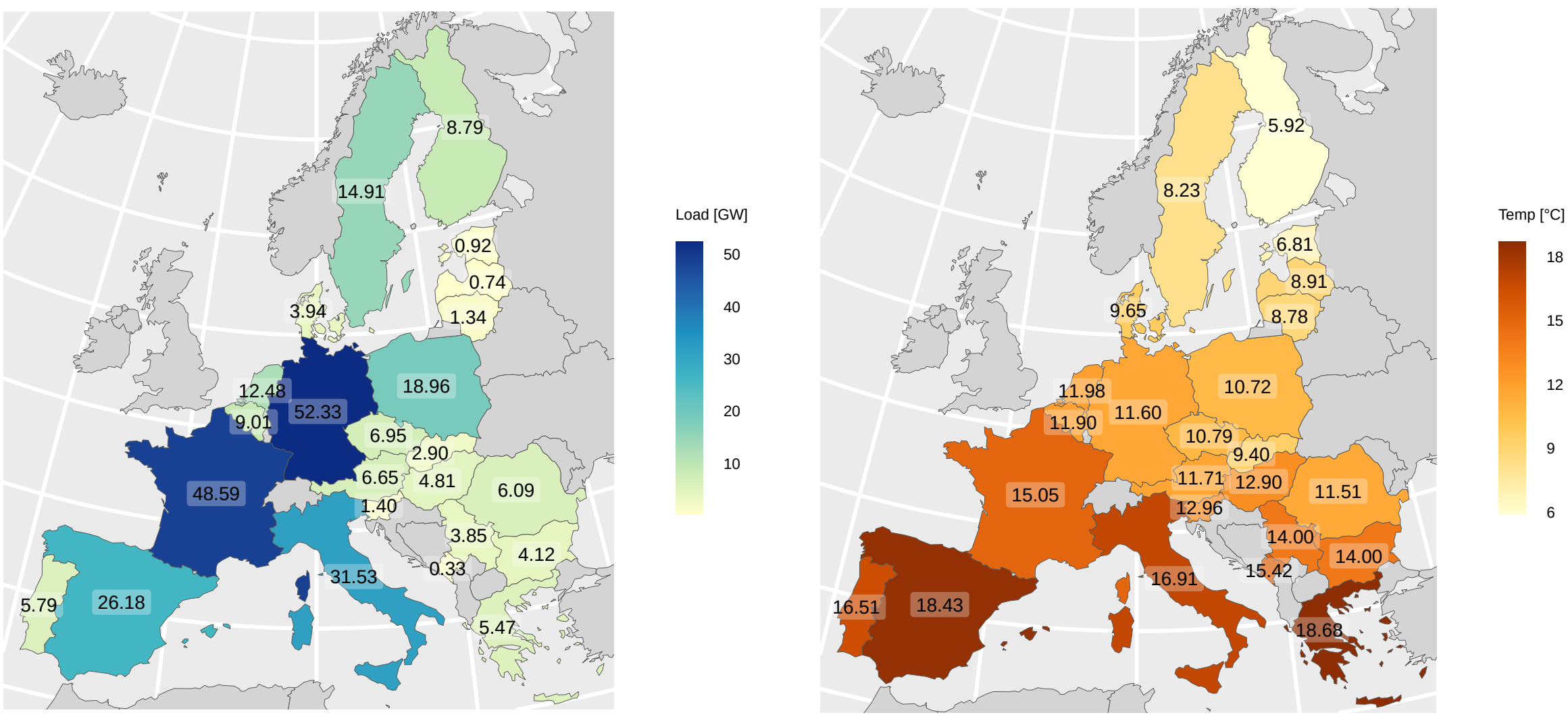

(a) Average load in GW

(b) Average temperature in °C

Figure 1: Average load and temperature of 2023 in the 24 European countries considered in the study.

***Results.*** In the evaluation study using over 9 years of data (2015-2024) our forecasting methodology demonstrated significantly improved CRPS compared to benchmarks based on seasonal random walks and FNN. Comparing three types of models for the socio-economic and political state with differing assumptions on an underlying unit root and cross-country dependencies, a model assuming a single cointegrated socio-economic and political unit root across Europe performed best for a majority of considered countries.

[1] While data from all European countries participating in ENTSOE was initially collected, some countries were ultimately excluded from the analysis due to data quality issues.

***Paper structure .*** The remainder of this paper is organized as follows: Section 2 details the multifaceted deterministic and stochastic effects on load. Section 3 introduces the applied statistical models, with a concise theoretical background on GAMs and probabilistic vector autoregressive and state models in 3.1 and 3.2, respectively. Section 4 begins with a conceptual overview of our model in 4.1 and details our modeling, estimation and forecasting specifications in 4.2. Benchmark models are detailed in Section 5. The forecasting study and evaluation design are explained in Section 6. The results of the comprehensive forecasting study are discussed and risk scenarios are interpreted in Section 7. Finally, Section 8 concludes and outlines further research to be built upon this work. For an elaborate explanation of the corresponding point forecasting model, we refer to Zimmermann & Ziel (2024).

## 2. Deterministic and Stochastic Effects on Load

Adapting a classification proposed by Pierrot & Goude (2011), the various facets explaining load can be consolidated into four main groups:

(i) **Deterministic Calender Effects**: These encompass the calender-based behavior of modern western societies entailing repetitive patterns of different seasonalities in electricity load, e.g. yearly patterns (higher load levels during winter vs. lower load levels during summer), weekly patterns (higher load levels on weekdays vs. lower load levels on weekends), and daily patterns (higher load during the day vs. lower load during the night) and holiday patterns.

(ii) **Stochastic Meteorological Effects**: These include air temperature, humidity, cloud cover, wind speed and climate change that affect load due to their impact on electric heating, cooling and lighting.

(iii) **Stochastic Socio-Economic and Political Effects**: These involve macroeconomic, socio-economic and energy variables along with political incentives, e.g. economic growth, population size, fossil fuel prices affecting industrial production or government subsidies for decarbonization. These influence, in particular in their transnational unit root behavior, mid to long-term (several months to years) levels of load.

(iv) **Stochastic Autoregressive Effects**: These include remaining short to mid-term (hours to several weeks) autoregressive deviations in load time series.

In the following sections, we illustrate these load drivers for Germany and France, selected due to their high electricity demand (see Fig. 1). Additionally, the strong dependence of load on temperature in France provides initial insights into the potential impact of meteorological risk on load.

The load data used in the subsequent were retrieved from ENTSOE, cover the period from January 1st, 2015 to February 17th, 2024 and are available for 24 countries[2] in an hourly resolution. Hourly temperature observations are collected for each considered country from meteostat.net (Meteostat Developers (2024)) starting from January 1st, 1990, to February 17th, 2024, at weather stations located within a 100 km geodesic radius of the five largest cities in each country. Holiday information was collected from Nager.Date, see Hager (2024), is available from January 1st to December 31st of 2000 to 2030 and encompasses both the label and the day of occurrence of the holiday. Calendar and time information, including daily, weekly, and annual seasonalities, changes due to daylight saving time and leap years are incorporated implicitly in the load time series. In all time series we adjust for daylight saving time changes.

[2]While data from all European countries participating in ENTSOE was initially collected, some countries were ultimately excluded from the analysis due to data quality issues. This resulted in a final dataset encompassing data from 24 European countries: Austria, Belgium, Bulgaria, Czech Republic, Germany, Denmark, Estonia, Spain, Finland, France, Greece, Hungary, Italy, Lithuania, Latvia, Montenegro, Netherlands, Poland, Portugal, Romania, Serbia, Sweden, Slovenia, Slovakia.

| Load Effect | Spatial Level | Type of Uncertainty |
|---|---|---|
| calendar | national | deterministic |
| meteorological | **national-**(transnational) | stochastic (multiple seasonalities) |
| socio-economic and political | (national)**-transnational** | stochastic (non-stationary with unit root ) |
| autoregressive | **national-**(transnational) | stochastic |

Table 1: The various facets explaining load, categorized in terms of their spatial level and type of uncertainty.

### *2.1. Calender Effects*

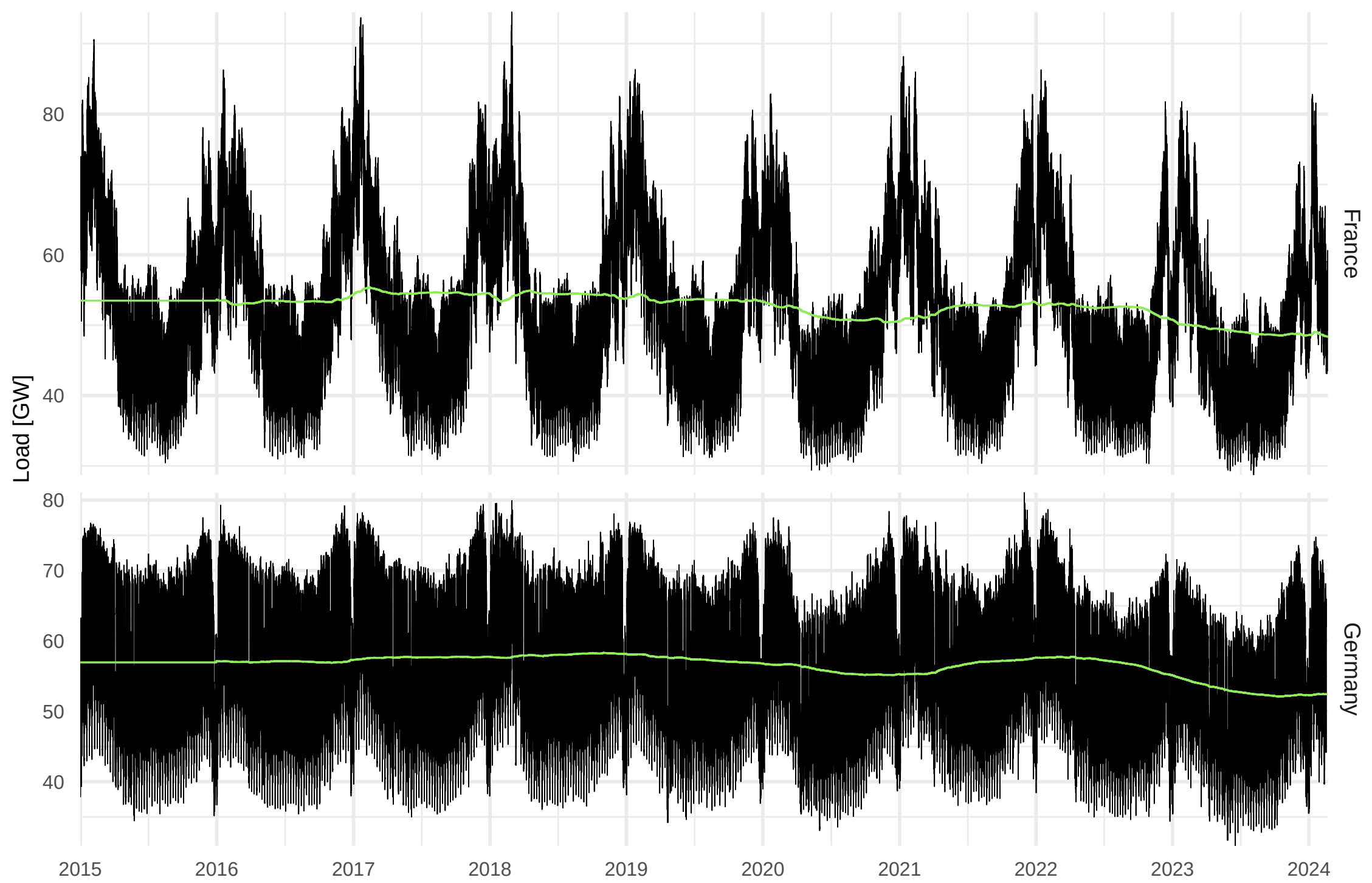

Figure 2: Hourly load (black) and yearly, i.e. 52 weeks, moving average load (green) from January 1st, 2015 to February 17th, 2024.

The load profiles exhibit strong seasonal patterns across daily, weekly, and annual cycles, as shown by Figures 2 and 3. These patterns include lower load levels during summer compared to winter months, higher loads on weekdays compared to weekends, and lower loads at night compared to daytime. Additionally, on holidays load is reduced due to decreased work and industrial activity. This disruption to the typical weekly load pattern is evident in Figure 3, which shows the load time series for the Easter holiday time in Germany. Moreover, from Figure 2 we observe, that during the entire Christmas period load levels drop. Thus, besides single-day holidays, a winter holiday period encompassing December 18th to January 6th is considered in every country.

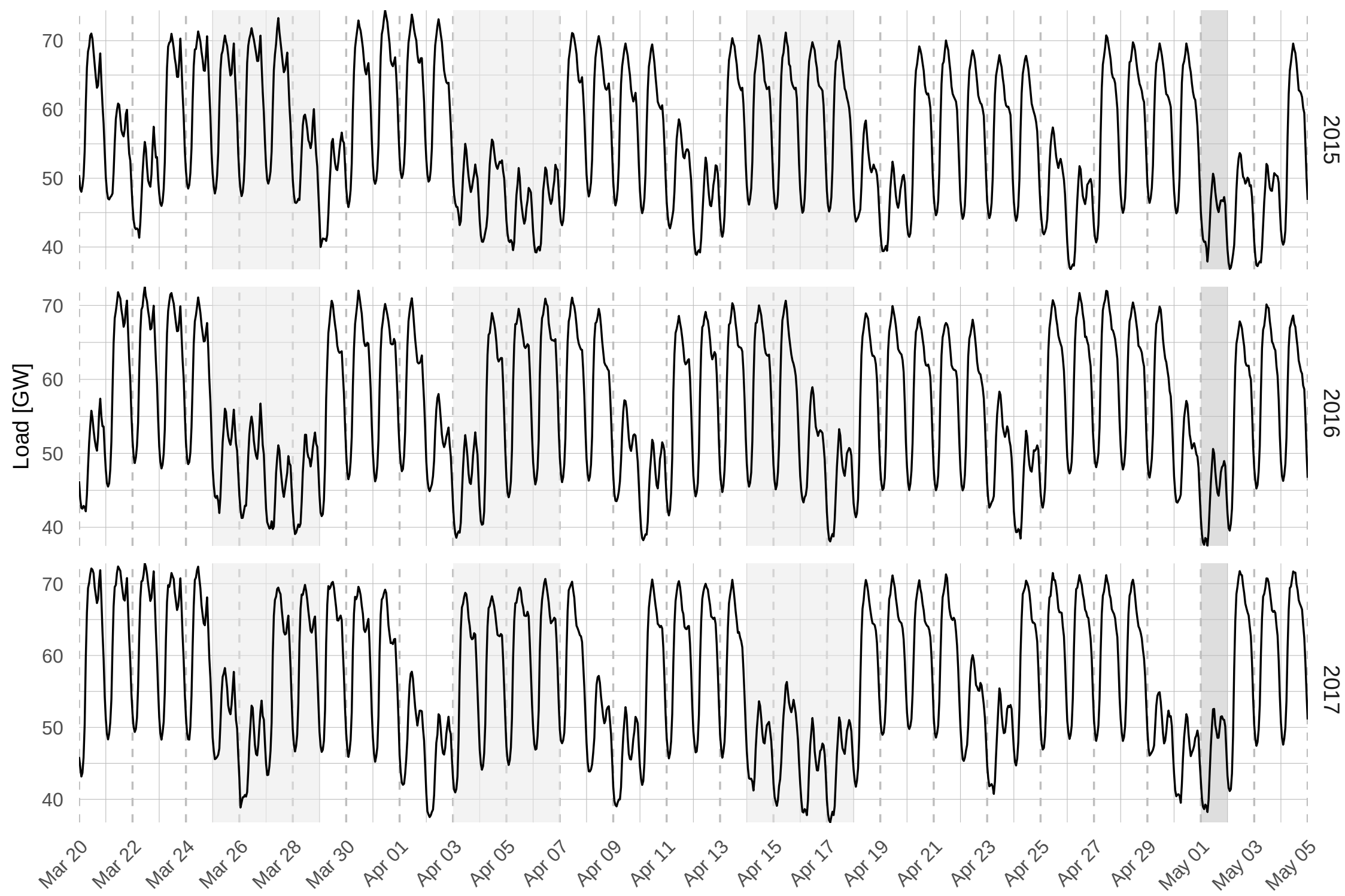

Figure 3: Hourly load in Germany in the Easter holiday time (March 1st to May 10th in 2016, 2017 and 2018) with holidays shaded in grey.

All calendar-based effects on load are known in advance, classifying them as deterministic. Their spatial level is national, as holidays are typically defined on the country level (see Tab. 1, l. 1).

### *2.2. Meteorological Effects*

Meteorological conditions, in particular, temperature significantly impact load by affecting electric heating and cooling, with delayed effects due to the thermal inertia of buildings and human response time, see Tian et al. (2024). Typically, during summer, higher temperatures lead to increased load as cooling demands rise. Conversely, in winter, lower temperatures result in higher loads due to heating demands. This non-linear relationship between temperature and load has been confirmed through various studies, see e.g. the overview by Verwiebe et al. (2021); Davis et al. (2016) and the comprehensive European study by Bessec & Fouquau (2008) or Bashiri Behmiri et al. (2023); Moral-Carcedo & Pérez-García (2019); Ziel (2018); Xie et al. (2018) and is evident in Figures 4 for France. Notably, the sensitivity of load to temperature is low in Germany.

The uncertainty in meteorological time series is stochastic with pronounced seasonal patterns, includ-ing weekly and annual cycles, and an increasing trend due to climate change see Figures 4-5. While short-term temperatures are primarily determined by regional weather conditions, their long-term underlying trend due to climate change evolves on transnational levels. Furthermore, mid-term prevailing weather systems, such as high and low-pressure areas, can influence weather patterns across multiple countries in Europe. Consequently, meteorological load effects are considered to have a national spatial level with transnational components (see Tab 1, l. 2).

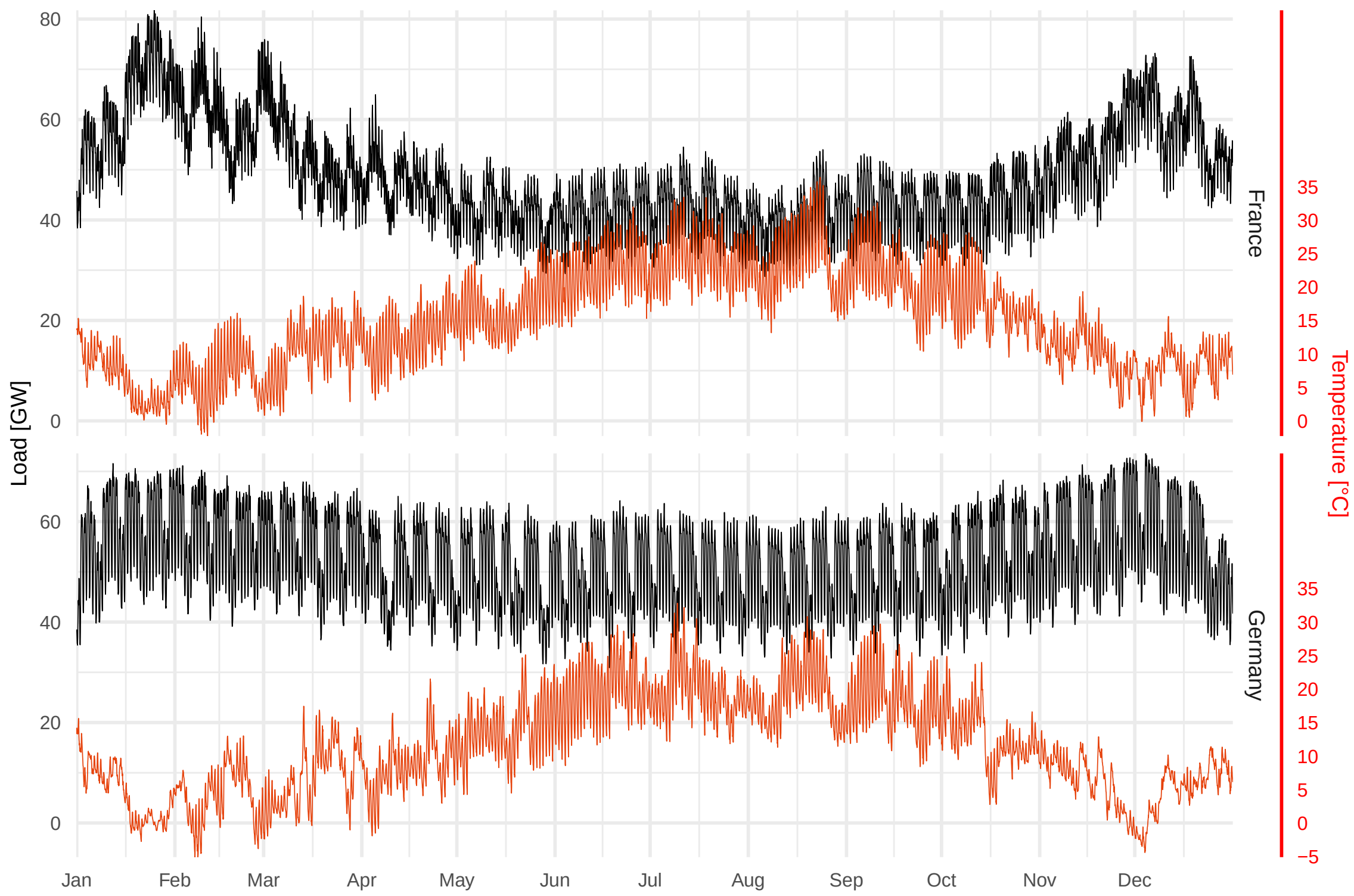

Figure 4: Hourly load (black) and temperature (red) in 2023.

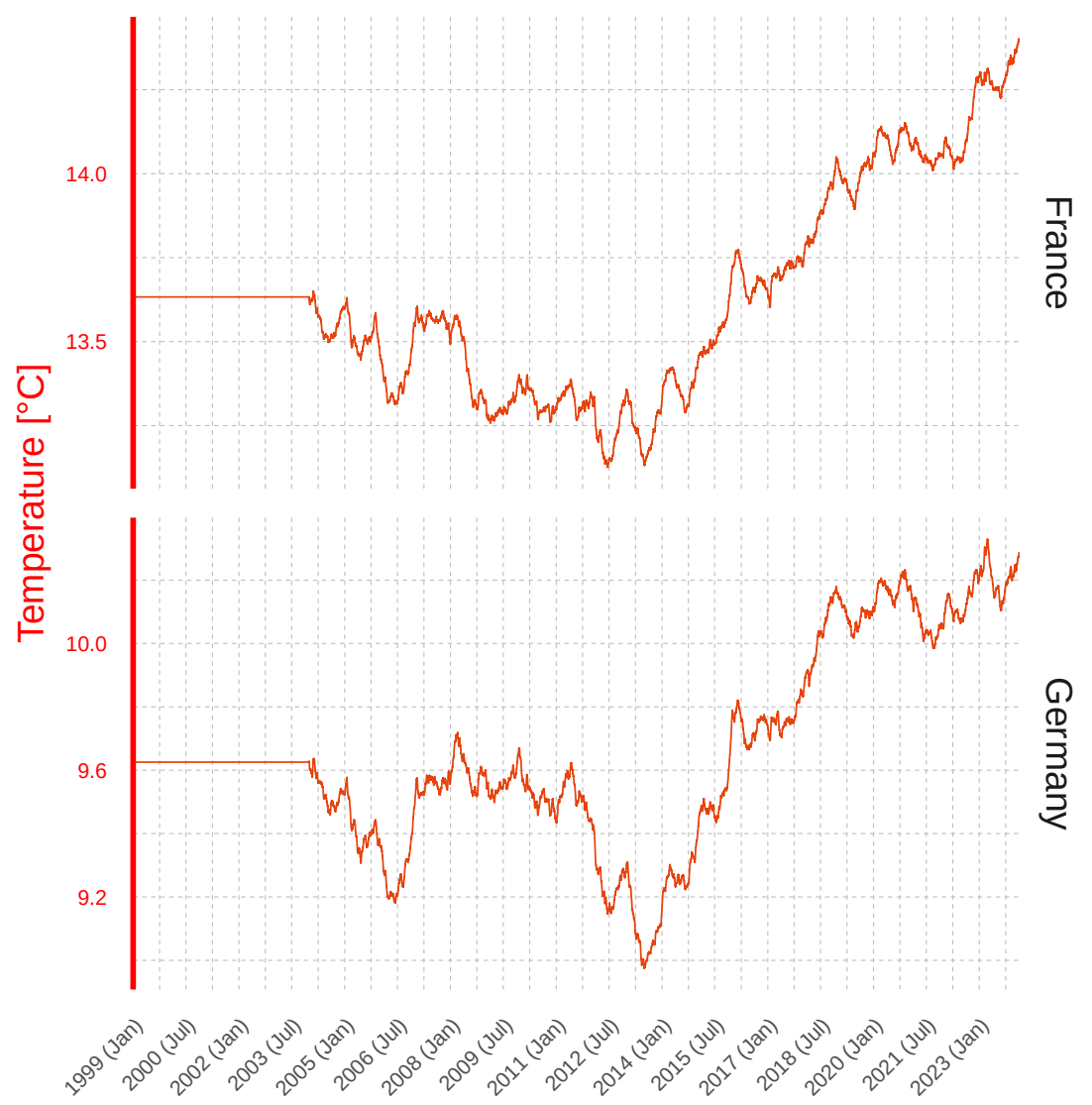

Figure 5: Five year, i.e. $5 \times 52$ weeks, moving average temperature calculated from Jan. 1st 1990 to Feb. 17th, 2024.

In our probabilistic modeling approach, we capture this spatial property of meteorological effects by estimating point forecast models for temperatures at the country level but incorporating cross-country correlations to simulate forecasts. By this, we account for the risk of Pan-European extreme weather scenarios on load.

### *2.3. Socio-Economic and Political Effects*

Load is influenced by past observations in terms of its previous mid-term level. More precisely, random shocks on load, e.g. due to transnational factors like a financial crisis, carbon tax, decarbonization incentives, technological advancements, supply shortages, or large-scale migra-tion caused by war, can cause permanent deviations from a predetermined equilibrium level, see González Grandón et al. (2024). Since the stochastic component in electricity demand is strongly tied to the economy, load inherits the prevalence of these shocks via transmission flows from key macroeconomic, socio-economic and energy variables. Additionally, political legislation and incentives inducing prevailing structural changes influence equilibrium load levels either directly or through the beforementioned variables Schneider & Strielkowski (2023); Narayan & Liu (2015); Hendry & Juselius (2000); Smyth (2013).

Key macroeconomic drivers of electricity demand are, for instance, Gross Domestic Product, employment rate, Consumer Price Index and Industrial Production Index measuring economic well-being, purchasing power, and production output of manufacturing and utilities, see e.g. González Grandón et al. (2024). These macroeconomic variables, with GDP being a prime example, see Kalhori et al. (2022); Harvey et al. (2007), are generally recognized to exhibit persistent shifts in their level component. This has been, firstly, evidenced by the seminal work of Nelson & Plosser (1982) and mostly supported by revisiting studies, see e.g. Gil-Alaña & Robinson (1997); Perron (1997) among others. Socio-economic and energy variables affecting load are, for instance, population size and fossil fuel prices or carbon taxes through industrial production, see e.g. González Grandón et al. (2024); Moral-Carcedo & Pérez-García (2019). Similar to macroeconomic variables, these factors are predominantly recognized to exhibit a permanent shift in levels.

Recent examples of political impacts on equilibrium load include load level increases driven by incentives promoting electrified heating and transportation to meet decarbonization goals, and load level decreases due to politically mandated energy-saving measures resulting from the Russian invasion of Ukraine and COVID-19 lockdowns Campagne et al. (2024).

Following the illustration of load levels in Ziel (2019), Figure 2 depicts hourly load beside the yearly moving average load in France and Germany. In both countries, the averaged load data exhibit gradual shifts, in particular, in the last four years.

Prevailing shifts in the level component of load make this process non-stationary. Recall that a non-stationary process exhibits fluctuations in its statistical properties, such as the mean, over time. By differencing the level time series, it can be stabilized. This form of non-stationarity is known as a unit root[3].

Despite strong cross-border economic interdependence and a rise in pan-European legislation, national differences in the socio-economic and political state reflected in load time series prevail. For instance, European countries handled the COVID-19 pandemic with varying approaches.

In our methodology, we account for this spatial property (see Table 1, l. 3) of the socio-economic and political effect on load by modeling it multivariate in all considered countries. Specifically, the models applied and compared make different assumptions about the cross-country dependence of the underlying unit root. In addition to a stationary model used as a baseline for comparison, we evaluate a cointegrated model and an individual unit root model. The cointegrated model assumes a single underlying pan-European socio-economic and political unit root, while the individual model assumes a country-specific unit root with a pan-European memory of past observations.

[3] Our study compares unit root and non-unit root modeling of the socio-economic and political levels in load, deliberately bypassing formal unit root testing due to the ongoing research in this area, see e.g. Schneider & Strielkowski (2023), and the statistical complexities inherent in applying such tests to electricity time series. For a comprehensive assessment of unit root behavior in electricity time series, we recommend the literature reviews by Schneider & Strielkowski (2023); Smyth (2013). For methodological approaches to measure the effect of economic and political shocks on energy variables, we refer to Schneider & Strielkowski (2023) and Narayan & Liu (2015).

### *2.4. Autoregressive Effects*

With weather and previous mid-term socio-economic and political effects on load explained, short to mid-term (hours to several weeks) autoregressive deviations persist in the calendar-based load pattern. Commonly, instantaneous external influences can impact load beyond their timeframe. For instance, even if weather conditions affecting heating or cooling subside, their impact on human behavior persists for several days, causing deviations in load patterns for the most recent lags, as well as lags of neighboring hours on the previous day and the previous week. Furthermore, major sporting events like the Football World Cup or the Olympics can cause short-term deviations in load patterns during match times and the hours surrounding these events due to increased viewership and associated activities.

Autoregressive effects are stochastic and mostly national in their spatial level. Thus point forecast autoregressive models are estimated at the country level. However, since the initiating causes of short-term external load effects can transcend national borders, inter-country correlations are incorporated to simulate forecasts.

### *2.5. The Confluence of Load Effects*

Disentangling and interpreting the combined effect of the various factors impacting load presents a significant challenge in mid-term hourly load forecasting. Figure 6 provides an intuition of this complexity by illustrating hourly load and temperatures in France during the first COVID-19 lockdown. We observe a significant drop in load levels coinciding with a slight temperature increase in the days leading up to and at the beginning of the lockdown. During this time it is difficult to disentangle the load-reducing effect of rising temperatures from the socio-economic lockdown effect, which likely caused the primary load reduction.

Further complicating the picture are several holidays within the lockdown period: Good Friday (April 10th), Easter Monday (April 13th), Labor Day (May 1st), and Victory in Europe Day (May 8th) (shaded in dark grey). These holidays presumably caused additional load reductions, making it even harder to isolate the individual effects. For example, on Easter Monday, the temperature decrease, which typically increases the load, might have been masked by the combined effects of reduced activity due to the holiday and the ongoing lockdown.

## 3. Primer On Statistical Models

### *3.1. Generalized Additive Models*

Generalized Additive Models (GAM) are the main building block to combine the multifaceted stochastic and deterministic characteristics of load and form our forecasting model. Specifically, we choose GAMs that describe a response variable $Y_t$ as an additive combination $\mu$ of smooth cubic B-splines $f_i : \mathbb{R}^m \longrightarrow \mathbb{R}$ $f_i(\boldsymbol{X}) = \sum_{l=1}^{k_1} ... \sum_{j=1}^{k_m} \beta_{l,...,j} b_l^1(x^1) \cdot ... \cdot b_j^m(x^m)$ in the covariates $\boldsymbol{X}_t = (X_t^1, ..., X_t^m) \in \mathbb{R}^m$, $t \in \{1, .., T\}$:

$$Y_t = o + \mu(\boldsymbol{X}_t) + \varepsilon_t, \quad \mu(\boldsymbol{X}_t) = f_1(\boldsymbol{X}_t) + f_2(\boldsymbol{X}_t) + ... + f_M(\boldsymbol{X}_t) \tag{1}$$

where $o \in \mathbb{R}$, $\varepsilon_t \sim \mathcal{N}(0, \sigma_t)$.

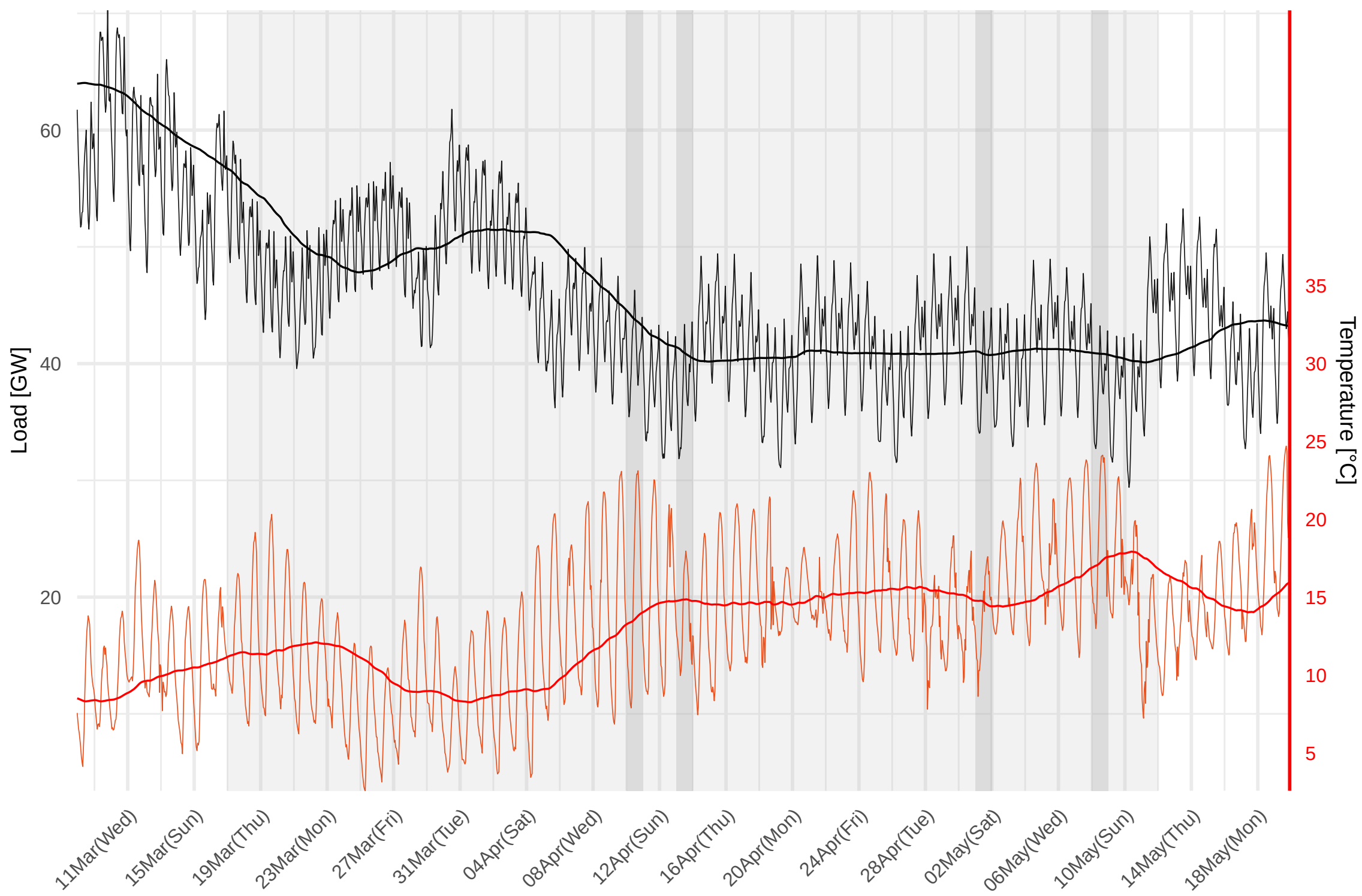

Figure 6: Hourly load (black) and temperature (red) and their weekly moving average in France from March 14th to May 19th, 2020 with the time of the first lockdown due to COVID-19 shaded in grey and holidays shaded in dark grey.

For estimation[4] of $\beta_i$ we choose a penalized least-squares objective with smoothing parameters $\lambda_i \in \mathbb{R}^m_{\geq 0}$ and a difference-based penalty on parameter deviation:

$$||\boldsymbol{Y} - \mathbb{X}\boldsymbol{\beta}||^2 + \sum_{i=1}^{M} \lambda_i \mathcal{P}(f_i), \tag{2}$$

$$\mathcal{P}\left(\sum_{j=1}^{k} \beta_j b_j(x)\right) = \sum_{j=1}^{k-p} (\Delta^p \beta_j)^2 = \boldsymbol{\beta}^T S \boldsymbol{\beta}, \tag{3}$$

where $\mathbb{X}_t = (\mathbb{X}_{t,1}^1 \otimes ... \otimes \mathbb{X}_{t,m}^1, ..., \mathbb{X}_{t,1}^M \otimes ... \otimes \mathbb{X}_{t,m}^M)$ for the Kronecker product $\otimes$ on matrix space, $\mathbb{X}_{t,j}^l = (b_1^j(X_t^j), ..., b_{k_j}^j(X_t^j))$ and $\Delta$ is the difference operator. The resulting smooth terms are named *P-splines*. For a so-called cyclic P-spline, terms are added to penalize deviation between the first and last coefficients, e.g. $(\beta_1 - 2\beta_n + \beta_{n-1})^2$ and $(\beta_2 - 2\beta_1 + \beta_n)^2$ for $p = 2$. The generalization of $\mathcal{P}$ to multivariate splines penalizes parameter deviation individually for each marginal smooth, where remaining covariates are integrated out and integrals are approximated or derived as sums.

[4]Note that smooth functions $f_1, ..., f_M$ in (1) can only be identified and thus estimated up to a constant. Consequently, practical applications employ identifiability constraints, commonly $\sum_{t=1}^{T} f_i(\boldsymbol{X}_t) = 0$ or $f_i(0) = 0$, see Wood (2017) for details.

To estimate the smoothing parameters $\lambda_i$, we choose to optimize the REML criterion. For this differently efficient iteration schemes with estimation of $\beta_i$ by minimization of (2) can be applied. For further insights on GAM estimation, we refer to Wood (2017) or Lepore et al. (2022). GAM modeling and estimation methods are implemented in the R-package mgcv, see Wood (2017).

### *3.2. Probabilistic Vector Additive Autoregressive and State Models*

To probabilistically forecast the stochastic drivers of load, three types of models are applied: Vector Autoregression (VAR), Vector Error Correction (VECM) and Vector Error-Trend-Seasonal (VETS) Models, whereby VAR and VETS models are used in their univariate and multivariate form. In all three models, forecasts are generated by the dependence on past observations, specifically as weighted sums of autoregressive, differenced autoregressive and initial state components.

| VAR($\boldsymbol{\phi}; p_{\max}$) | VECM($\boldsymbol{\Gamma}, \boldsymbol{\Pi}; r < n$) |
|---|---|
| $\boldsymbol{Y}_t = \sum_{k\in\mathcal{S}} \boldsymbol{\phi}_k \boldsymbol{Y}_{t-k} + \boldsymbol{\epsilon}_t$ | $\Delta\boldsymbol{Y}_t = \boldsymbol{\Gamma}\Delta\boldsymbol{Y}_{t-1} + \boldsymbol{\Pi}\boldsymbol{Y}_{t-1} + \boldsymbol{\epsilon}_t$ |
| $\widehat{\boldsymbol{Y}}_{T+h\mid T} = \sum_{k\in\mathcal{S}} \boldsymbol{\phi}_k \widehat{\boldsymbol{Y}}_{T+h-k\mid T} + \boldsymbol{\epsilon}_{T+h}$ | $\Delta\widehat{\boldsymbol{Y}}_{T+h\mid T} = \boldsymbol{\Gamma}\Delta\widehat{\boldsymbol{Y}}_{T+h-1\mid T} + \boldsymbol{\Pi}\widehat{\boldsymbol{Y}}_{T+h-1\mid T} + \boldsymbol{\epsilon}_{T+h}$ |
| for $\boldsymbol{\epsilon}_{T+h} \sim \mathcal{N}(0, \boldsymbol{\Sigma})$ | for $\boldsymbol{\epsilon}_{T+h} \sim \mathcal{N}(0, \boldsymbol{\Sigma})$ |
| $\widehat{\boldsymbol{Y}}_{T+h-k\mid T} = \boldsymbol{Y}_{T+h-k}$ for $h \leq k$ | $\Delta\widehat{\boldsymbol{Y}}_{T+1\mid T} = \boldsymbol{\Gamma}\Delta\boldsymbol{Y}_T + \boldsymbol{\Pi}\boldsymbol{Y}_T$ |

Table 2: VAR model for a set of lags $\mathcal{S} \subset \{1, ..., p_{\max}\}$. VECM for $\boldsymbol{\Pi} = \alpha\beta^\top$ and $\beta^\top \boldsymbol{Y}_{t-1}$ a $r \times 1$ vector of stationary cointegrated relations.

For the multivariate autoregressive, thus VAR model (see Tab. 2, col. 1), maximum likelihood is applied for estimation and the set of lags is fixed to $\mathcal{S} = \{1, 2\}$. For the univariate autoregressive model (see Tab. 2, col. 1, for $\boldsymbol{Y}_t, \boldsymbol{\epsilon}_t, \boldsymbol{\phi}_t^\top \in \mathbb{R}^n, n = 1$) Post-Lasso OLS is applied, i.e. lags $p \in \mathcal{S} \subset \{1, ..., p_{\max}\}$ are chosen by Lasso and OLS estimation is carried out for the resulting non-zero lags for bias reduction, see Lee et al. (2016).

For the VECM model (see Tab. 2, col. 2) maximum likelihood is applied for estimation. The reduced rank $r < n$ of the matrix $\boldsymbol{\Pi}$ captures the cointegrated unit root behavior of $\boldsymbol{Y}_t \in \mathbb{R}^n$. Cointegration vectors are represented by the columns of the matrix $\beta$. The remaining unit roots $n - r$ can be interpreted as joint underlying drivers of the non-stationary trend behaviors in $\boldsymbol{Y}_t$, see Jusélius (2009).

| VETS($\alpha$) | VETS($\alpha, \gamma; m$) |
|---|---|
| $\boldsymbol{Y}_t = \boldsymbol{l}_{t-1} + \boldsymbol{\epsilon}_t$ | $\boldsymbol{Y}_t = \boldsymbol{l}_{t-1} + \boldsymbol{s}_{t-m} + \boldsymbol{\epsilon}_t$ |
| $\boldsymbol{l}_t = \boldsymbol{l}_{t-1} + \alpha\boldsymbol{\epsilon}_t$ | $\boldsymbol{l}_t = \boldsymbol{l}_{t-1} + \alpha\boldsymbol{\epsilon}_t$ |
| $\widehat{\boldsymbol{Y}}_{T+h\mid t} = \boldsymbol{l}_T + \boldsymbol{\epsilon}_{T+h}$ for $\boldsymbol{\epsilon}_{T+h} \sim \mathcal{N}(0, \boldsymbol{\Sigma})$ | $\boldsymbol{s}_t = \boldsymbol{s}_{t-m} + \gamma\boldsymbol{\epsilon}_t$ |
| | $\widehat{\boldsymbol{Y}}_{T+h\mid t} = \boldsymbol{l}_T + \boldsymbol{s}_{T+h-m\lceil \frac{h}{m} \rceil} + \boldsymbol{\epsilon}_{T+h}$ for $\boldsymbol{\epsilon}_{T+h} \sim \mathcal{N}(0, \boldsymbol{\Sigma})$ |

Table 3: Additive VETS models with level states $\boldsymbol{l}_t \in \mathbb{R}^n$, seasonal states $\boldsymbol{s}_t \in \mathbb{R}^n$, common smoothing parameters $\alpha, \gamma \in (0, 1)$ and periodicity $m$.

While for VAR models no assumption is imposed on the weights assigned to past observations, additive

VETS models (see Tab. 3, col. 1-2), formulated by De Silva et al. (2010) and specified by Svetunkov et al. (2023), impose exponentially diminishing weights. This is obtained by applying exponential smoothing with smoothing equations for the additive decomposition of the univariate time series $Y_t^1, \ldots, Y_t^n$ into unobserved states, i.e. level $l_t^1, \ldots, l_t^n$ and seasonal $s_t^1, \ldots, s_t^n$ component[5]. From the model formulation, it directly follows that each resulting univariate time series exhibits an individual non-stationary behavior of unit root type, i.e. by first-order and seasonal differencing a stationary white noise process remains. Thereby, the exponential decay of weights assigned to past observations is governed for all univariate time series by common smoothing parameters $\alpha, \gamma \in (0, 1)$. This can be interpreted as a common memory of deviations from equilibrium levels. For estimation, the implementation by Svetunkov et al. (2023) is applied and the trace of the covariance matrix of errors is minimized.

Probabilistic forecasts from the three models are obtained iteratively by sampling multivariate normal error terms $\boldsymbol{\epsilon}_t \sim \mathcal{N}(0, \boldsymbol{\Sigma}_s)$ assuming zero mean. For this, hour-specific covariance matrices $\boldsymbol{\Sigma}_s \in \{\boldsymbol{\Sigma}_0, \ldots, \boldsymbol{\Sigma}_{23}\}$ are estimated from the corresponding in-sample residuals in the first-step estimation

$$\widehat{\boldsymbol{\Sigma}}_s = \frac{1}{|T^s| - 1} \sum_{i \in T^s} \widehat{\boldsymbol{\epsilon}}_i \widehat{\boldsymbol{\epsilon}}_i^\top \text{ for } T^s = \{i \in \{1, \ldots, T\} | i \bmod 24 = s\}. \tag{4}$$

For models with resolution $\tau$ reduced to a weekly frequency, constant covariance is assumed: $\boldsymbol{\epsilon}_\tau \sim \mathcal{N}(0, \boldsymbol{\Sigma})$ with estimated covariance

$$\widehat{\boldsymbol{\Sigma}} = \frac{1}{|T^\tau| - 1} \sum_{i \in T^\tau} \widehat{\boldsymbol{\epsilon}}_i \widehat{\boldsymbol{\epsilon}}_i^\top. \tag{5}$$

For the univariate VAR, the LASSO-based selection $\mathcal{S}$ of relevant lags in $\{1, \ldots, p_{\max}\}$ enables to implement each forecasting iteration with sparse matrix algebra, which greatly reduces calculation times.

## 4. Methodology

### *4.1. Modelling Concept*

As detailed in Section 2.1, disentangling the combined influence of the various deterministic or stochastic and national or transnational load effects and obtaining interpretable models, in particular, in stochastic load effects, presents a significant challenge in mid-term hourly load forecasting. To address this challenge, we propose an interpretable model based on GAMs that integrates probabilistic VETS, VAR and VECM models to capture stochastic effects. The general process of the model is illustrated by the flowchart in Figure 9 and described in the subsequent. In this section, we will focus on the probabilistic components of the model. A detailed explanation of the underlying point forecasting model and parameter definitions can be found in our previous work, see Zimmermann & Ziel (2024).

Let $\boldsymbol{Y}_t \in \mathbb{R}^n$ and $\boldsymbol{X}_t^{\text{Temp}} \in \mathbb{R}^n$ (see Tab 4, l. 2) be the observed load and temperature in $n$ considered countries at time $t = 1, \ldots, T$ and $t = -C, \ldots, T$, respectively. Additionally, we have access to deterministic calender-based information about seasonalities and holidays $\boldsymbol{X}_t^{\text{Cal-based}}, \boldsymbol{X}_t^{\text{Season}} \in \mathbb{R}^n$ (see Tab 4, l. 1, l. 3) for the extended horizon up to $t = T + H$. It is the objective to probabilistically forecast load over a mid-term prediction horizon $H$ of several weeks to one year by the modeling equations (6)-(8) for countries $i = 1, \ldots, n$:

$$Y_t^i = o + \underbrace{\mu^{i,\text{Cal-based}}(X_t^{i,\text{Cal-based}})}_{\text{smooth terms in deterministic calendar effects}} + \underbrace{\mu^{i,\text{Temp}}(X_t^{i,\text{Temp}})}_{\text{smooth terms in stochastic meteorological effects (i)}} \tag{6}$$

$$+ \underbrace{\mu^{i,\text{SocEconPol-State}}(X_t^{i,\text{SocEconPol-State}})}_{\text{smooth terms in stochastic (socio-)economic and political effects (ii)}} \tag{7}$$

$$+ \underbrace{\mathcal{E}_t^i}_{\text{stochastic autoregressive effects (iii)}} \tag{8}$$

[5]In all VETS models considered in this work the trend component is assumed to be zero.

| Variable | Processed Information | Available Timeframe |
|---|---|---|
| $\boldsymbol{X}_t^{\text{Cal-based}}$ | hourly calender-based information, time of the day, week, year, interactions, holidays and holiday period | $t = 1, \ldots, T + H$ |
| $\boldsymbol{X}_t^{\text{Temp}}$ | hourly average temperatures in most populated cities | $t = -C, \ldots, T$ |
| $\boldsymbol{X}_t^{\text{Season}}$ | hourly time of the day, time of the year and interaction | $t = -C, \ldots, T + H$ |
| $\boldsymbol{X}_t^{\text{Trend}}$ | linear trend in the time index | $t = -C, \ldots, T + H$ |
| $\boldsymbol{X}_t^{\text{SocEconPol-State}}$ | underlying load level | $t = 1, \ldots, T$ |

Table 4: Input variables $\boldsymbol{X}_t^m \in \mathbb{R}^n$ of the proposed probabilistic load forecasting model.

The variable representing socio-economic and political states, $\boldsymbol{X}_t^{\text{SocEconPol-State}} \in \mathbb{R}^n$ (see Tab 4, l. 5), is obtained in step (ii) of the model and is therefore not available as initial information.

**(i) Probabilistic temperature forecast**: Since accurate weather and satellite imagery-based temperature forecasts are not available for mid-term horizons, we rely on deterministic seasonalities, a monotonous climatological trend and stochastic autoregressive components as inputs for probabilistic temperature modeling (see Figure 9 on the left-hand-side). To capture extreme meteorological events and climate change in observed temperatures $\boldsymbol{X}_t^{\text{Temp}}$, the in-sample data set of this model part comprises additional $C$ observations. Temperatures $\boldsymbol{X}_t^{\text{Temp}}$ are first smoothed by VETS models with fixed smoothing parameters. By this, we account for the thermal inertia of buildings and delayed human reactivity to heating demands. Additionally, smoothing dampens fluctuations not capturable by the available mid-term modeling inputs. To account for the primarily national spatial level of weather (see Tab. 1, l. 2), smoothed temperatures $\widetilde{\boldsymbol{X}}_t^{\text{Temp}}$ are then modeled by country-specific two-step GAMs in the deterministic seasonalities, i.e. the time of the day and the time of the year, and a monotonously non-decreasing climatological trend. To capture the stochastic autoregressive effect in temperatures, we employ univariate VAR models on the residuals of each country-specific two-step GAM. In resulting VAR residuals, high cross-country correlations, in particular, for neighboring countries (see Fig. 7) remain. This implies an underlying pan-European meteorological system affecting the temperatures in multiple countries. To account for this second spatial property of weather (see Tab. 1, l. 2), nationally modeled temperatures are probabilistically forecasted from sampled multivariate normal error terms with cross-country residual covariance.

**(ii) Probabilistic socio-economic and political state forecast**: To probabilistically forecast the

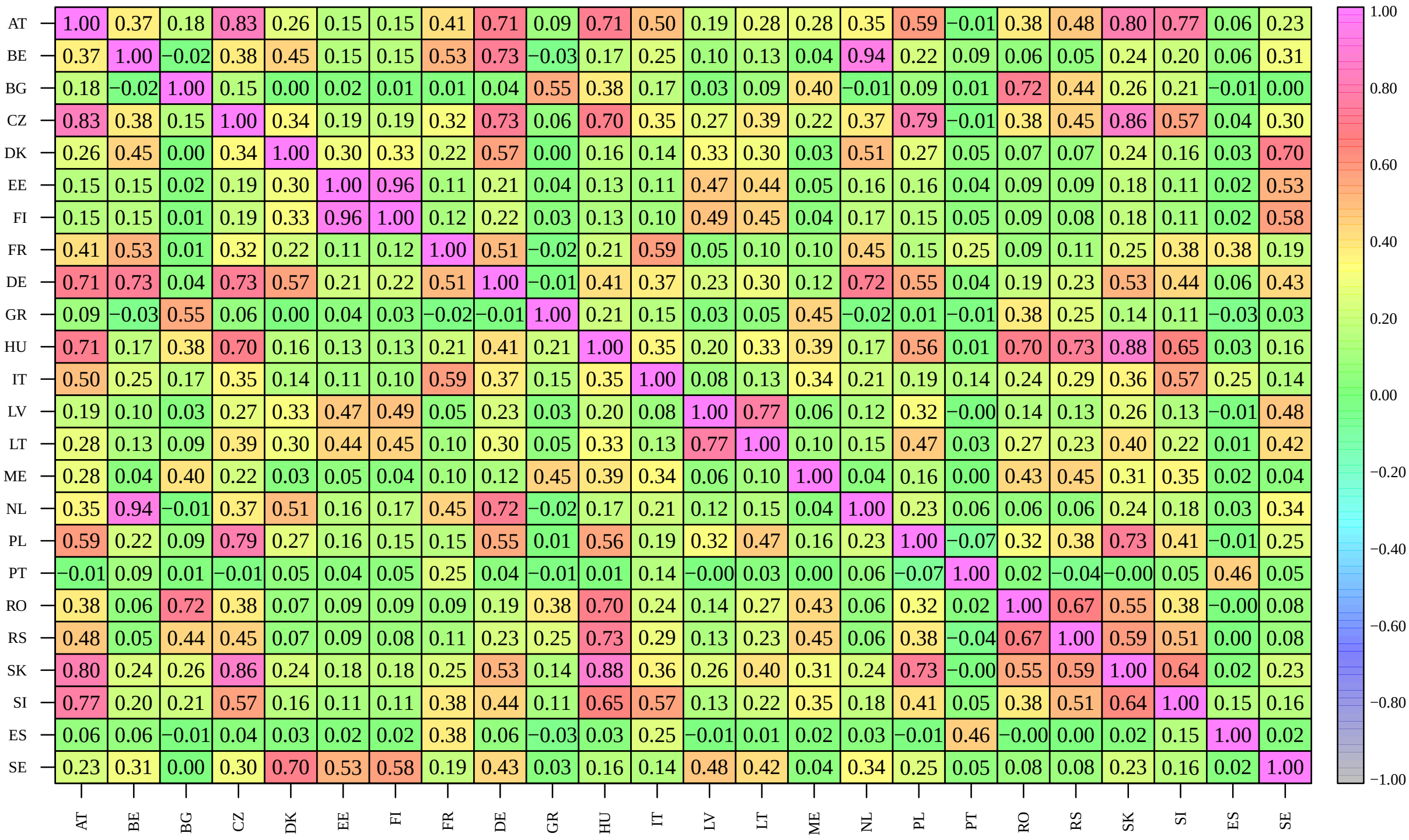

Figure 7: The in-sample correlation matrix $\widehat{\text{Corr}}(\boldsymbol{\epsilon}_t^{\text{Temp}})$ for $t \bmod 24 = 8$ from February 6th, 2019 to February 1st, 2023 and smoothing parameter $\alpha = 1/24$.

socio-economic and political state effect (see Figure 9 middle part), firstly, load is modeled country-specific by GAMs from the smoothed temperatures obtained in (i) and deterministic seasonal and holiday information. Since these GAMs do not capture the unit root socio-economic and political state effects, they remain in their residuals. Thus, secondary models are applied to extract these effects from the temporally aggregated residuals. To evaluate spatial and uncertainty properties of socio-economic and political state effects (see Tab. 1, l. 3), three models with varying assumptions on unit roots and cross-country dependencies are compared. As a base case for comparison, the VAR model (see Tab 2, col. 2) that assumes no unit root in temporally aggregated residuals is applied. In the VAR model cross-country effects are incorporated by multivariate modeling (i.e. the state effect of one country is affected by the state effects of all other countries). As the second model, the VECM (see Tab 2, col. 2) is applied. In our application of the model, we assume[6] that there is one joint underlying unit root for the $n$ country-specific state effect time series, i.e. that there is one pan-European socio-economic and political non-stationary trend fully explaining the non-stationarity of all country-specific state effect series. Like the VAR model, the VECM incorporates cross-country effects by multivariate modeling. As the third model, a VETS (see Table 3, col. 2) is applied. This model assumes a unit root in the state effect series of each country with a common smoothing parameter $\alpha$, i.e. a common memory of past observations. While the model is estimated jointly, it assumes no direct influence of socio-economic and political state effects between countries. Since for all three models, cross-country correlations of the residuals are high (see Fig. 8 for VECM and 21, 22 for VAR, VETS in Appendix A) they are incorporated for probabilistic forecasting of all three models.

[6]Different numbers of cointegration vectors have been tested in an in-sample hyperparameter study, with decreasing error measures for an increasing number of cointegration vectors. The lowest errors were observed with $n-1$ cointegration vectors, indicating the presence of one underlying unit root.

| | AT | BE | BG | CZ | DK | EE | FI | FR | DE | GR | HU | IT | LV | LT | ME | NL | PL | PT | RO | RS | SK | SI | ES | SE |
|---|---|---|---|---|---|---|---|---|---|---|---|---|---|---|---|---|---|---|---|---|---|---|---|---|
| AT | 1.00 | 0.67 | 0.54 | 0.61 | −0.22 | 0.18 | 0.55 | −0.10 | 0.29 | 0.33 | 0.15 | 0.55 | 0.34 | 0.43 | 0.61 | 0.77 | 0.47 | 0.12 | 0.45 | 0.59 | 0.69 | 0.79 | 0.59 | −0.06 |
| BE | 0.67 | 1.00 | 0.49 | 0.43 | −0.47 | 0.26 | 0.67 | −0.08 | 0.37 | 0.28 | −0.16 | 0.68 | 0.04 | 0.30 | 0.35 | 0.68 | 0.23 | 0.42 | 0.40 | 0.30 | 0.51 | 0.73 | 0.74 | −0.30 |
| BG | 0.54 | 0.49 | 1.00 | 0.24 | −0.20 | 0.27 | 0.40 | −0.18 | 0.18 | 0.63 | −0.19 | 0.36 | 0.19 | 0.13 | 0.48 | 0.53 | −0.01 | 0.12 | 0.11 | 0.48 | 0.32 | 0.51 | 0.47 | −0.21 |
| CZ | 0.61 | 0.43 | 0.24 | 1.00 | 0.30 | 0.27 | 0.14 | 0.22 | 0.69 | 0.01 | 0.40 | 0.18 | 0.34 | 0.23 | 0.28 | 0.23 | 0.53 | −0.03 | 0.17 | −0.02 | 0.72 | 0.28 | 0.14 | 0.15 |
| DK | −0.22 | −0.47 | −0.20 | 0.30 | 1.00 | 0.26 | −0.55 | 0.41 | 0.27 | −0.30 | 0.49 | −0.51 | 0.48 | −0.10 | −0.30 | −0.45 | 0.07 | −0.54 | −0.29 | −0.42 | 0.08 | −0.59 | −0.72 | 0.66 |
| EE | 0.18 | 0.26 | 0.27 | 0.27 | 0.26 | 1.00 | −0.11 | 0.19 | 0.04 | −0.22 | 0.23 | 0.21 | 0.56 | 0.23 | −0.08 | 0.03 | −0.02 | 0.13 | 0.05 | 0.01 | 0.07 | 0.07 | 0.13 | 0.47 |
| FI | 0.55 | 0.67 | 0.40 | 0.14 | −0.55 | −0.11 | 1.00 | −0.32 | 0.26 | 0.45 | −0.35 | 0.45 | −0.02 | 0.26 | 0.53 | 0.70 | 0.15 | 0.19 | 0.33 | 0.44 | 0.44 | 0.66 | 0.61 | −0.31 |
| FR | −0.10 | −0.08 | −0.18 | 0.22 | 0.41 | 0.19 | −0.32 | 1.00 | 0.21 | −0.16 | 0.48 | −0.10 | −0.02 | 0.14 | −0.35 | −0.35 | 0.08 | −0.22 | −0.01 | −0.22 | 0.01 | −0.24 | −0.21 | 0.34 |
| DE | 0.29 | 0.37 | 0.18 | 0.69 | 0.27 | 0.04 | 0.26 | 0.21 | 1.00 | 0.09 | 0.04 | 0.06 | 0.04 | 0.01 | 0.06 | 0.11 | 0.21 | −0.05 | −0.09 | −0.24 | 0.58 | 0.13 | 0.05 | −0.01 |
| GR | 0.33 | 0.28 | 0.63 | 0.01 | −0.30 | −0.22 | 0.45 | −0.16 | 0.09 | 1.00 | −0.28 | 0.30 | −0.04 | −0.07 | 0.49 | 0.42 | 0.03 | −0.02 | 0.17 | 0.43 | 0.19 | 0.35 | 0.37 | −0.35 |
| HU | 0.15 | −0.16 | −0.19 | 0.40 | 0.49 | 0.23 | −0.35 | 0.48 | 0.04 | −0.28 | 1.00 | 0.10 | 0.49 | 0.32 | 0.10 | −0.15 | 0.65 | −0.33 | 0.48 | −0.03 | 0.39 | −0.10 | −0.34 | 0.68 |
| IT | 0.55 | 0.68 | 0.36 | 0.18 | −0.51 | 0.21 | 0.45 | −0.10 | 0.06 | 0.30 | 0.10 | 1.00 | 0.05 | 0.33 | 0.49 | 0.47 | 0.35 | 0.46 | 0.68 | 0.38 | 0.39 | 0.66 | 0.75 | −0.18 |
| LV | 0.34 | 0.04 | 0.19 | 0.34 | 0.48 | 0.56 | −0.02 | −0.02 | 0.04 | −0.04 | 0.49 | 0.05 | 1.00 | 0.31 | 0.29 | 0.20 | 0.43 | −0.29 | 0.28 | 0.16 | 0.41 | −0.02 | −0.22 | 0.68 |
| LT | 0.43 | 0.30 | 0.13 | 0.23 | −0.10 | 0.23 | 0.26 | 0.14 | 0.01 | −0.07 | 0.32 | 0.33 | 0.31 | 1.00 | 0.09 | 0.28 | 0.55 | 0.02 | 0.28 | 0.05 | 0.36 | 0.37 | 0.18 | 0.16 |
| ME | 0.61 | 0.35 | 0.48 | 0.28 | −0.30 | −0.08 | 0.53 | −0.35 | 0.06 | 0.49 | 0.10 | 0.49 | 0.29 | 0.09 | 1.00 | 0.67 | 0.39 | 0.02 | 0.57 | 0.59 | 0.57 | 0.58 | 0.40 | −0.03 |
| NL | 0.77 | 0.68 | 0.53 | 0.23 | −0.45 | 0.03 | 0.70 | −0.35 | 0.11 | 0.42 | −0.15 | 0.47 | 0.20 | 0.28 | 0.67 | 1.00 | 0.20 | 0.15 | 0.39 | 0.65 | 0.46 | 0.85 | 0.61 | −0.21 |
| PL | 0.47 | 0.23 | −0.01 | 0.53 | 0.07 | −0.02 | 0.15 | 0.08 | 0.21 | 0.03 | 0.65 | 0.35 | 0.43 | 0.55 | 0.39 | 0.20 | 1.00 | −0.20 | 0.63 | 0.09 | 0.71 | 0.23 | −0.03 | 0.29 |
| PT | 0.12 | 0.42 | 0.12 | −0.03 | −0.54 | 0.13 | 0.19 | −0.22 | −0.05 | −0.02 | −0.33 | 0.46 | −0.29 | 0.02 | 0.02 | 0.15 | −0.20 | 1.00 | 0.03 | 0.04 | −0.05 | 0.33 | 0.63 | −0.41 |
| RO | 0.45 | 0.40 | 0.11 | 0.17 | −0.29 | 0.05 | 0.33 | −0.01 | −0.09 | 0.17 | 0.48 | 0.68 | 0.28 | 0.28 | 0.57 | 0.39 | 0.63 | 0.03 | 1.00 | 0.47 | 0.49 | 0.41 | 0.34 | 0.16 |
| RS | 0.59 | 0.30 | 0.48 | −0.02 | −0.42 | 0.01 | 0.44 | −0.22 | −0.24 | 0.43 | −0.03 | 0.38 | 0.16 | 0.05 | 0.59 | 0.65 | 0.09 | 0.04 | 0.47 | 1.00 | 0.23 | 0.58 | 0.49 | −0.08 |
| SK | 0.69 | 0.51 | 0.32 | 0.72 | 0.08 | 0.07 | 0.44 | 0.01 | 0.58 | 0.19 | 0.39 | 0.39 | 0.41 | 0.36 | 0.57 | 0.46 | 0.71 | −0.05 | 0.49 | 0.23 | 1.00 | 0.44 | 0.19 | 0.17 |
| SI | 0.79 | 0.73 | 0.51 | 0.28 | −0.59 | 0.07 | 0.66 | −0.24 | 0.13 | 0.35 | −0.10 | 0.66 | −0.02 | 0.37 | 0.58 | 0.85 | 0.23 | 0.33 | 0.41 | 0.58 | 0.44 | 1.00 | 0.79 | −0.31 |
| ES | 0.59 | 0.74 | 0.47 | 0.14 | −0.72 | 0.13 | 0.61 | −0.21 | 0.05 | 0.37 | −0.34 | 0.75 | −0.22 | 0.18 | 0.40 | 0.61 | −0.03 | 0.63 | 0.34 | 0.49 | 0.19 | 0.79 | 1.00 | −0.52 |
| SE | −0.06 | −0.30 | −0.21 | 0.15 | 0.66 | 0.47 | −0.31 | 0.34 | −0.01 | −0.35 | 0.68 | −0.18 | 0.68 | 0.16 | −0.03 | −0.21 | 0.29 | −0.41 | 0.16 | −0.08 | 0.17 | −0.31 | −0.52 | 1.00 |

Figure 8: The in-sample correlation matrix $\widehat{\text{Corr}}(\boldsymbol{\epsilon}^{\text{VECM}}_{\tau=1,\dots,T/(7\times 24)})$ for $\tau$ from February 6th, 2019 to February 1st, 2023.

**(iii) Probabilistic load forecast**:

Lastly, the stochastic autoregressive effect of load is captured (see Figure 9 right-hand-side). For this, load is modeled again by country-specific GAMs with deterministic seasonal and holiday information, smoothed temperature, and, additionally, with the fitted socio-economic and political state from step (ii) as inputs. By using country-specific GAMs, we regard the impact of these inputs on load as national. This approach is reasonable for calendar-based and temperature effects. For example, different nationalities have specific calendar-based behaviors, such as common siesta times, peak working hours, and holidays, which do not affect the load in other countries. Moreover, countries vary in their use of electric heating or cooling leading to different temperature sensitivities in load, as shown in Figure 4 comparing Germany and France. For the socio-economic and political states, cross-country influences on load are plausible due to the increasing interconnection of European power grids. Although we use country-specific GAMs, we account for this interconnectedness by modeling socio-economic and political effects multivariate, as explained in (ii). In these GAMs residuals, the short to mid-term stochastic autoregressive effects remain. Thus, as for temperature modeling, we employ univariate VAR models on the residuals of each country-specific GAM and capture the transnational spatial property of the autoregressive component (see Tab. 1, l. 4) by forecasting it probabilistically with sampled multivariate normal error terms with cross-country residual covariance[7]. Finally, forecasted load distributions are obtained as the sum of the GAM point and VAR probabilistic forecasts for each country with the temperature and state forecasts from (i) and (ii).

Note that a joint distribution of residuals from the three models (i) - (iii), i.e. a dependence between temperature, socio-economic and political and autoregressive scenarios, was not considered reasonable, as confirmed by estimated in-sample correlations (see Fig. 26 - 28 in Appendix A).

### *4.2. Modeling, Estimation and Forecasting Specifications*

In the subsequent, the specific modeling equations, parameter choices along with the estimation and forecasting methods applied to obtain the probabilistic temperature, socio-economic and political states and load forecasts, described in (i), (ii) and (iii), respectively, are defined. An overview of input variables is given in Tabel 4. For a precise definition of all variables, we refer to the detailed explanation of the corresponding point forecasting model in our previous work, see Zimmermann & Ziel (2024).

#### *4.2.1. Probabilistic Temperature Model:*

To probabilistically forecast smoothed temperatures, the following modeling equation is applied for countries $i = 1, \ldots, n$:

$$\widetilde{X}_t^{i,\text{Temp}} = o^{\text{Temp}} + \mu^{i,\text{Season}}(X_t^{i,\text{Season}}) + \mu^{i,\text{Trend}}(X_t^{i,\text{Trend}}) + \varepsilon_t^{i,\text{Temp}}, \tag{9}$$

$$\varepsilon_t^{i,\text{Temp}} \sim \text{VAR}_{\text{Lasso}}(\boldsymbol{\phi}^i; p_{\max}) \tag{10}$$

where $\mu^{i,\text{Season}}(X_t^{i,\text{Season}})$ are smooth terms in the daily and yearly seasonal information and $\mu^{i,\text{Trend}}(X_t^{i,\text{Trend}})$ is a smooth trend in the time steps (see Tab. 4, l. 3-4).

Smooth terms $\mu^{i,\text{Season}}$ are estimated as penalized cyclic cubic B-spline in (11) according to (2) with details specified in Zimmermann & Ziel (2024). The smooth trend $\mu^{i,\text{Trend}}$ is estimated by a second-step GAM on the residuals of (11) as penalized cubic regression splines with side constraints to be monotonically non-decreasing and fixed smoothing parameter from the corresponding unconstraint GAM, see Wood (2006) for details:

$$\widetilde{X}_t^{i,\text{Temp}} = o^{\text{Temp}} + \mu^{i,\text{Season}}(X_t^{i,\text{Season}}) + T_t^i \tag{11}$$

$$T_t^i = \mu^{i,\text{Trend}}(X_t^{i,\text{Trend}}) + \varepsilon_t^{i,\text{Trend}} \tag{12}$$

Parameter estimation and probabilistic forecasting of (10) are carried out as specified in Tab. 2, col. 1 by Post-Lasso with $p_{\max} = 4 \times 7 \times 24$. To account for interday variations of the covariance, hour-specific matrices are estimated according to (4).

For each country, the above-described modeling procedure is applied to two smoothed temperature time series $\widetilde{X}_t^{i,\text{Temp}_{1,2}}$:

$$\widetilde{\boldsymbol{X}}_t^{i,\text{Temp}_{1,2}} \sim \text{VETS}(\alpha_{1,2}) \tag{13}$$

[7] Corresponding correlation matrices can be found in Figures 24, 23, 25 in Appendix A

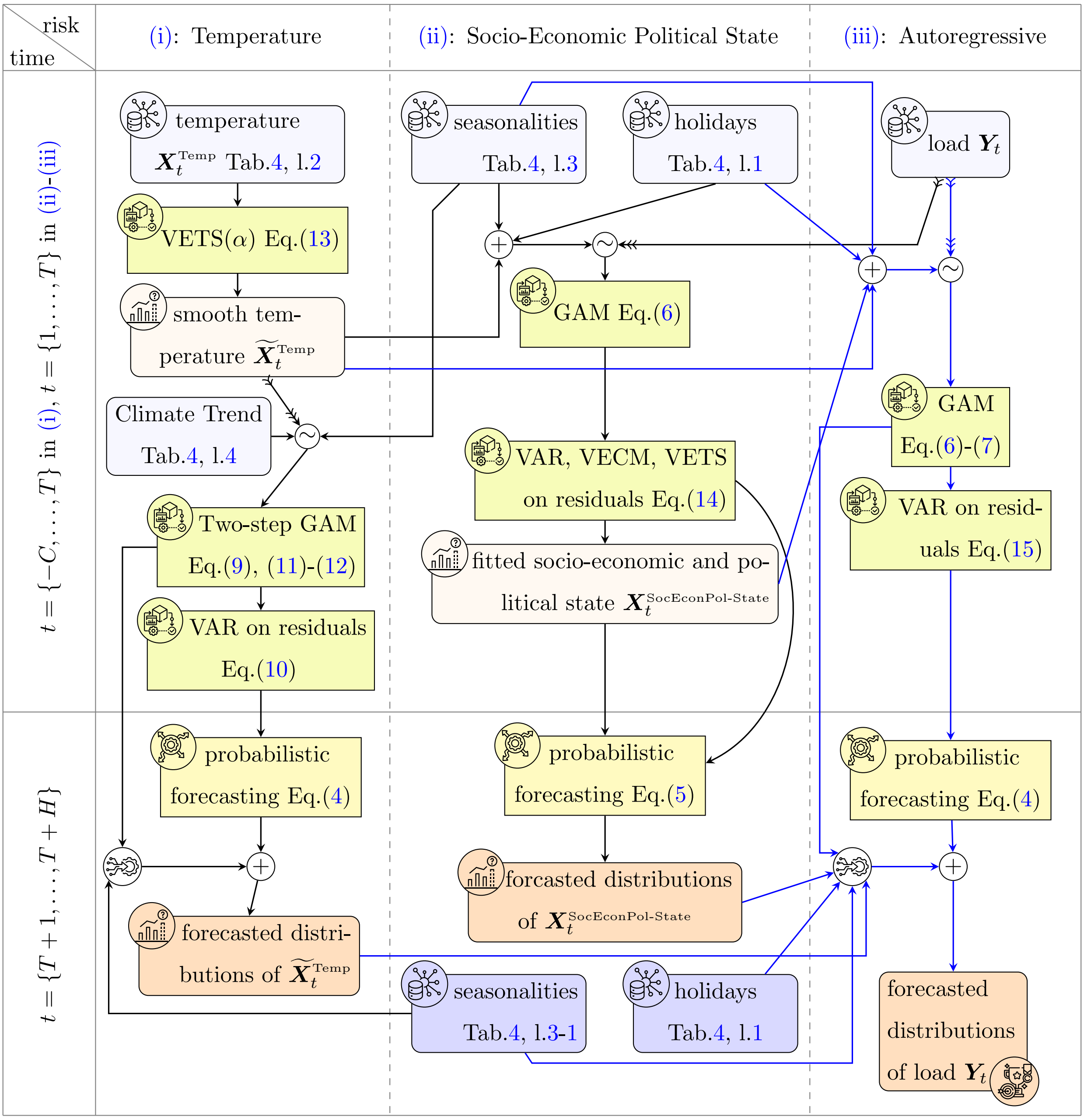

Figure 9: Flowchart representing the general process of the probabilistic load forecasting model with temperature, socio-economic and political state and autoregressive sources of risk resulting in multivariate distributions $\widehat{F_{T+1}}(\boldsymbol{Y}_{T+1}), ..., \widehat{F_{T+H}}(\boldsymbol{Y}_{T+H})$. The external input information is framed by blue boxes and the output of models by orange boxes. Thereby, lighter shades are used for in-sample times and darker shades for out-of-sample times. Models are framed by yellow boxes. By blue lines, the inputs and models leading to the final load forecasts are marked.

for $\alpha_1 = 1/6$ and $\alpha_2 = 1/24$ and VETS according to Tab. 3, col. 1. By this, a temperature time series with more variation[8] and a shorter dependence on past temperatures ($\alpha_1 = 1/6$) and a smoother temperature time series with a longer dependence on past temperatures are considered ($\alpha_2 = 1/24$).

[8]Figure 7 only depicts residual correlations for $\alpha_2 = 1/24$, since a stronger load effect is attributed to the smoothed temperatures $\widetilde{\boldsymbol{X}}_t^{\text{Temp}}$ in this $\alpha$ in France and Germany. A similar Figure can be found for the smoothing parameter $\alpha_1 = 1/6$ in Appendix A (see Fig. 20).

#### 4.2.2. Probabilistic Socio-Economic and Political State Model:

Three models, the VAR, VECM and VETS are applied and compared to probabilistically forecast the socio-economic and political state effect:

$$\boldsymbol{r}_\tau \sim \{\text{VAR}(\boldsymbol{\phi}; \mathcal{S}), \text{VECM}(\boldsymbol{\Gamma}, \boldsymbol{\Pi}; r < n), \text{VETS}(\alpha, \gamma; m)\} \tag{14}$$

for the temporal aggregation $\boldsymbol{r}_\tau = 1/|C| \sum_{c \in C} \boldsymbol{\varepsilon}_{7S(\tau-1)+c}$ of residuals $\boldsymbol{\varepsilon}_t$ of GAMs in (6). By this aggregation, residuals for hours from 8:00 to 19:00 for each working day of the week[9], i.e. $C = \bigcup_{n=0}^{4} \{8, ..., 19\} + nS$, are averaged and the frequency of $\varepsilon_t$ is reduced to a weekly interval.

For estimation and probabilistic forecasting of (14), specifications as detailed in Section 3.2 with $\mathcal{S} = \{1, 2\}, r = n - 1$, an annual seasonality $m = 52$ and constant covariance matrix estimated according to (4) are applied. The temporal aggregation renders computation times feasible for estimation and forecasting. It is, furthermore, sensible since the socio-economic and political state effect is represented by the impact of previous load levels. The information loss from this aggregation is mitigated by a specific setup of the smooth terms $\mu^{i,\text{SocEconPol-State}}$ in (7), as explained in Zimmermann & Ziel (2024).

For VAR and VECM, the socio-economic and political state $\boldsymbol{X}_t^{\text{SocEconPol-State}}$ is retrieved as linear interpolation from $\boldsymbol{r}_\tau$ with extended boundaries, i.e. $\boldsymbol{X}_t^{\text{SocEconPol-State}} = \boldsymbol{r}_{\tau_1}$ and $\boldsymbol{X}_t^{\text{SocEconPol-State}} = \boldsymbol{r}_{\tau_T}$ for $t < \min(C)$ and $t > \max(C) + (7 \times 24)\tau_T$, respectively. For the VETS model the fit and forecast of the level component $\boldsymbol{l}_\tau$ such that $\boldsymbol{r}_\tau \sim \text{VETS}(\alpha, \gamma; m)$ are retrieved and linearly interpolated with extended boundaries to obtain $\boldsymbol{X}_t^{\text{SocEconPol-State}}$.

#### 4.2.3. Probabilistic Load Model:

Finally, load is forecasted probabilistically by the modeling equations (6)-(8), where in (8) the remaining uncertainty, i.e. the stochastic autoregressive effect, is incorporated by:

$$\varepsilon_t^{i,\text{AR}} \sim \text{VAR}_{\text{Lasso}}(\boldsymbol{\phi}^i; p_{\max}) \tag{15}$$

Estimation and forecasting are carried out as specified in Tab. 2, col. 1 with Post-Lasso, $p_{\max} = 8 \times 24 \times 7$ and covariance estimated according to (4).

## 5. Benchmarks

Two benchmark models are considered for comparison in a forecasting study described in Section 6. Both models result from point forecasting with autoregressive postprocessing where probabilistic forecasts are obtained by ensembling iterative error term simulations. Specifically, as point forecasting models the two best-performing models from Zimmermann & Ziel (2024) are considered: a Seasonal Random Walk model with an Annual period (SRWA) and a Feed Forward Neural Network (FNN) with one hidden layer and linear output.

- **SRWA**: $Y_t^i = Y^i_{t-m\lceil\frac{t-T}{m}\rceil} + \varepsilon_t^i, m = 365S$

[9]This choice aligns with the peak load definition commonly used in European electricity markets for trading electricity futures.

- **FNN**: $Y_t^i = \upsilon + \sum_{j=1}^{k_{\text{FNN}}} v_j g\left(\sum_{j=1}^{m} w_{ij} \boldsymbol{X}_t^i\right) + \varepsilon_t^i$, where $\boldsymbol{X}_t^i = (\widetilde{X}_t^{i,\text{Temp}_1}, \widetilde{X}_t^{i,\text{Temp}_2}, X_t^{i,\text{Cal-based}})$ for $\widetilde{X}_t^{i,\text{Temp}_i}$ the fitted values of (13), $m = |\boldsymbol{X}_t^i|$, $k_{\text{FNN}} = m/2 + 1$ and sigmoid activation function $g(x) = 1/(1 + \exp^{-x})$.

for countries $i = 1, \ldots, n$ and $\varepsilon_t^i \sim \text{VAR}_{\text{Lasso}}(\boldsymbol{\phi}^i; p_{\max})$ estimated and forecasted as specified in Tab. 2, col. 1 with Post-Lasso, $p_{\max} = 8 \times 24 \times 7$ and covariance estimated by (4). For the FNN temperatures simulated according to (9)-(10) are used as inputs in the forecasting horizon.

## 6. Forecasting Study and Evaluation Design

To evaluate the proposed probabilistic load forecasting method we conducted a rolling window forecasting study comparing all three model options for the socio-economic and political state (VAR, VECM, VETS) and benchmarks SRWA, FNN. For the study, more than 9 years of load data, from January 2015 to February 2024 were used. In-sample data in each forecasting experiment comprised 4 years of data, resulting in $4 \times 365 \times 24$ hourly observations. Following Staffell & Pfenninger (2018), historic data to estimate the temperature model (9)-(10) covered approximately 20 years, i.e. $C = 20 \times 52 \times 168 - 1$. By this, we guarantee to capture extreme meteorological events and a trend in temperatures due to climate change. As emphasized by Mosquera-López et al. (2024) in a study across six European countries, changing climate conditions amplify risks in electricity markets.

The forecasting horizon is chosen as 52 weeks, i.e. $H = 52 \times 168$. To reduce overlap and consequent serial correlation of forecasting experiments, with each experiment the window of in-sample observations is rolled forward by one month, with forecasts always starting at 9 a.m. on the first day of each month. This results in $N = 50$ experiments. In each experiment, 200 simulations are carried out for probabilistic forecasting. A data preprocessing of single outliers in the load data is applied as described in Zimmermann & Ziel (2024).

To assess the predictive performance and applicability of the probabilistic forecasting models in guiding financial and industry stakeholders in their decision-making processes, we evaluate calibration and sharpness. While a well-calibrated probabilistic forecast, i.e. its predictive distribution aligns with the sample distribution, is essential for deriving quantitative decisions from the forecast, achieving optimal sharpness, i.e. the uncertainty of the predictive distribution is as low as possible, is crucial for reducing uncertainties in decision-making based on the forecast, see Browell & Fasiolo (2021); Gneiting & Katzfuss (2014).

Common measures jointly evaluating calibration and sharpness are the proper scoring rules *Continuous Ranked Probability Score* (CRPS) and *Pinnball loss*, see Koenker & Bassett (1978); Gneiting & Raftery (2007). Both scoring rules are frequently employed in the field, as demonstrated by GEFCom Hong et al. (2016). For our application, scoring rules are averaged over the one-year forecasting horizons $\{1, \ldots, H\}$, see (16), (17). To assess if differences in the CRPS are significant, the Diebold-Mariano test is applied, see Diebold & Mariano (1994).

Additionally, we evaluate model calibration through reliability plots and quantile fan plots. To analyze cross-country dependencies, trajectories of forecasted load are plotted. To assess weather, socio-economic, political and autoregressive risks, we analyze plots depicting the load effect of extreme scenarios for each of these factors.

$$\mathrm{CRPS}^i = \frac{1}{NH}\sum_{n=1}^{N}\sum_{h=1}^{H}\int_{-\infty}^{\infty}\left(\widehat{F}_{Y^i_{T+h;n}}(z) - \mathbb{1}_{\{Y^i_{T+h}\leq z\}}\right)^2 dz \tag{16}$$

$$\Lambda^i_q = \frac{1}{NH}\sum_{n=1}^{N}\sum_{h=1}^{H}\left[q\cdot(Y^i_{T+h} - \widehat{Y}^i_{T+h;n,q})\cdot\mathbb{1}_{\{Y^i_{T+h}\geq\widehat{Y}^i_{T+h;n,q}\}} + (1-q)\cdot(\widehat{Y}^i_{T+h;n,q} - Y^i_{T+h})\cdot\mathbb{1}_{\{Y^i_{T+h}<\widehat{Y}^i_{T+h;n,q}\}}\right] \tag{17}$$

for $q\in(0,1)$ and countries $i = 1,\dots,n$.

## 7. Results and Interpretation

### *7.1. Calibration and Sharpness*

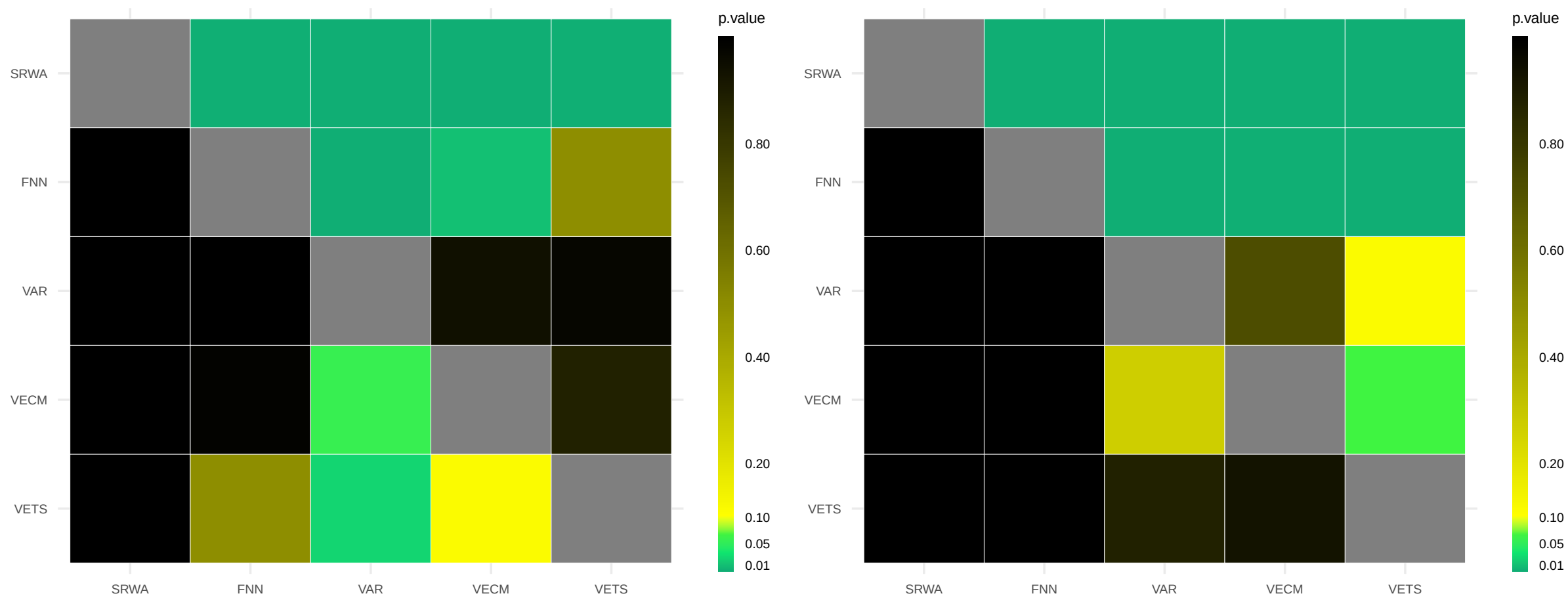

(a) France

(b) Germany

Figure 10: P-values of the DM-test for France and Germany.

Evaluating predictive performance in terms of averaged CRPS (see Tab. 5), we observe that the VECM performs best in most countries, narrowly followed by VAR. Among the three proposed models, VETS performs the worst, being surpassed in several countries by FNN. The SRWA model demonstrates the poorest performance across all countries, except in the Netherlands.

These findings imply that the assumptions underlying the VECM and VAR models, namely, a single European socio-economic and political trend or a stationary process, respectively, are best suited for capturing the mid- to long-term dynamics of socio-economic and political factors influencing European load. In contrast, the VETS model, which assumes country-specific unit roots in load combined with shared transnational memory of past socio-economic and political shocks, is not supported by the results.

In France, the VAR model achieves the highest performance, while in Germany, the VETS model performs best. Notably, in France, the VAR model significantly outperforms all other models (see Fig. 10). In contrast, for Germany, we fail to reject the null hypothesis of a zero series of CRPS differences among the VAR and VETS model at a 5% significance level. However, in Germany, all three models VAR, VECM and VETS are significantly superior to the two benchmark models.

|  | SRWA | FNN | VAR | VECM | VETS |
|---|---|---|---|---|---|
| Austria | 0.333 | 0.277 | 0.252 | **0.252** | 0.257 |
| Belgium | 0.404 | 0.372 | 0.340 | 0.340 | **0.334** |
| Bulgaria | 0.253 | 0.210 | 0.205 | **0.205** | 0.228 |
| Czech Rep. | 0.343 | 0.260 | **0.247** | 0.248 | 0.253 |
| Denmark | 0.181 | 0.172 | 0.163 | **0.154** | 0.165 |
| Estonia | 0.049 | 0.040 | 0.040 | **0.039** | 0.043 |
| Finland | 0.541 | 0.427 | 0.422 | **0.408** | 0.421 |
| France | 5.052 | 2.834 | **2.708** | 2.748 | 2.834 |
| Germany | 2.422 | 2.043 | 1.789 | 1.814 | **1.695** |
| Greece | 0.419 | 0.335 | 0.326 | **0.317** | 0.362 |
| Hungary | 0.237 | 0.201 | 0.199 | **0.192** | 0.202 |
| Italy | 1.856 | 1.600 | 1.519 | **1.470** | 1.789 |
| Latvia | 0.036 | 0.034 | 0.031 | **0.029** | 0.031 |
| Lithuania | 0.063 | 0.054 | **0.054** | 0.058 | 0.060 |
| Montenegro | 0.030 | 0.028 | **0.027** | 0.029 | 0.027 |
| Netherlands | **0.633** | 0.751 | 0.734 | 0.723 | 0.669 |
| Poland | 0.802 | 0.590 | 0.590 | **0.538** | 0.575 |
| Portugal | 0.242 | 0.210 | 0.191 | **0.187** | 0.214 |
| Romania | 0.340 | 0.294 | 0.288 | **0.269** | 0.270 |
| Serbia | 0.256 | 0.287 | 0.258 | **0.246** | 0.252 |
| Slovakia | 0.186 | 0.152 | 0.146 | **0.130** | 0.143 |
| Slovenia | 0.086 | 0.078 | **0.072** | 0.079 | 0.086 |
| Spain | 1.249 | 1.185 | 1.091 | **1.052** | 1.091 |
| Sweden | 0.865 | 0.679 | **0.674** | 0.683 | 0.678 |
| Sum | 16.877 | 13.113 | 12.365 | **12.208** | 12.679 |

Table 5: Forecasting accuracy in terms of CRPS averaged over the forecasting horizon $1, \ldots, H$ and experiments $1, \ldots, N$ in GW for 24 European countries. The color scheme transitioning from red to yellow and green indicates models ranging from low to high accuracy. The best model for each country is marked by bolt writing.

The comparatively low performance of the SRWA benchmark can be attributed to two key factors. Firstly, as a baseline model, it is the only model not including weather variables resulting in particularly worse performance in countries with pronounced electric heating. For instance, in France the SRWA model exhibits an average CRPS more than 85% higher than that of the best-performing model. Secondly, estimation of $\varepsilon_t^i \sim \text{VAR}_{\text{Lasso}}(\boldsymbol{\phi}^i; p_{\max})$ as specified in Tab. 2, col. 1 with Post-Lasso allows for non-stationary solutions, which occur only in the SRWA model. Consequently, while the simple SRWA performed well in the point forecasting study, it proves unsuitable for this probabilistic forecasting methodology.

The high CRPS of the FNN compared to VAR, VECM and VETS in most countries can be attributed to a fundamental difference between neural networks and GAMs. As Chris Kolb et al. (2023) explains, while GAMs are typically designed with low-dimensional interactions between inputs, FNNs allow for high-order interactions. Consequently, neural networks tend to generalize poorly to unobserved data, as noted by Baviera & Manzoni (2024); Guo et al. (2017).

Since temperature trajectories incorporated in the models are derived from a longer historical dataset containing more extreme weather events, which were unobserved during model fitting, this limitation of the FNN becomes particularly apparent. The extrapolation issue of the FNN is even enhanced by small-sample-induced misspecifications. In several forecasting experiments, unreasonable interaction estimates, particularly involving infrequently observed variables such as holidays, led to sudden and unrealistic spikes in the forecasted load. These spikes were amplified when extreme temperature trajectories were used as inputs (see Fig. 29-30 in Appendix B).

In a forecasting study on probabilistic mid-term hourly load forecasting using recurrent neural networks with a single hidden layer by Baviera & Manzoni (2024), poor extrapolating properties and consequent overly narrow predictive intervals around the point forecast were handled by adjusting the loss function applied in training.

An evaluation of predictive performance using the averaged Pinball loss (see Fig. 11), reveals that in France, the VETS model performs worse compared to VAR and VECM for quantiles $q > 0.5$. In contrast, in Germany, the VETS model outperforms the other models for the lower quantiles and is not notably worse for the higher quantiles.

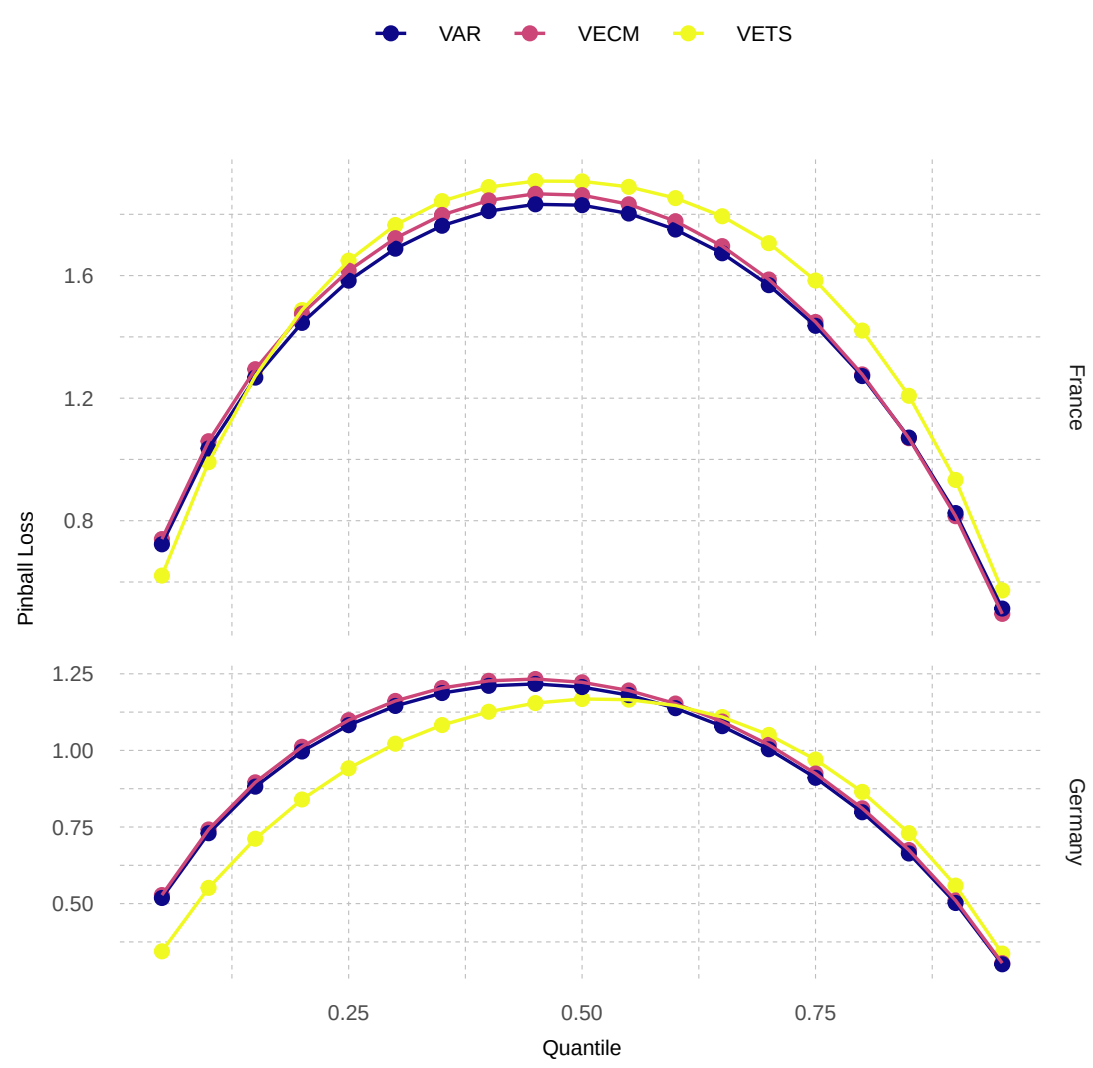

Figure 11: Pinball loss averaged over the forecasting horizon $1, \ldots, H$ and experiments $1, \ldots, N$.

The subsequent reliability plots[10] show the nominal quantile $q \in (0, 1)$ against frequencies observed in the forecasting study: $N^{-1} \sum_{n=1}^{N} \mathbb{1}_{\{Y^i_{T+h} \leq \widehat{Y}^i_{T+h;q,n}\}}$.

Comparing the reliability of the VAR, VECM and VETS models averaged for each hour of the day in Figure 12 (left column), we observe that VAR and VECM overestimate for quantiles $q \leq 75\%$, while VETS is better calibrated. For the French data, although the calibration of VETS is superior, its CRPS is significantly higher than that of VAR. This indicates that when VETS overestimates or underestimates, the deviations from the actual values, averaged over the entire horizon, are larger. This can be attributed to the following: For the VETS model, a constant state forecast is incorporated into the GAM models (see Eq. (7)). If this state

[10] We do not account for serial correlation in reliability assessment, as recommended by Browell & Fasiolo (2021); Pinson et al. (2010).

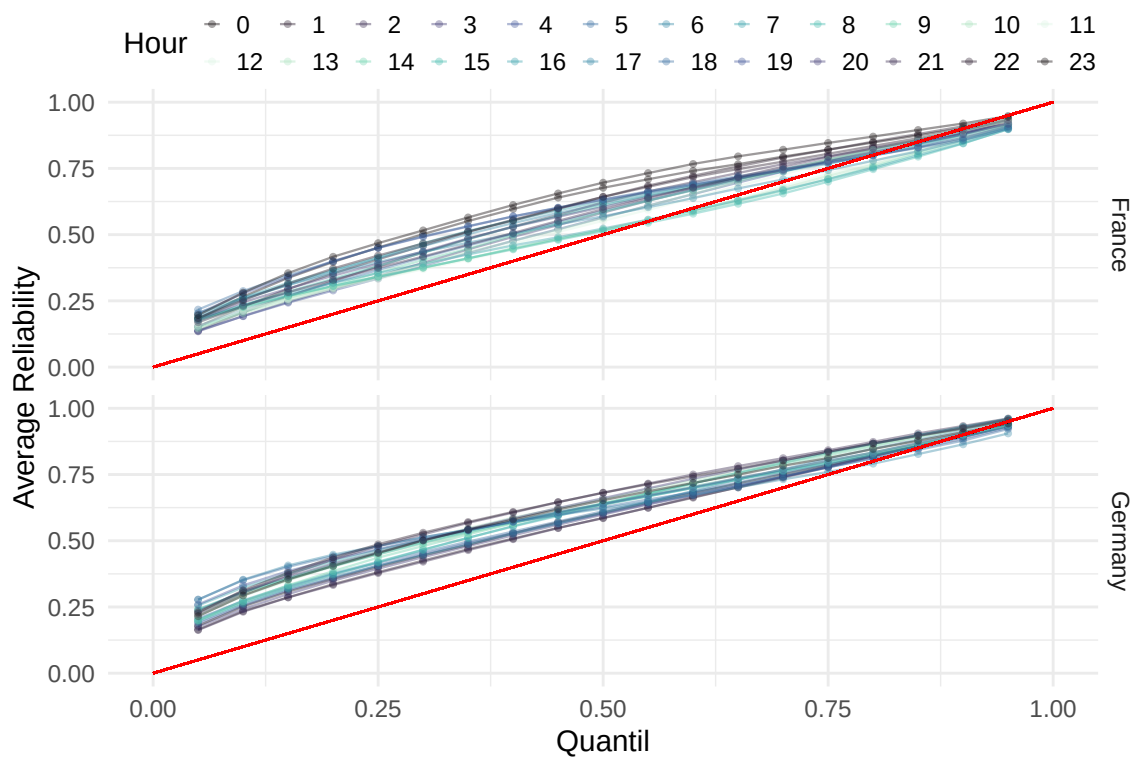

(a) VAR

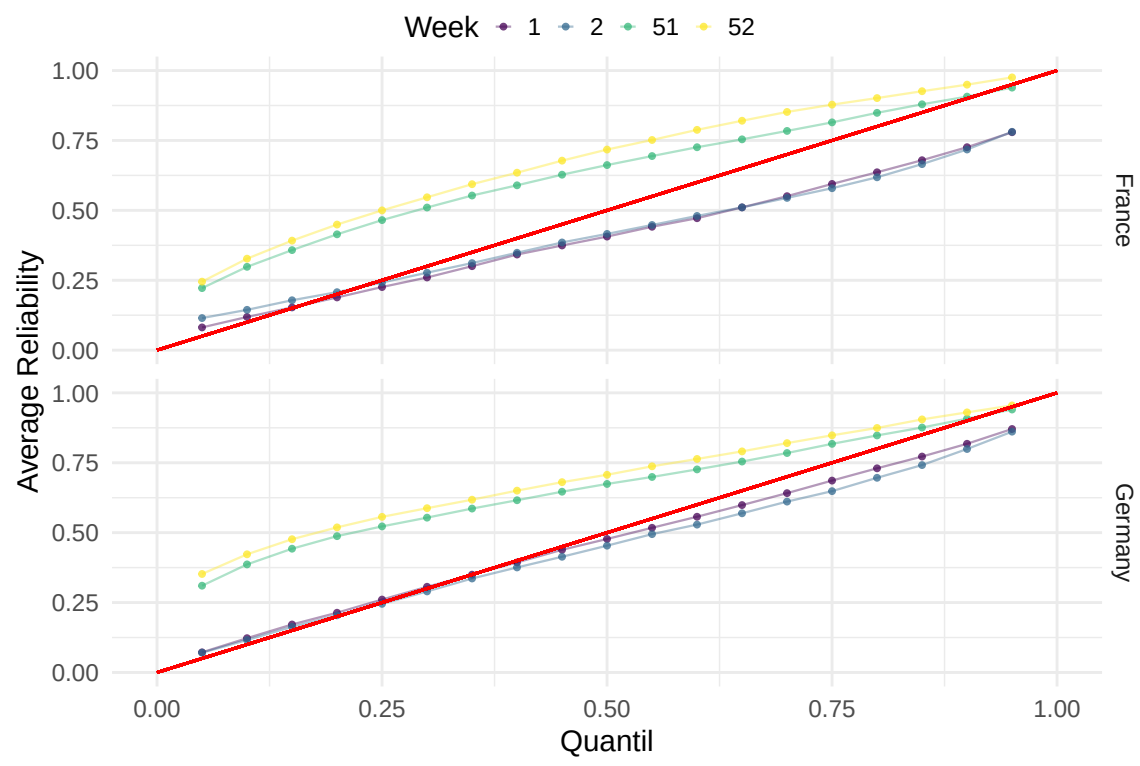

(b) VAR

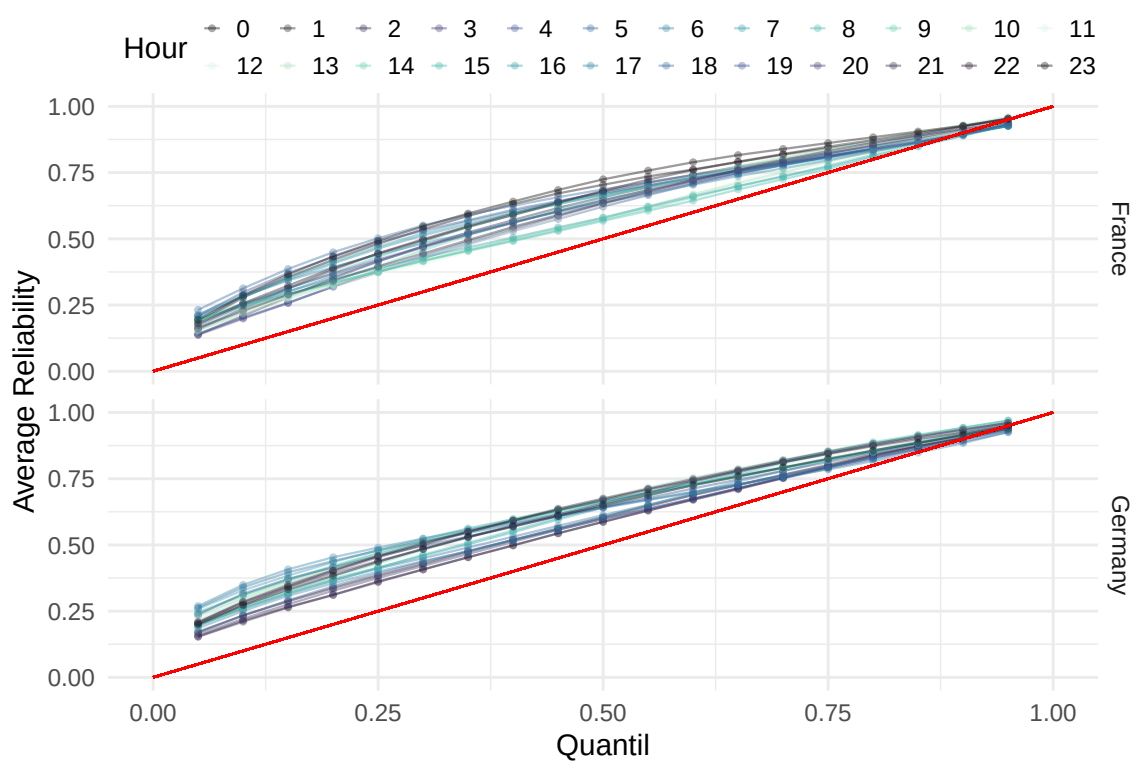

(c) VECM

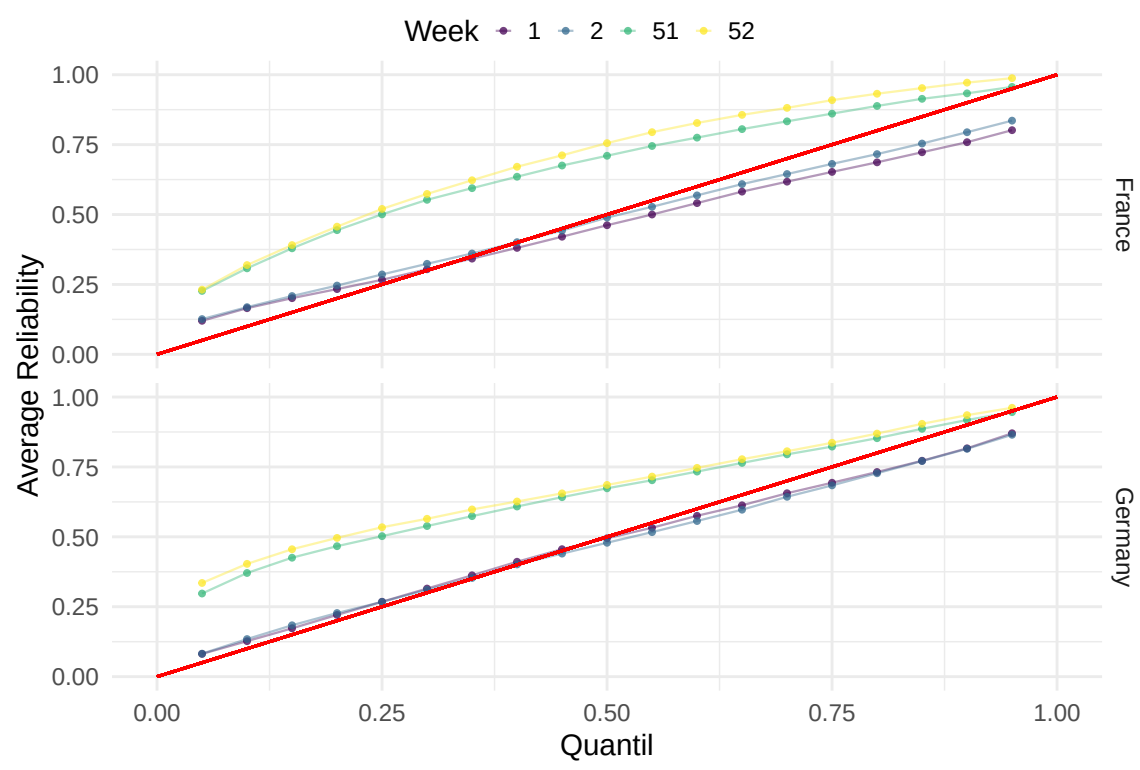

(d) VECM

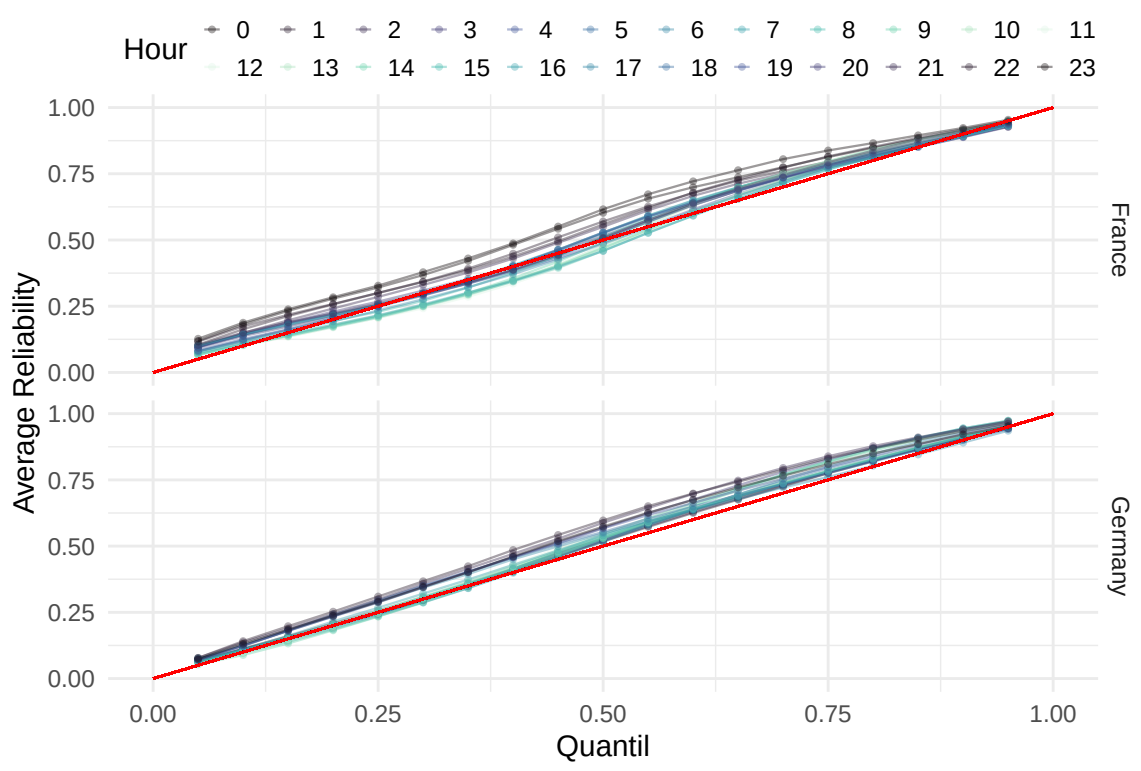

(e) VETS

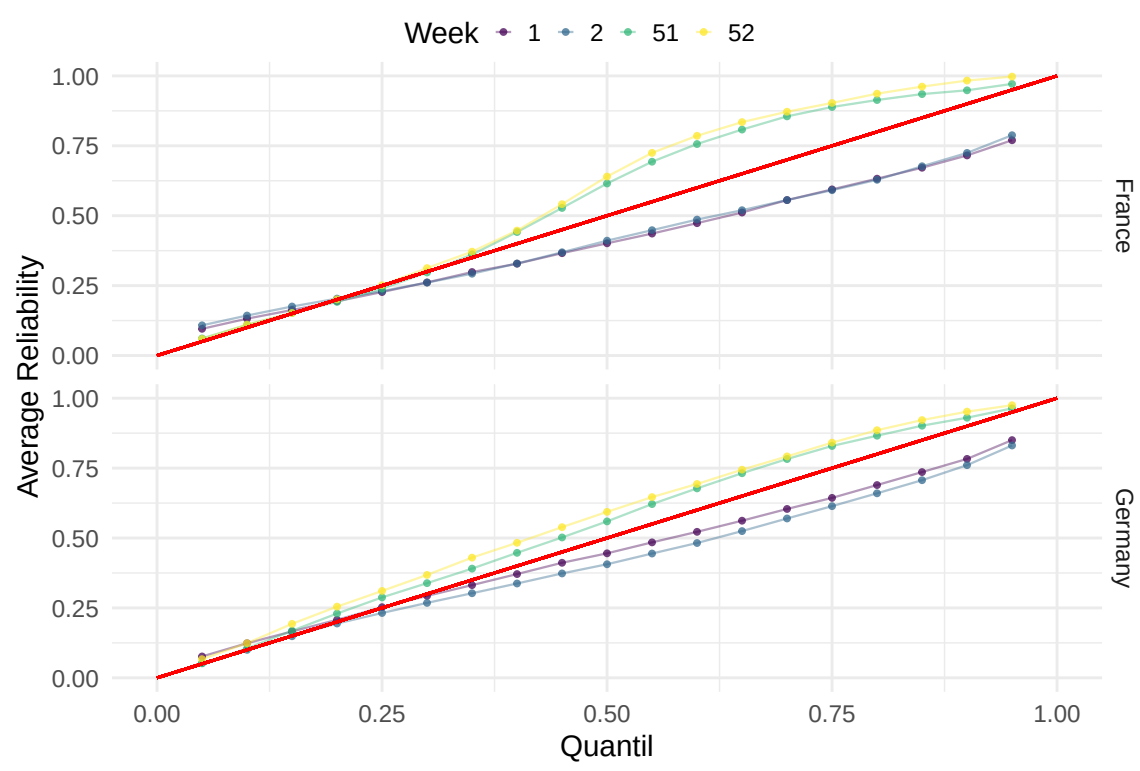

(f) VETS

Figure 12: Reliability plots for $N = 50$ forecasting experiments. Left column: results averaged for each hour of the day. Right column: results averaged for the first, middle, and last two weeks in the forecasting horizon.

forecast over- or underestimates, deviations will accumulate with each forecasting step throughout the horizon. In contrast, the state forecast of the VAR model, which assumes no unit root, converges to zero throughout the horizon.

When comparing the reliability of predictions across different hours of the day, the morning hours (7–10 AM) in France are better calibrated across all three models. In Germany no clear pattern is visible. From the weekly calibration plot (see Fig. 12, right column), we observe, as expected for iterative probabilistic forecasting, that the VAR and VECM models exhibit better calibration at the beginning of the forecasting horizon but overestimate in the end. In contrast, the VETS model, while well-calibrated for all considered weeks for quantiles $q < 0.25$, underestimates at the beginning of the forecasting horizon and overestimates in the end for higher quantiles.

*7.2. Trajectory and Quantile Plots*

Considering the calibration results for France and Germany (see Section 7.1) and the interest in the unit root effect of the mid-term socio-economic and political state, i.e. a persistent continuation of state levels in the prediction horizon, for risk assessment, the figures in this and the subsequent section will focus on the VETS model. Similar figures for the VAR model are available in Appendix B (see Fig. 31 - 37). In all figures, the forecasting experiment for $N = 50$ with in-sample data from February 6th, 2019 to February 1st, 2023 and the one-year ahead forecasting until January 31st, 2024 is depicted.

Figures 13 - 14 illustrate the ability of our model to capture transnational dependencies. These figures display three randomly sampled load forecast trajectories in France and Germany in the first two weeks and the last two weeks of the one-year horizon. Especially, from Figure 14 the alignment of higher and lower load levels in one country with similar trends in the other country is evident. Notably, the variation between trajectories increases towards the end of the horizon as expected for our iterative forecasting approach.

Figures 15 - 16 show the corresponding forecast quantiles. Consistent with the trajectory plots, we observe an increase in variation, particularly for France, towards the end of the forecasting horizon. The range between the 5th and 95th percentile for France expands to approximately 35 GW, compared to 25 GW for Germany. This difference is due to a higher sensitivity of load to stochastic temperatures in France.

An underestimation of all quantiles of the actual load in France during the week of February 7th to February 13th (see Fig. 15) could be attributed to two factors. Firstly, the forecasting period began with temperatures that were relatively moderate for winter in France but then fell to negative values, reaching a yearly low for 2023 around this week (see Fig. 4). This drop in temperatures could not be captured by a week of iteratively simulated trajectories with sampled errors. Secondly, the GAM model in equations (6) - (7) might underestimate temperature effects by incorporating them into the deterministic yearly seasonal component, i.e. by assigning them to the variation in load between winter and summer.

*7.3. Risk Scenario Analysis and Interpretation*

A key strength of our load forecasting model lies in its interpretability. Specifically, we can disentangle the combined effect of the multifaceted factors influencing load along the forecasting horizon.

To deconstruct the combined load effect of extreme risk scenarios during the Christmas vacation time, we determined minimal and maximal trajectories for the stochastic temperature, socio-economic and political state and autoregressive components in our forecasting model. These trajectories were selected based on the smallest and largest integrals over the analyzed timeframe. Figure 17 shows these trajectories alongside a median trajectory.

For the minimal temperature and maximal socio-economic and political state and autoregressive trajectories, Figures 18-19 visualize the contribution of each deterministic and stochastic modeling input to the load forecast. Additionally, the forecasted load scenario (see Eq. (6)-(7)) resulting from these

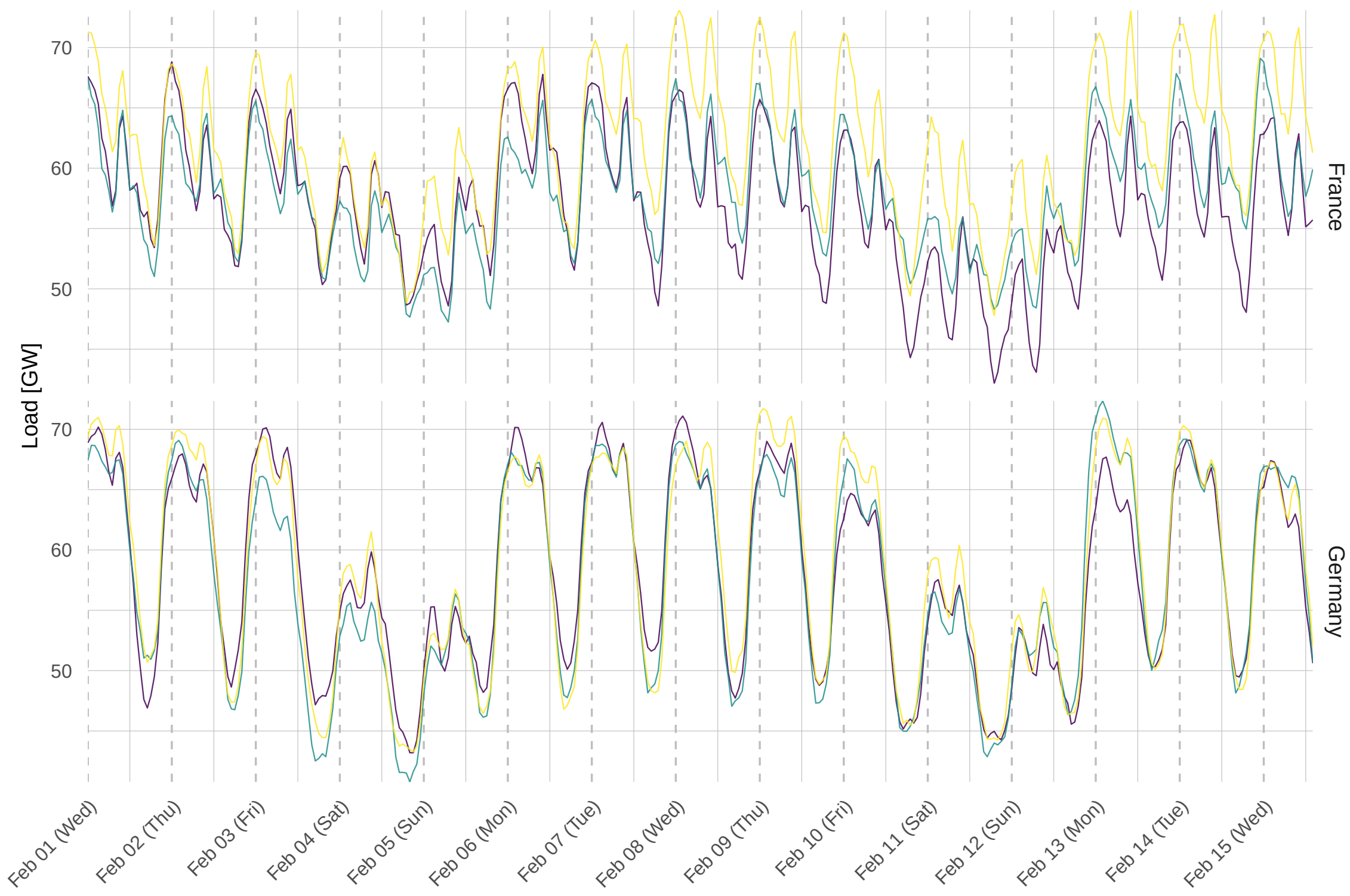

Figure 13: Three sampled load trajectories for the VETS model (forecasting experiment $N = 50$) in France and Germany forecasting the first two weeks of the one-year forecasting horizon (February 1st, 2023 to February 15th, 2023).

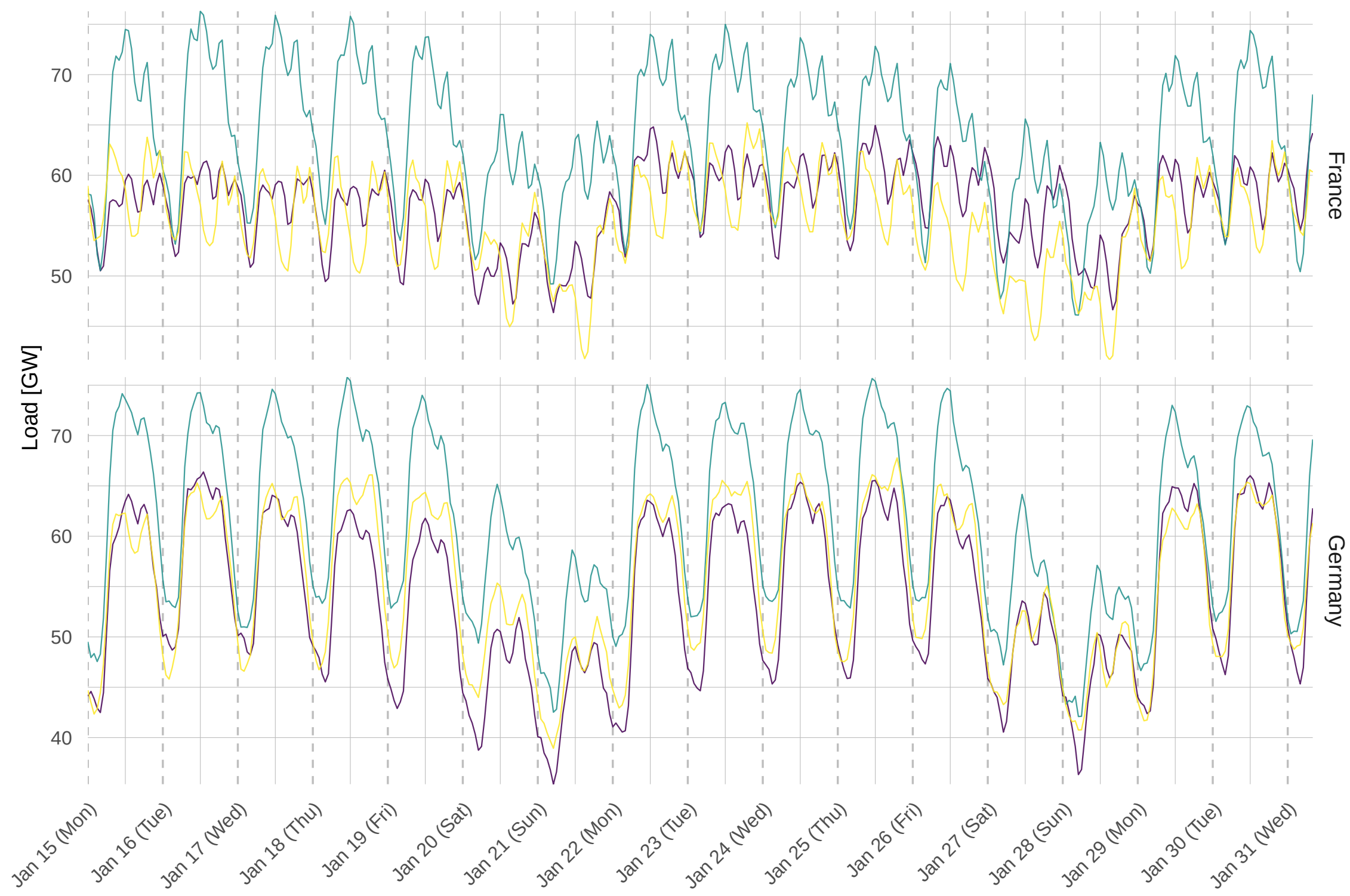

Figure 14: Three sampled load trajectories from the VETS model (forecasting experiment $N = 50$) in France and Germany forecasting the last two weeks of the one-year forecasting horizon (January 15th, 2024 to January 31st, 2024).

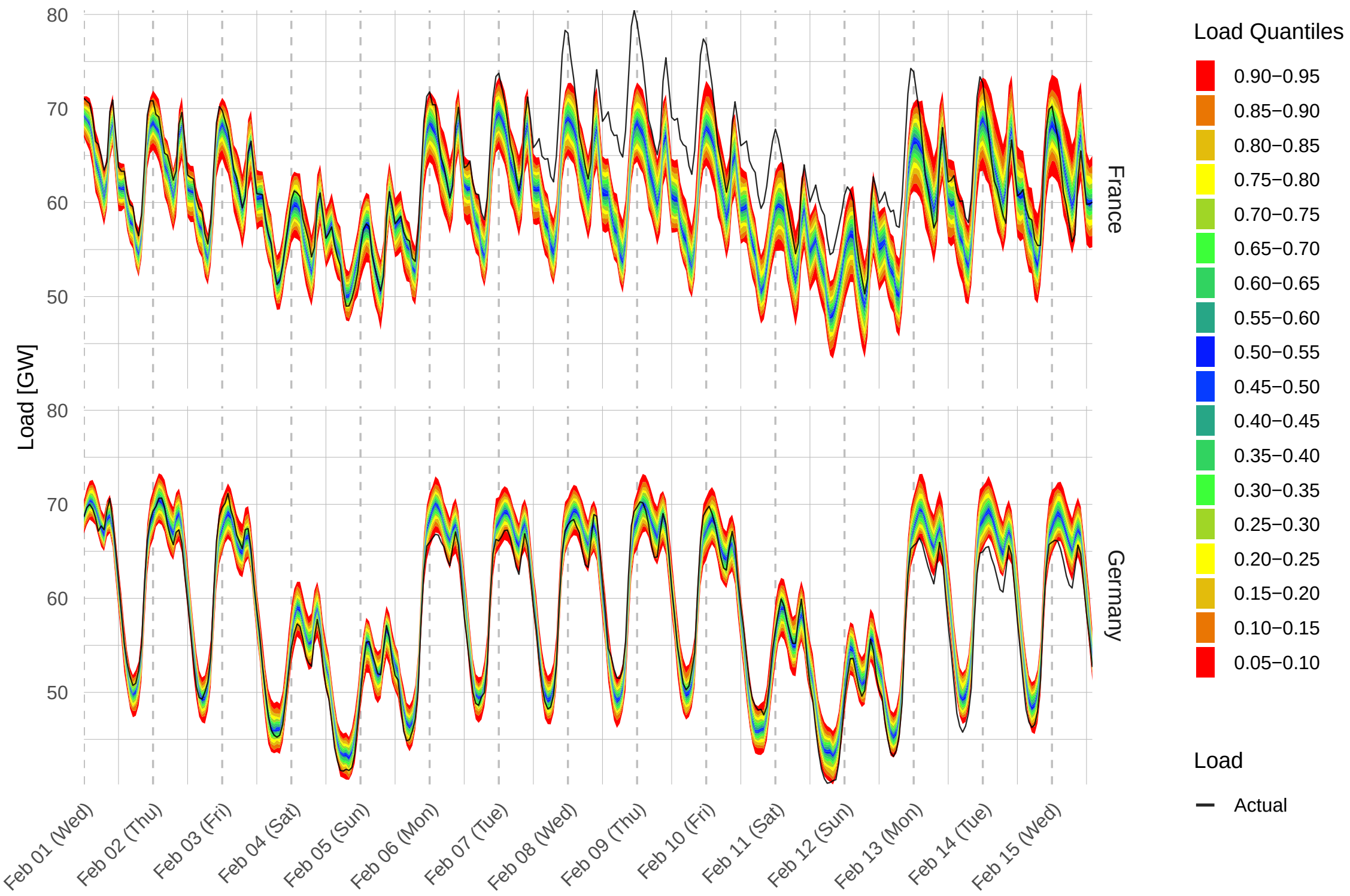

Figure 15: Quantiles of the VETS model (forecasting experiment $N = 50$) in France and Germany forecasting the first two weeks of the one-year forecasting horizon (February 1st, 2023 to February 15th, 2023).

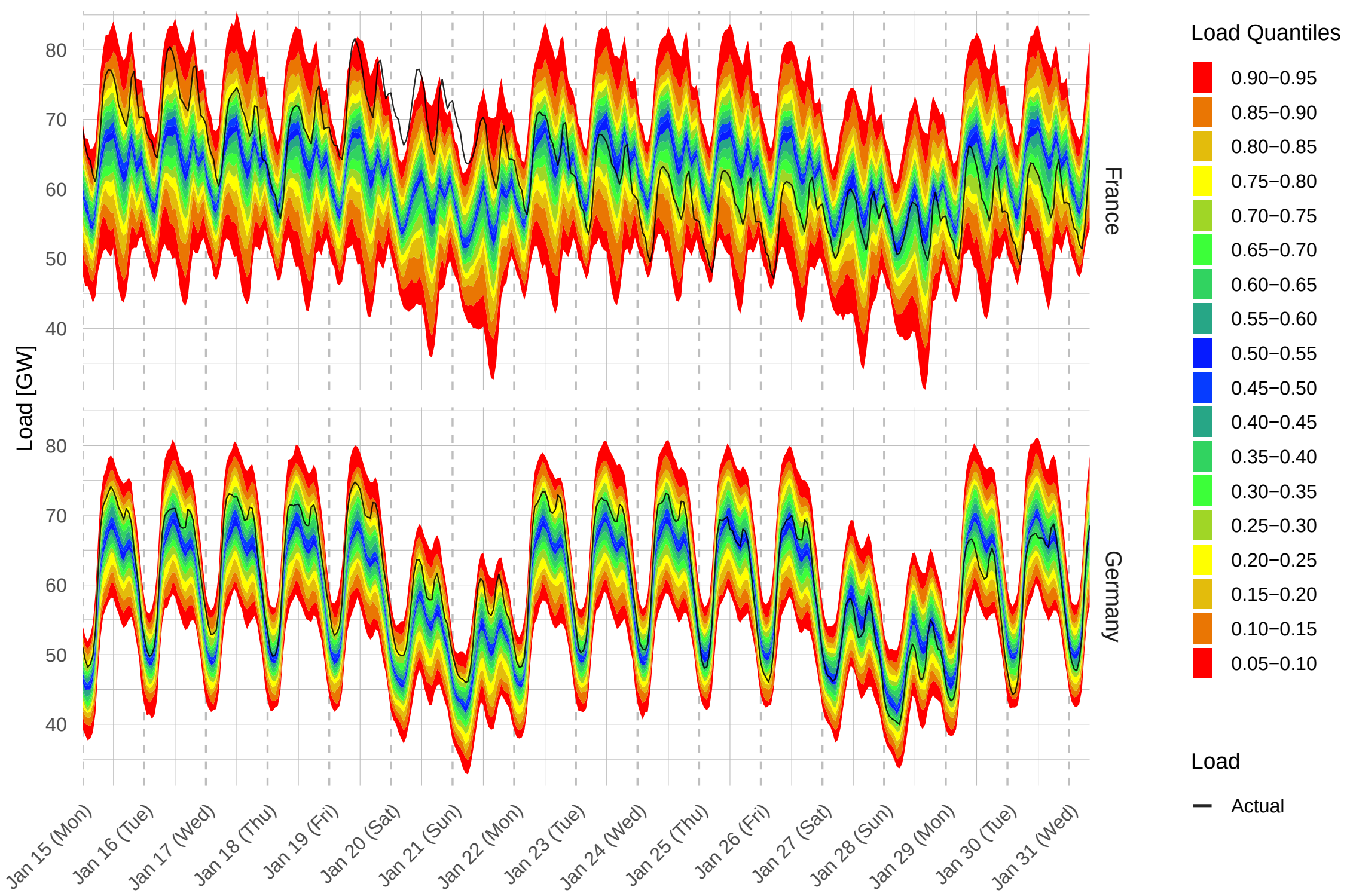

Figure 16: Quantiles of the VETS model (forecasting experiment $N = 50$) in France and Germany forecasting the last two weeks of the one-year forecasting horizon (January 15th, 2024 to January 31st, 2024).

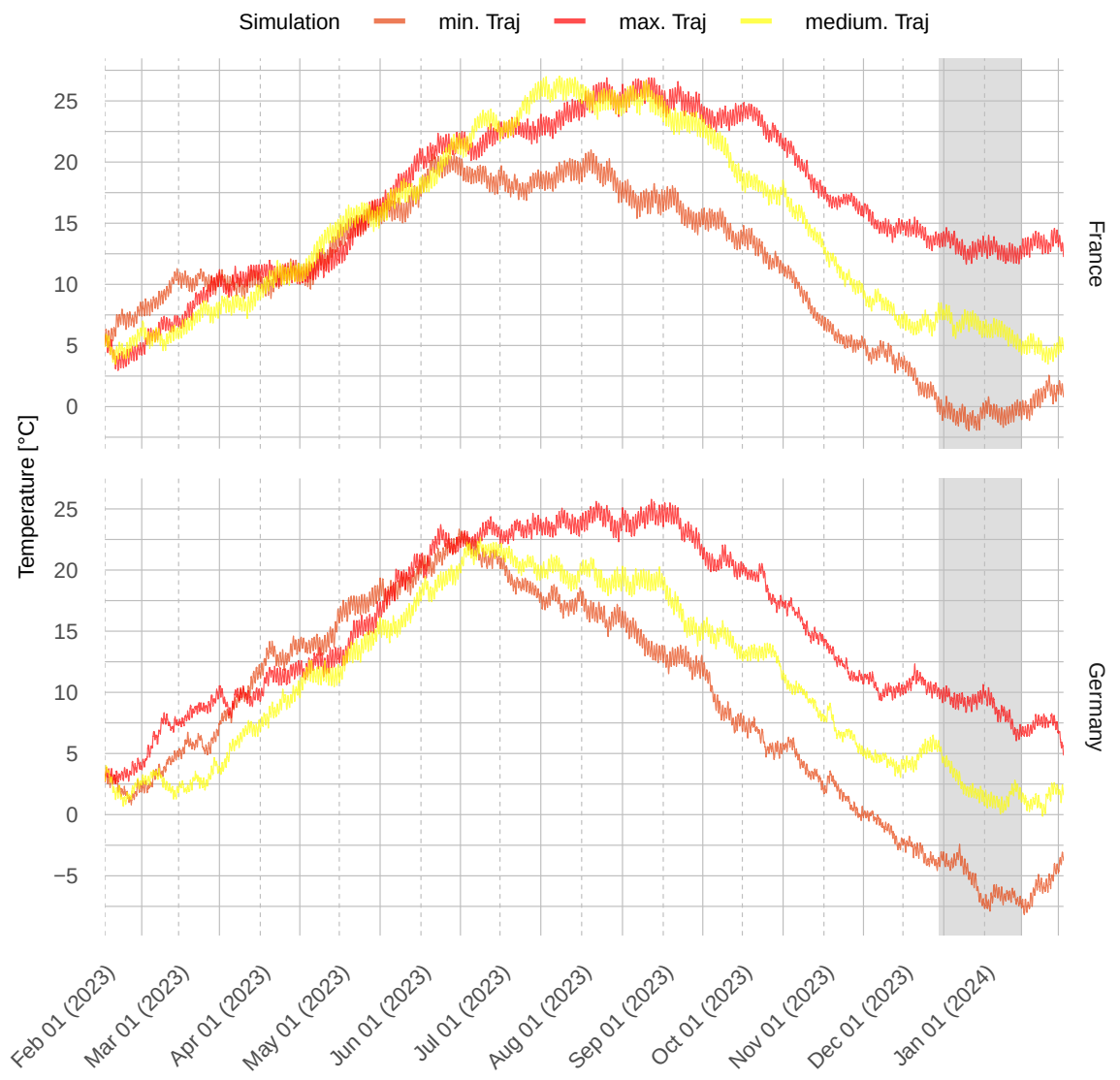

(a) Temperature for $\alpha = 1/24$ in °C

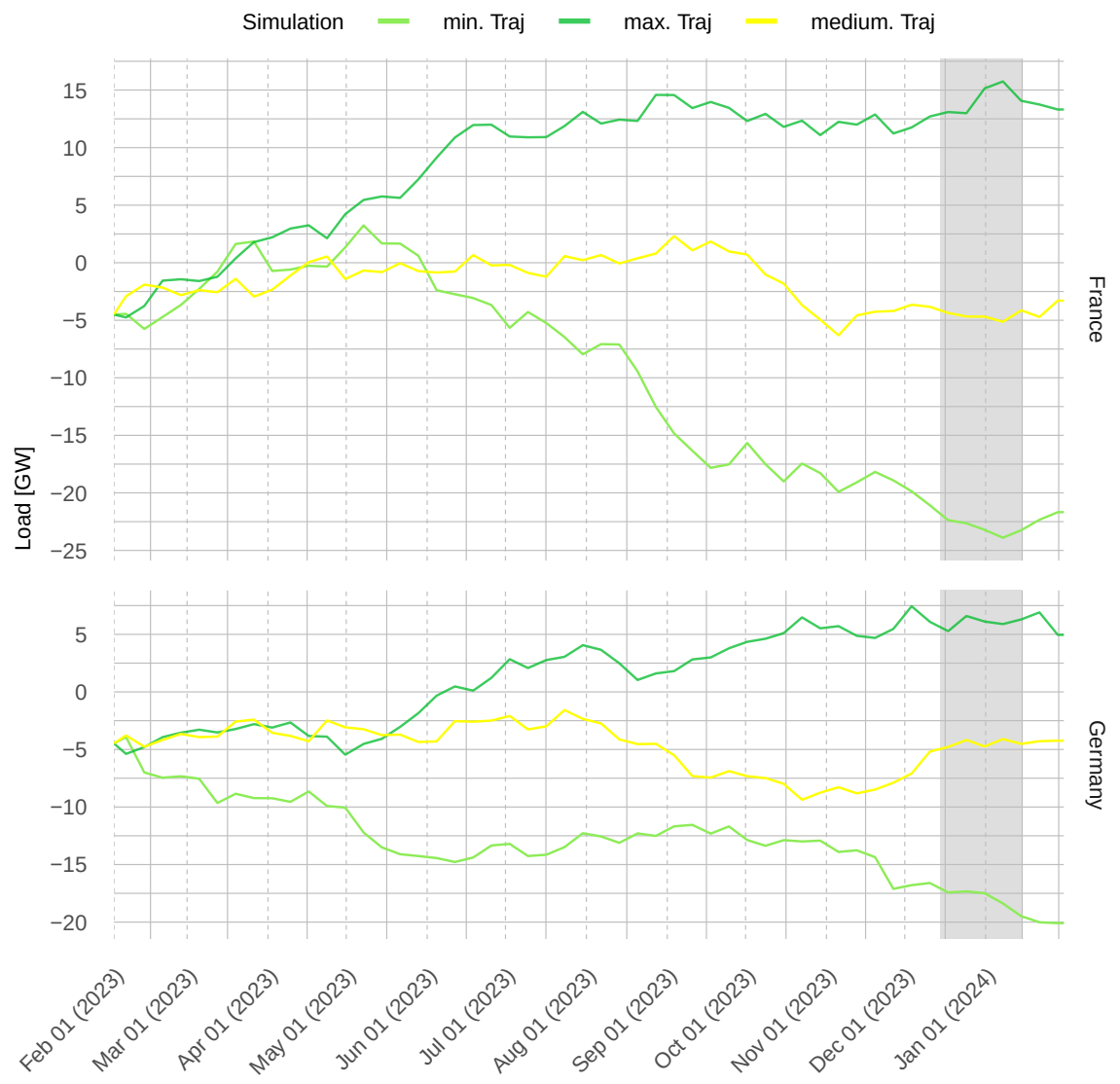

(b) socio-economic and political State in GW

Figure 17: Minimum, median and maximum trajectories based on the integral during December 15th, 2023 to January 14th of the temperature and VETS socio-economic and political state in France and Germany for the one-year forecasting horizon of forecasting experiment $N = 50$.

trajectories and the point forecasted load (see Zimmermann & Ziel (2024)) are depicted. Both load trajectories were reduced by the estimated intercept.

From these figures, we observe a difference in the deviation between scenario and point forecasts between France and Germany. France exhibits a larger discrepancy (approx. 33 GW) due to a stronger combined effect from extreme temperature, socio-economic and political scenarios. These risky scenarios contribute roughly 22 GW and 13 GW, respectively, to the load scenario in France, in contrast to 4 GW and 6 GW in Germany. For comparison, the corresponding point forecast model explained approximately 1-2 GW in French and German in-sample load by the socio-economic and political state through several years of winter months. In the same period, temperature effects on French load varied between 2 GW and 20 GW and were mostly negligible for German load, see Zimmermann & Ziel (2024) for details.

Consequently, the temperature effects induced by the extreme scenario are realistic real-world events, making them relevant for risk assessment. In France, the temperature scenarios represent more than 40% of the yearly moving average load (between 50-55 GW, see Fig. 2). This load impact provides a baseline for scenarios where similar temperature sensitivity is present across many European countries due to the widespread adoption of electric heating. For instance, one could apply a capacity-scaled temperature effect estimated for France in (6) to other European countries and evaluate the pan-European electricity demand under different temperature scenarios.

A comparison of the VETS to the VAR modeling approach (see Fig. 36-37, Appendix B) under extreme scenarios reveals lower impacts estimated for the VAR model (approximately -2 to 3 GW in France, -1 to 1 GW in Germany). Given that the Christmas holiday period falls near the end of the one-year forecasting horizon (January 31st, 2024 for experiment N=50) and that due to the stationary property of the VAR model state levels converge towards zero, this lower impact is reasonable. Conversely, recall that in the VETS model state levels are continued persistently in the prediction horizon.

The extreme autoregressive scenario has a minimal impact in both countries, contributing between 1 and 5 GW (positive or negative) to the load forecast.

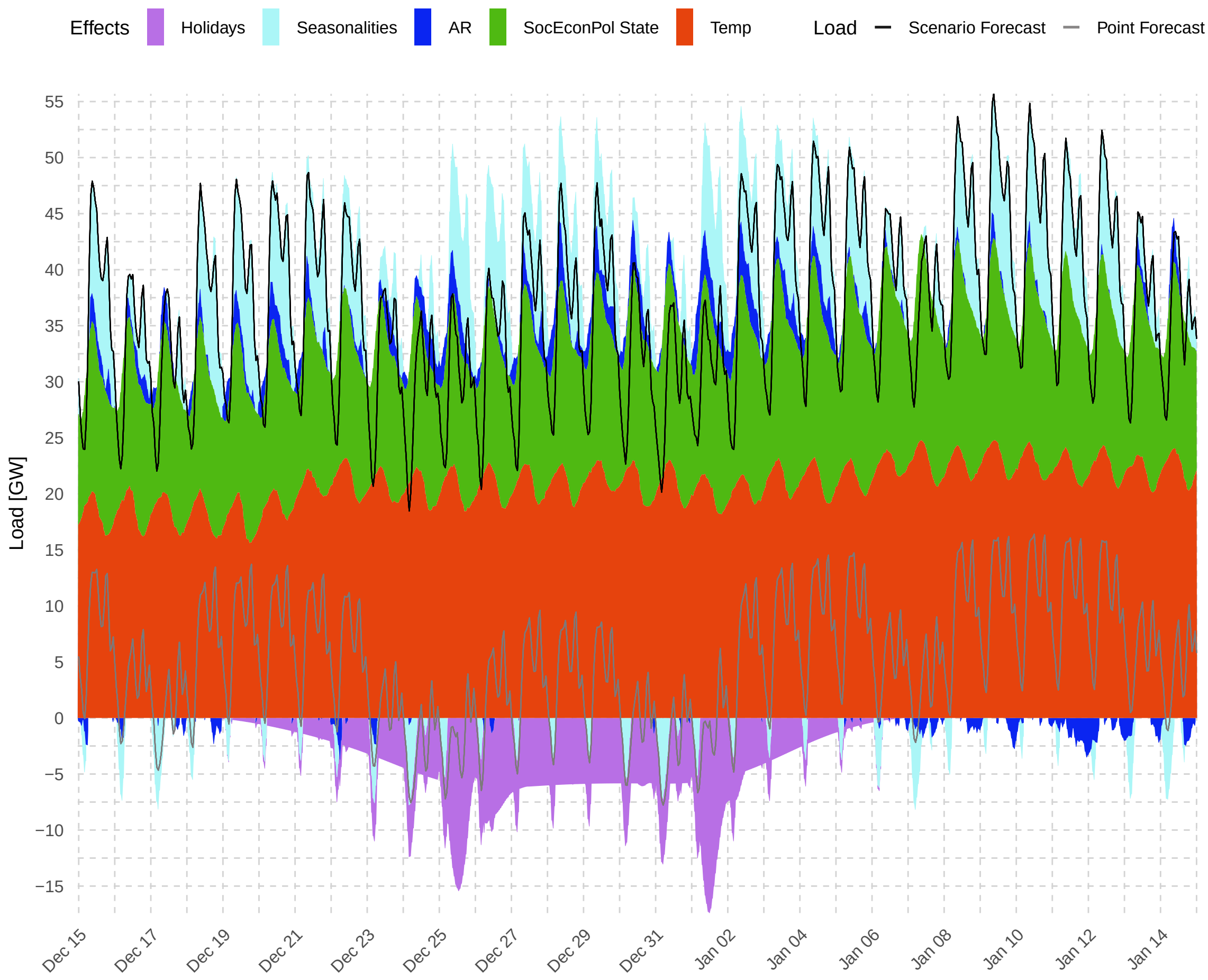

Figure 18: Forecasted load scenario (see (6) - (8)) decomposed in its modeling components for min. temperature, max. socio-economic and political state and max. autoregressive trajectories along with point forecasted load reduced by the estimated intercept for the VETS model (forecasting experiment $N = 50$) in France from December 15th, 2023 to January 14th, 2024.

## 8. Conclusion and Discussion

This paper presents an interpretable probabilistic and multivariate mid-term forecasting model for the hourly load that captures, besides all deterministic effects, uncertain meteorological, socio-economic and political, and autoregressive effects on load. The interpretability of the model allows for clear decomposition of transmission under any given scenario for uncertain load drivers, making it highly suitable for risk assessment. In a scenario analysis, extreme temperatures explained more than 40% of the French yearly moving average load.

The proposed probabilistic forecasting method may be improved further: Firstly, assuming homoscedasticity of residuals beyond hourly variation may not be appropriate, in particular, when electricity consumption is becoming more volatile due to widespread electric heating. As Haben et al. (2023); Ziel & Liu (2016) argue, variation in load is higher during peak hours and seasons. While we address this in an initial approach by calculating hour-specific sample covariance matrices, more sophisticated methods such as GARCH models applied to residuals, as summarized in Haben et al. (2023); Davis et al. (2016), or GAMs modeling time-varying distribution parameters, as proposed in Gioia et al. (2024); Browell & Fasiolo (2021), could be employed.

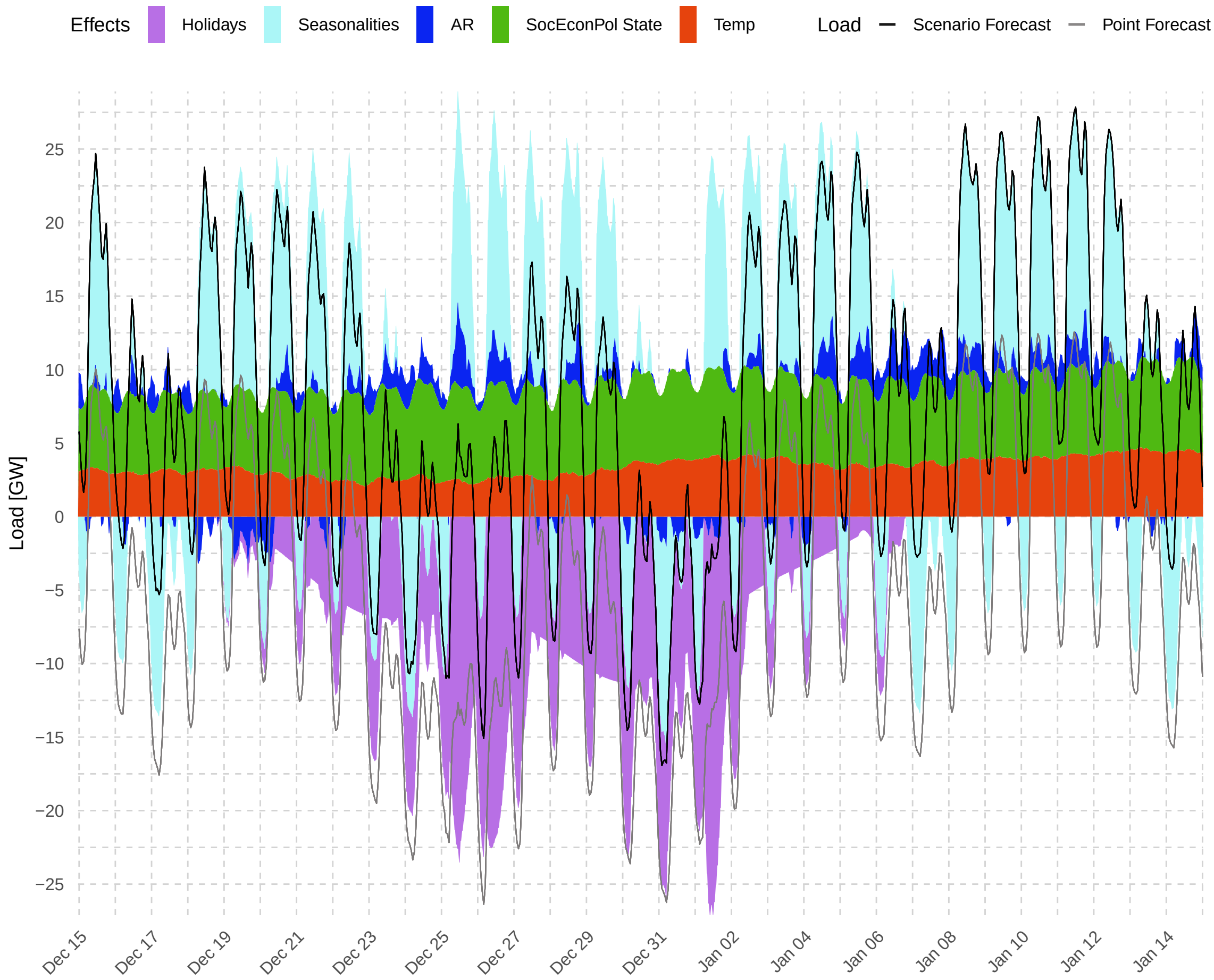

Figure 19: Forecasted load scenario (see (6) - (8)) decomposed in its modeling components for min. temperature, max. socio-economic and political state and max. autoregressive trajectories along with point forecasted load reduced by the estimated intercept for the VETS model (forecasting experiment $N = 50$) in Germany from December 15th, 2023 to January 14th, 2024.

Secondly, long-term socio-economic and political indicators could be included in the model, as implemented by González Grandón et al. (2024). Examples encompass GDP, population growth, country-specific conflict indices or a COVID-Lockdown dummy for in-sample handling of the COVID-19 pandemic. By incorporating these exogenous variables into the VETS model, see Svetunkov (2023) for modeling details, over or underestimation of persistent state forecasts and the consequent error accumulation could be mitigated. Alternatively, they could be incorporated in regression frameworks like our proposed VAR and VECM models on seasonally aggregated load values as suggested for the long-term trend model in González Grandón et al. (2024). However, a challenge in using exogenous variables is the need for reliable mid-term forecasts, as emphasized by González Grandón et al. (2024); Zimmermann & Ziel (2024). This was the primary reason we decided not to incorporate them. Alternatively, forecasting combinations of models like VETS assuming a persistent trend and models like VAR assuming a trend converging to zero could be considered for the socio-economic and political state.

### Author Contributions

M. Z.: Conceptualization; methodology; software; validation; formal analysis; investigation; data curation; writing - original draft; writing review and editing; visualization.

F. Z.: Conceptualization; methodology; validation; formal analysis; funding acquisition; resources; writing - review and editing; supervision.

### Funding Statement

This research has been funded by the Deutsche Forschungsgemeinschaft (DFG, German Research Foundation) – 505565850.

# Appendices

## A. Correlation Tables

| | AT | BE | BG | CZ | DK | EE | FI | FR | DE | GR | HU | IT | LV | LT | ME | NL | PL | PT | RO | RS | SK | SI | ES | SE |
|---|---|---|---|---|---|---|---|---|---|---|---|---|---|---|---|---|---|---|---|---|---|---|---|---|
| AT | 1.00 | 0.23 | 0.07 | 0.74 | 0.09 | 0.03 | 0.01 | 0.27 | 0.61 | 0.06 | 0.62 | 0.33 | 0.03 | 0.11 | 0.18 | 0.21 | 0.52 | −0.09 | 0.23 | 0.31 | 0.72 | 0.68 | 0.03 | 0.06 |
| BE | 0.23 | 1.00 | −0.08 | 0.24 | 0.29 | 0.04 | 0.04 | 0.39 | 0.62 | −0.05 | 0.07 | 0.12 | −0.01 | 0.02 | −0.02 | 0.90 | 0.15 | 0.00 | −0.04 | −0.02 | 0.12 | 0.10 | 0.04 | 0.15 |
| BG | 0.07 | −0.08 | 1.00 | 0.02 | −0.08 | −0.06 | −0.07 | −0.09 | −0.06 | 0.39 | 0.23 | 0.03 | −0.08 | −0.06 | 0.28 | −0.08 | 0.02 | −0.04 | 0.61 | 0.32 | 0.10 | 0.09 | 0.01 | −0.09 |
| CZ | 0.74 | 0.24 | 0.02 | 1.00 | 0.14 | 0.04 | 0.03 | 0.17 | 0.61 | 0.02 | 0.57 | 0.17 | 0.08 | 0.19 | 0.11 | 0.23 | 0.75 | −0.08 | 0.21 | 0.26 | 0.79 | 0.42 | 0.03 | 0.10 |
| DK | 0.09 | 0.29 | −0.08 | 0.14 | 1.00 | 0.15 | 0.17 | 0.08 | 0.41 | −0.05 | 0.00 | 0.00 | 0.18 | 0.13 | −0.05 | 0.35 | 0.14 | −0.03 | −0.06 | −0.03 | 0.06 | 0.02 | 0.03 | 0.57 |
| EE | 0.03 | 0.04 | −0.06 | 0.04 | 0.15 | 1.00 | 0.93 | 0.01 | 0.08 | −0.03 | −0.01 | −0.00 | 0.32 | 0.25 | −0.03 | 0.06 | 0.06 | −0.02 | −0.04 | −0.01 | 0.03 | 0.01 | 0.01 | 0.35 |
| FI | 0.01 | 0.04 | −0.07 | 0.03 | 0.17 | 0.93 | 1.00 | 0.00 | 0.07 | −0.06 | −0.03 | −0.02 | 0.36 | 0.26 | −0.05 | 0.06 | 0.06 | −0.02 | −0.06 | −0.03 | 0.01 | −0.01 | 0.02 | 0.41 |
| FR | 0.27 | 0.39 | −0.09 | 0.17 | 0.08 | 0.01 | 0.00 | 1.00 | 0.36 | −0.06 | 0.07 | 0.49 | −0.07 | −0.03 | 0.00 | 0.30 | 0.06 | 0.21 | −0.06 | 0.00 | 0.11 | 0.30 | 0.14 | 0.04 |
| DE | 0.61 | 0.62 | −0.06 | 0.61 | 0.41 | 0.08 | 0.07 | 0.36 | 1.00 | −0.04 | 0.28 | 0.21 | 0.06 | 0.12 | 0.03 | 0.61 | 0.48 | −0.07 | 0.05 | 0.10 | 0.39 | 0.30 | 0.04 | 0.25 |
| GR | 0.06 | −0.05 | 0.39 | 0.02 | −0.05 | −0.03 | −0.06 | −0.06 | −0.04 | 1.00 | 0.12 | 0.08 | −0.07 | −0.06 | 0.32 | −0.04 | 0.02 | −0.05 | 0.25 | 0.19 | 0.06 | 0.05 | −0.01 | −0.03 |
| HU | 0.62 | 0.07 | 0.23 | 0.57 | 0.00 | −0.01 | −0.03 | 0.07 | 0.28 | 0.12 | 1.00 | 0.16 | 0.02 | 0.13 | 0.28 | 0.06 | 0.50 | −0.06 | 0.57 | 0.56 | 0.81 | 0.52 | 0.02 | −0.02 |
| IT | 0.33 | 0.12 | 0.03 | 0.17 | 0.00 | −0.00 | −0.02 | 0.49 | 0.21 | 0.08 | 0.16 | 1.00 | −0.06 | −0.03 | 0.19 | 0.08 | 0.07 | 0.14 | 0.04 | 0.12 | 0.18 | 0.43 | 0.08 | −0.00 |
| LV | 0.03 | −0.01 | −0.08 | 0.08 | 0.18 | 0.32 | 0.36 | −0.07 | 0.06 | −0.07 | 0.02 | −0.06 | 1.00 | 0.65 | −0.03 | 0.01 | 0.18 | −0.06 | −0.03 | −0.01 | 0.08 | −0.01 | −0.00 | 0.33 |
| LT | 0.11 | 0.02 | −0.06 | 0.19 | 0.13 | 0.25 | 0.26 | −0.03 | 0.12 | −0.06 | 0.13 | −0.03 | 0.65 | 1.00 | −0.01 | 0.03 | 0.32 | −0.04 | 0.06 | 0.05 | 0.21 | 0.05 | 0.02 | 0.22 |
| ME | 0.18 | −0.02 | 0.28 | 0.11 | −0.05 | −0.03 | −0.05 | 0.00 | 0.03 | 0.32 | 0.28 | 0.19 | −0.03 | −0.01 | 1.00 | −0.03 | 0.09 | −0.03 | 0.32 | 0.32 | 0.19 | 0.23 | −0.02 | −0.05 |
| NL | 0.21 | 0.90 | −0.08 | 0.23 | 0.35 | 0.06 | 0.06 | 0.30 | 0.61 | −0.04 | 0.06 | 0.08 | 0.01 | 0.03 | −0.03 | 1.00 | 0.16 | −0.03 | −0.04 | −0.02 | 0.11 | 0.07 | 0.04 | 0.18 |
| PL | 0.52 | 0.15 | 0.02 | 0.75 | 0.14 | 0.06 | 0.06 | 0.06 | 0.48 | 0.02 | 0.50 | 0.07 | 0.18 | 0.32 | 0.09 | 0.16 | 1.00 | −0.08 | 0.23 | 0.25 | 0.68 | 0.29 | 0.02 | 0.13 |
| PT | −0.09 | 0.00 | −0.04 | −0.08 | −0.03 | −0.02 | −0.02 | 0.21 | −0.07 | −0.05 | −0.06 | 0.14 | −0.06 | −0.04 | −0.03 | −0.03 | −0.08 | 1.00 | −0.05 | −0.06 | −0.07 | 0.04 | 0.17 | −0.02 |
| RO | 0.23 | −0.04 | 0.61 | 0.21 | −0.06 | −0.04 | −0.06 | −0.06 | 0.05 | 0.25 | 0.57 | 0.04 | −0.03 | 0.06 | 0.32 | −0.04 | 0.23 | −0.05 | 1.00 | 0.53 | 0.38 | 0.21 | 0.03 | −0.08 |
| RS | 0.31 | −0.02 | 0.32 | 0.26 | −0.03 | −0.01 | −0.03 | 0.00 | 0.10 | 0.19 | 0.56 | 0.12 | −0.01 | 0.05 | 0.32 | −0.02 | 0.25 | −0.06 | 0.53 | 1.00 | 0.39 | 0.33 | 0.02 | −0.04 |
| SK | 0.72 | 0.12 | 0.10 | 0.79 | 0.06 | 0.03 | 0.01 | 0.11 | 0.39 | 0.06 | 0.81 | 0.18 | 0.08 | 0.21 | 0.19 | 0.11 | 0.68 | −0.07 | 0.38 | 0.39 | 1.00 | 0.49 | 0.02 | 0.04 |
| SI | 0.68 | 0.10 | 0.09 | 0.42 | 0.02 | 0.01 | −0.01 | 0.30 | 0.30 | 0.05 | 0.52 | 0.43 | −0.01 | 0.05 | 0.23 | 0.07 | 0.29 | 0.04 | 0.21 | 0.33 | 0.49 | 1.00 | 0.05 | 0.01 |
| ES | 0.03 | 0.04 | 0.01 | 0.03 | 0.03 | 0.01 | 0.02 | 0.14 | 0.04 | −0.01 | 0.02 | 0.08 | −0.00 | 0.02 | −0.02 | 0.04 | 0.02 | 0.17 | 0.03 | 0.02 | 0.02 | 0.05 | 1.00 | 0.03 |
| SE | 0.06 | 0.15 | −0.09 | 0.10 | 0.57 | 0.35 | 0.41 | 0.04 | 0.25 | −0.03 | −0.02 | −0.00 | 0.33 | 0.22 | −0.05 | 0.18 | 0.13 | −0.02 | −0.08 | −0.04 | 0.04 | 0.01 | 0.03 | 1.00 |

Figure 20: The in-sample correlation matrix $\widehat{\text{Corr}}(\boldsymbol{\epsilon}_t^{\text{Temp}})$ for $t \bmod 24 = 8$ from February 6th, 2019 to February 1st, 2023 and smoothing parameter $\alpha = 1/6$.

| | AT | BE | BG | CZ | DK | EE | FI | FR | DE | GR | HU | IT | LV | LT | ME | NL | PL | PT | RO | RS | SK | SI | ES | SE |
|---|---|---|---|---|---|---|---|---|---|---|---|---|---|---|---|---|---|---|---|---|---|---|---|---|
| AT | 1.00 | 0.43 | 0.27 | 0.57 | 0.23 | 0.16 | 0.06 | 0.41 | 0.46 | 0.23 | 0.57 | 0.44 | 0.35 | 0.36 | 0.24 | 0.48 | 0.59 | 0.10 | 0.39 | 0.27 | 0.60 | 0.59 | 0.19 | 0.16 |
| BE | 0.43 | 1.00 | 0.07 | 0.50 | 0.29 | 0.31 | 0.23 | 0.35 | 0.46 | 0.02 | 0.20 | 0.18 | 0.23 | 0.17 | −0.04 | 0.62 | 0.40 | 0.02 | 0.21 | 0.01 | 0.48 | 0.32 | 0.22 | 0.29 |
| BG | 0.27 | 0.07 | 1.00 | 0.19 | −0.04 | 0.11 | −0.08 | 0.00 | 0.05 | 0.57 | 0.38 | 0.17 | 0.14 | 0.05 | 0.46 | 0.01 | 0.23 | −0.12 | 0.55 | 0.40 | 0.16 | 0.21 | 0.04 | −0.12 |
| CZ | 0.57 | 0.50 | 0.19 | 1.00 | 0.27 | 0.29 | 0.11 | 0.39 | 0.51 | 0.03 | 0.53 | 0.16 | 0.28 | 0.33 | 0.16 | 0.42 | 0.61 | 0.10 | 0.36 | 0.14 | 0.71 | 0.50 | 0.24 | 0.22 |
| DK | 0.23 | 0.29 | −0.04 | 0.27 | 1.00 | 0.32 | 0.27 | 0.06 | 0.50 | −0.03 | 0.13 | −0.08 | 0.40 | 0.21 | 0.01 | 0.29 | 0.16 | −0.14 | −0.02 | −0.05 | 0.25 | 0.07 | −0.12 | 0.52 |
| EE | 0.16 | 0.31 | 0.11 | 0.29 | 0.32 | 1.00 | 0.33 | 0.17 | 0.26 | 0.00 | 0.20 | 0.07 | 0.56 | 0.38 | 0.08 | 0.22 | 0.21 | 0.02 | 0.15 | −0.02 | 0.25 | 0.14 | 0.11 | 0.46 |
| FI | 0.06 | 0.23 | −0.08 | 0.11 | 0.27 | 0.33 | 1.00 | 0.05 | 0.23 | 0.04 | 0.02 | −0.01 | 0.25 | 0.18 | −0.00 | 0.16 | −0.04 | −0.08 | −0.04 | 0.03 | 0.17 | 0.01 | 0.11 | 0.44 |
| FR | 0.41 | 0.35 | 0.00 | 0.39 | 0.06 | 0.17 | 0.05 | 1.00 | 0.37 | 0.09 | 0.28 | 0.40 | 0.13 | 0.26 | 0.03 | 0.31 | 0.30 | 0.28 | 0.29 | 0.04 | 0.39 | 0.45 | 0.53 | 0.12 |
| DE | 0.46 | 0.46 | 0.05 | 0.51 | 0.50 | 0.26 | 0.23 | 0.37 | 1.00 | 0.02 | 0.26 | 0.21 | 0.29 | 0.13 | 0.10 | 0.54 | 0.28 | 0.05 | 0.18 | 0.02 | 0.49 | 0.34 | 0.19 | 0.40 |
| GR | 0.23 | 0.02 | 0.57 | 0.03 | −0.03 | 0.00 | 0.04 | 0.09 | 0.02 | 1.00 | 0.23 | 0.36 | 0.08 | −0.00 | 0.43 | 0.01 | 0.12 | −0.09 | 0.40 | 0.35 | 0.10 | 0.20 | 0.18 | −0.07 |
| HU | 0.57 | 0.20 | 0.38 | 0.53 | 0.13 | 0.20 | 0.02 | 0.28 | 0.26 | 0.23 | 1.00 | 0.32 | 0.30 | 0.31 | 0.26 | 0.29 | 0.55 | 0.07 | 0.61 | 0.37 | 0.56 | 0.56 | 0.15 | 0.21 |
| IT | 0.44 | 0.18 | 0.17 | 0.16 | −0.08 | 0.07 | −0.01 | 0.40 | 0.21 | 0.36 | 0.32 | 1.00 | 0.13 | 0.20 | 0.24 | 0.11 | 0.24 | 0.30 | 0.38 | 0.08 | 0.25 | 0.39 | 0.49 | −0.03 |
| LV | 0.35 | 0.23 | 0.14 | 0.28 | 0.40 | 0.56 | 0.25 | 0.13 | 0.29 | 0.08 | 0.30 | 0.13 | 1.00 | 0.51 | 0.20 | 0.28 | 0.34 | −0.07 | 0.25 | 0.10 | 0.30 | 0.16 | −0.02 | 0.47 |
| LT | 0.36 | 0.17 | 0.05 | 0.33 | 0.21 | 0.38 | 0.18 | 0.26 | 0.13 | −0.00 | 0.31 | 0.20 | 0.51 | 1.00 | −0.02 | 0.18 | 0.50 | 0.11 | 0.21 | 0.01 | 0.34 | 0.24 | 0.13 | 0.24 |
| ME | 0.24 | −0.04 | 0.46 | 0.16 | 0.01 | 0.08 | −0.00 | 0.03 | 0.10 | 0.43 | 0.26 | 0.24 | 0.20 | −0.02 | 1.00 | 0.09 | 0.09 | −0.02 | 0.30 | 0.32 | 0.13 | 0.29 | 0.11 | 0.02 |
| NL | 0.48 | 0.62 | 0.01 | 0.42 | 0.29 | 0.22 | 0.16 | 0.31 | 0.54 | 0.01 | 0.29 | 0.11 | 0.28 | 0.18 | 0.09 | 1.00 | 0.35 | −0.01 | 0.09 | 0.08 | 0.40 | 0.41 | 0.12 | 0.26 |
| PL | 0.59 | 0.40 | 0.23 | 0.61 | 0.16 | 0.21 | −0.04 | 0.30 | 0.28 | 0.12 | 0.55 | 0.24 | 0.34 | 0.50 | 0.09 | 0.35 | 1.00 | 0.00 | 0.42 | 0.19 | 0.55 | 0.40 | 0.14 | 0.15 |
| PT | 0.10 | 0.02 | −0.12 | 0.10 | −0.14 | 0.02 | −0.08 | 0.28 | 0.05 | −0.09 | 0.07 | 0.30 | −0.07 | 0.11 | −0.02 | −0.01 | 0.00 | 1.00 | 0.07 | −0.22 | 0.09 | 0.17 | 0.49 | −0.03 |
| RO | 0.39 | 0.21 | 0.55 | 0.36 | −0.02 | 0.15 | −0.04 | 0.29 | 0.18 | 0.40 | 0.61 | 0.38 | 0.25 | 0.21 | 0.30 | 0.09 | 0.42 | 0.07 | 1.00 | 0.37 | 0.37 | 0.39 | 0.30 | 0.07 |
| RS | 0.27 | 0.01 | 0.40 | 0.14 | −0.05 | −0.02 | 0.03 | 0.04 | 0.02 | 0.35 | 0.37 | 0.08 | 0.10 | 0.01 | 0.32 | 0.08 | 0.19 | −0.22 | 0.37 | 1.00 | 0.19 | 0.24 | 0.02 | 0.02 |
| SK | 0.60 | 0.48 | 0.16 | 0.71 | 0.25 | 0.25 | 0.17 | 0.39 | 0.49 | 0.10 | 0.56 | 0.25 | 0.30 | 0.34 | 0.13 | 0.40 | 0.55 | 0.09 | 0.37 | 0.19 | 1.00 | 0.46 | 0.21 | 0.27 |
| SI | 0.59 | 0.32 | 0.21 | 0.50 | 0.07 | 0.14 | 0.01 | 0.45 | 0.34 | 0.20 | 0.56 | 0.39 | 0.16 | 0.24 | 0.29 | 0.41 | 0.40 | 0.17 | 0.39 | 0.24 | 0.46 | 1.00 | 0.31 | 0.13 |
| ES | 0.19 | 0.22 | 0.04 | 0.24 | −0.12 | 0.11 | 0.11 | 0.53 | 0.19 | 0.18 | 0.15 | 0.49 | −0.02 | 0.13 | 0.11 | 0.12 | 0.14 | 0.49 | 0.30 | 0.02 | 0.21 | 0.31 | 1.00 | −0.03 |
| SE | 0.16 | 0.29 | −0.12 | 0.22 | 0.52 | 0.46 | 0.44 | 0.12 | 0.40 | −0.07 | 0.21 | −0.03 | 0.47 | 0.24 | 0.02 | 0.26 | 0.15 | −0.03 | 0.07 | 0.02 | 0.27 | 0.13 | −0.03 | 1.00 |

Figure 21: The in-sample correlation matrix $\widehat{\text{Corr}}(\boldsymbol{\epsilon}^{\text{VAR}}_{\tau=1,\ldots,T/(7\times24)})$ for $\tau$ from February 6th, 2019 to February 1st, 2023.

| | AT | BE | BG | CZ | DK | EE | FI | FR | DE | GR | HU | IT | LV | LT | ME | NL | PL | PT | RO | RS | SK | SI | ES | SE |
|---|---|---|---|---|---|---|---|---|---|---|---|---|---|---|---|---|---|---|---|---|---|---|---|---|
| AT | 1.00 | 0.46 | 0.37 | 0.61 | 0.43 | 0.37 | 0.23 | 0.15 | 0.29 | 0.18 | 0.60 | −0.02 | 0.39 | 0.42 | 0.37 | 0.43 | 0.58 | 0.22 | 0.46 | 0.39 | 0.55 | 0.51 | −0.03 | 0.30 |
| BE | 0.46 | 1.00 | 0.30 | 0.59 | 0.51 | 0.48 | 0.35 | 0.06 | 0.23 | −0.02 | 0.42 | −0.27 | 0.46 | 0.46 | 0.43 | 0.51 | 0.41 | 0.20 | 0.33 | 0.32 | 0.56 | 0.48 | −0.03 | 0.43 |
| BG | 0.37 | 0.30 | 1.00 | 0.47 | 0.43 | 0.65 | 0.28 | −0.31 | −0.27 | 0.53 | 0.61 | −0.37 | 0.66 | 0.64 | 0.68 | 0.05 | 0.27 | 0.22 | 0.71 | 0.67 | 0.61 | 0.65 | −0.23 | 0.10 |
| CZ | 0.61 | 0.59 | 0.47 | 1.00 | 0.59 | 0.64 | 0.42 | −0.10 | 0.15 | 0.08 | 0.64 | −0.42 | 0.64 | 0.66 | 0.62 | 0.34 | 0.56 | 0.31 | 0.48 | 0.46 | 0.80 | 0.69 | −0.11 | 0.40 |
| DK | 0.43 | 0.51 | 0.43 | 0.59 | 1.00 | 0.80 | 0.58 | −0.29 | 0.11 | 0.12 | 0.56 | −0.56 | 0.79 | 0.80 | 0.76 | 0.26 | 0.22 | 0.31 | 0.37 | 0.50 | 0.74 | 0.72 | −0.34 | 0.60 |
| EE | 0.37 | 0.48 | 0.65 | 0.64 | 0.80 | 1.00 | 0.60 | −0.40 | −0.22 | 0.23 | 0.68 | −0.59 | 0.99 | 0.98 | 0.97 | 0.11 | 0.21 | 0.46 | 0.52 | 0.65 | 0.88 | 0.91 | −0.30 | 0.41 |
| FI | 0.23 | 0.35 | 0.28 | 0.42 | 0.58 | 0.60 | 1.00 | −0.22 | 0.01 | 0.08 | 0.35 | −0.38 | 0.57 | 0.58 | 0.54 | 0.12 | 0.06 | 0.20 | 0.23 | 0.34 | 0.54 | 0.50 | −0.12 | 0.50 |
| FR | 0.15 | 0.06 | −0.31 | −0.10 | −0.29 | −0.40 | −0.22 | 1.00 | 0.40 | −0.17 | −0.18 | 0.46 | −0.41 | −0.38 | −0.42 | 0.24 | 0.14 | 0.05 | −0.11 | −0.33 | −0.29 | −0.29 | 0.54 | −0.07 |
| DE | 0.29 | 0.23 | −0.27 | 0.15 | 0.11 | −0.22 | 0.01 | 0.40 | 1.00 | −0.20 | −0.08 | 0.19 | −0.24 | −0.23 | −0.27 | 0.49 | 0.18 | −0.04 | −0.11 | −0.21 | −0.09 | −0.19 | 0.20 | 0.34 |
| GR | 0.18 | −0.02 | 0.53 | 0.08 | 0.12 | 0.23 | 0.08 | −0.17 | −0.20 | 1.00 | 0.27 | 0.08 | 0.25 | 0.24 | 0.27 | −0.02 | 0.09 | 0.02 | 0.36 | 0.39 | 0.21 | 0.24 | −0.07 | −0.04 |
| HU | 0.60 | 0.42 | 0.61 | 0.64 | 0.56 | 0.68 | 0.35 | −0.18 | −0.08 | 0.27 | 1.00 | −0.29 | 0.68 | 0.69 | 0.68 | 0.19 | 0.47 | 0.33 | 0.68 | 0.64 | 0.75 | 0.74 | −0.16 | 0.32 |
| IT | −0.02 | −0.27 | −0.37 | −0.42 | −0.56 | −0.59 | −0.38 | 0.46 | 0.19 | 0.08 | −0.29 | 1.00 | −0.58 | −0.55 | −0.58 | −0.08 | −0.01 | −0.01 | −0.19 | −0.40 | −0.52 | −0.50 | 0.51 | −0.34 |
| LV | 0.39 | 0.46 | 0.66 | 0.64 | 0.79 | 0.99 | 0.57 | −0.41 | −0.24 | 0.25 | 0.68 | −0.58 | 1.00 | 0.98 | 0.99 | 0.09 | 0.21 | 0.47 | 0.54 | 0.67 | 0.89 | 0.93 | −0.32 | 0.37 |
| LT | 0.42 | 0.46 | 0.64 | 0.66 | 0.80 | 0.98 | 0.58 | −0.38 | −0.23 | 0.24 | 0.69 | −0.55 | 0.98 | 1.00 | 0.97 | 0.10 | 0.27 | 0.49 | 0.55 | 0.65 | 0.90 | 0.92 | −0.29 | 0.38 |
| ME | 0.37 | 0.43 | 0.68 | 0.62 | 0.76 | 0.97 | 0.54 | −0.42 | −0.27 | 0.27 | 0.68 | −0.58 | 0.99 | 0.97 | 1.00 | 0.08 | 0.18 | 0.48 | 0.55 | 0.69 | 0.88 | 0.93 | −0.32 | 0.33 |
| NL | 0.43 | 0.51 | 0.05 | 0.34 | 0.26 | 0.11 | 0.12 | 0.24 | 0.49 | −0.02 | 0.19 | −0.08 | 0.09 | 0.10 | 0.08 | 1.00 | 0.33 | −0.01 | 0.06 | 0.07 | 0.21 | 0.16 | −0.05 | 0.33 |
| PL | 0.58 | 0.41 | 0.27 | 0.56 | 0.22 | 0.21 | 0.06 | 0.14 | 0.18 | 0.09 | 0.47 | −0.01 | 0.21 | 0.27 | 0.18 | 0.33 | 1.00 | 0.12 | 0.42 | 0.28 | 0.40 | 0.29 | 0.06 | 0.21 |
| PT | 0.22 | 0.20 | 0.22 | 0.31 | 0.31 | 0.46 | 0.20 | 0.05 | −0.04 | 0.02 | 0.33 | −0.01 | 0.47 | 0.49 | 0.48 | −0.01 | 0.12 | 1.00 | 0.28 | 0.21 | 0.42 | 0.46 | 0.30 | 0.14 |
| RO | 0.46 | 0.33 | 0.71 | 0.48 | 0.37 | 0.52 | 0.23 | −0.11 | −0.11 | 0.36 | 0.68 | −0.19 | 0.54 | 0.55 | 0.55 | 0.06 | 0.42 | 0.28 | 1.00 | 0.62 | 0.59 | 0.60 | −0.06 | 0.18 |
| RS | 0.39 | 0.32 | 0.67 | 0.46 | 0.50 | 0.65 | 0.34 | −0.33 | −0.21 | 0.39 | 0.64 | −0.40 | 0.67 | 0.65 | 0.69 | 0.07 | 0.28 | 0.21 | 0.62 | 1.00 | 0.64 | 0.67 | −0.27 | 0.24 |
| SK | 0.55 | 0.56 | 0.61 | 0.80 | 0.74 | 0.88 | 0.54 | −0.29 | −0.09 | 0.21 | 0.75 | −0.52 | 0.89 | 0.90 | 0.88 | 0.21 | 0.40 | 0.42 | 0.59 | 0.64 | 1.00 | 0.89 | −0.25 | 0.39 |
| SI | 0.51 | 0.48 | 0.65 | 0.69 | 0.72 | 0.91 | 0.50 | −0.29 | −0.19 | 0.24 | 0.74 | −0.50 | 0.93 | 0.92 | 0.93 | 0.16 | 0.29 | 0.46 | 0.60 | 0.67 | 0.89 | 1.00 | −0.24 | 0.33 |
| ES | −0.03 | −0.03 | −0.23 | −0.11 | −0.34 | −0.30 | −0.12 | 0.54 | 0.20 | −0.07 | −0.16 | 0.51 | −0.32 | −0.29 | −0.32 | −0.05 | 0.06 | 0.30 | −0.06 | −0.27 | −0.25 | −0.24 | 1.00 | −0.17 |
| SE | 0.30 | 0.43 | 0.10 | 0.40 | 0.60 | 0.41 | 0.50 | −0.07 | 0.34 | −0.04 | 0.32 | −0.34 | 0.37 | 0.38 | 0.33 | 0.33 | 0.21 | 0.14 | 0.18 | 0.24 | 0.39 | 0.33 | −0.17 | 1.00 |

Figure 22: The in-sample correlation matrix $\widehat{\text{Corr}}(\boldsymbol{\epsilon}^{\text{VETS}}_{\tau=1,\ldots,T/(7\times24)})$ for $\tau$ from February 6th, 2019 to February 1st, 2023.

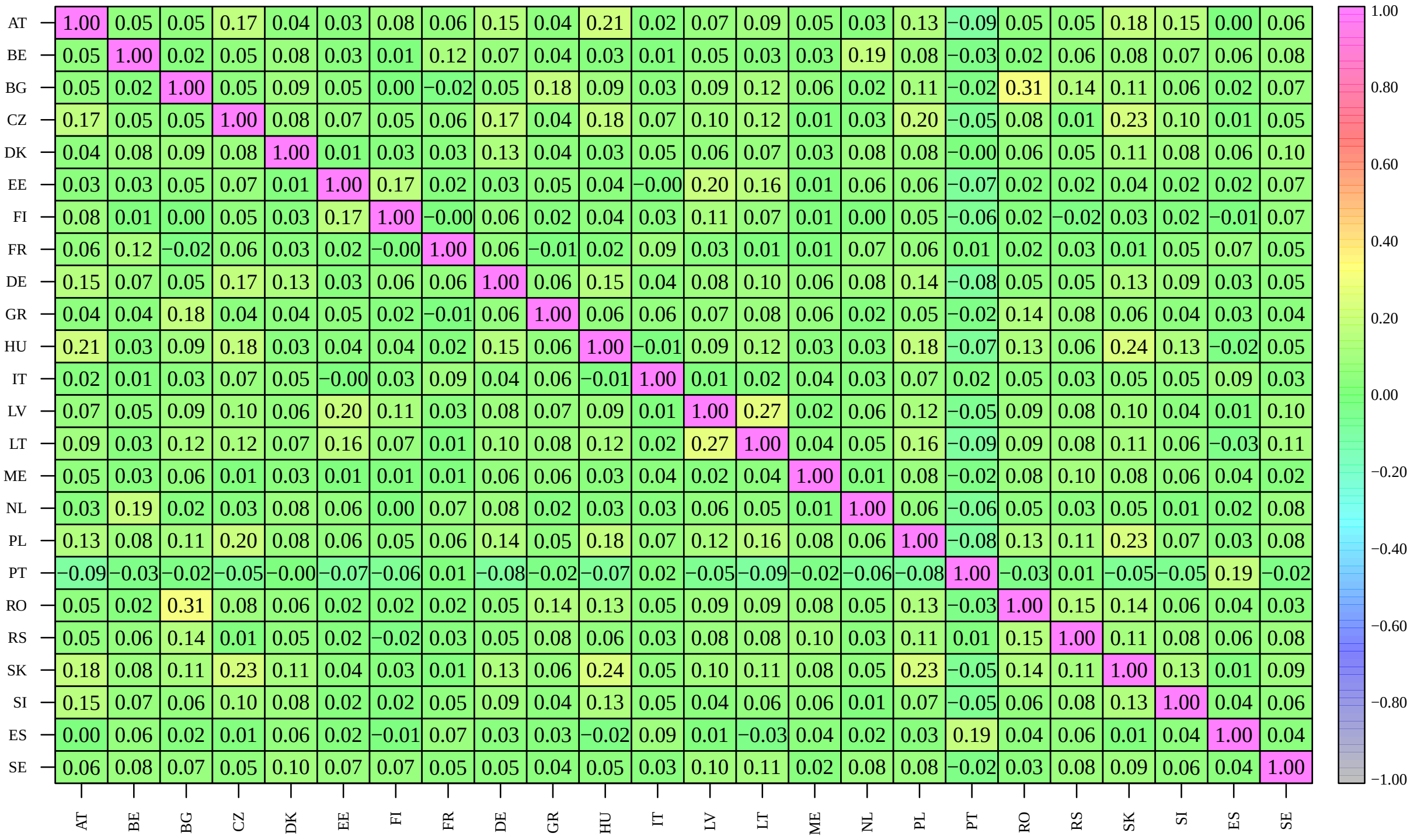

Figure 23: The in-sample correlation matrix $\widehat{\text{Corr}}(\boldsymbol{\epsilon}_t^{\text{AR}})$ for $t \bmod 24 = 8$ from February 6th, 2019 to February 1st, 2023 and the VAR model for the socio-economic and political state.

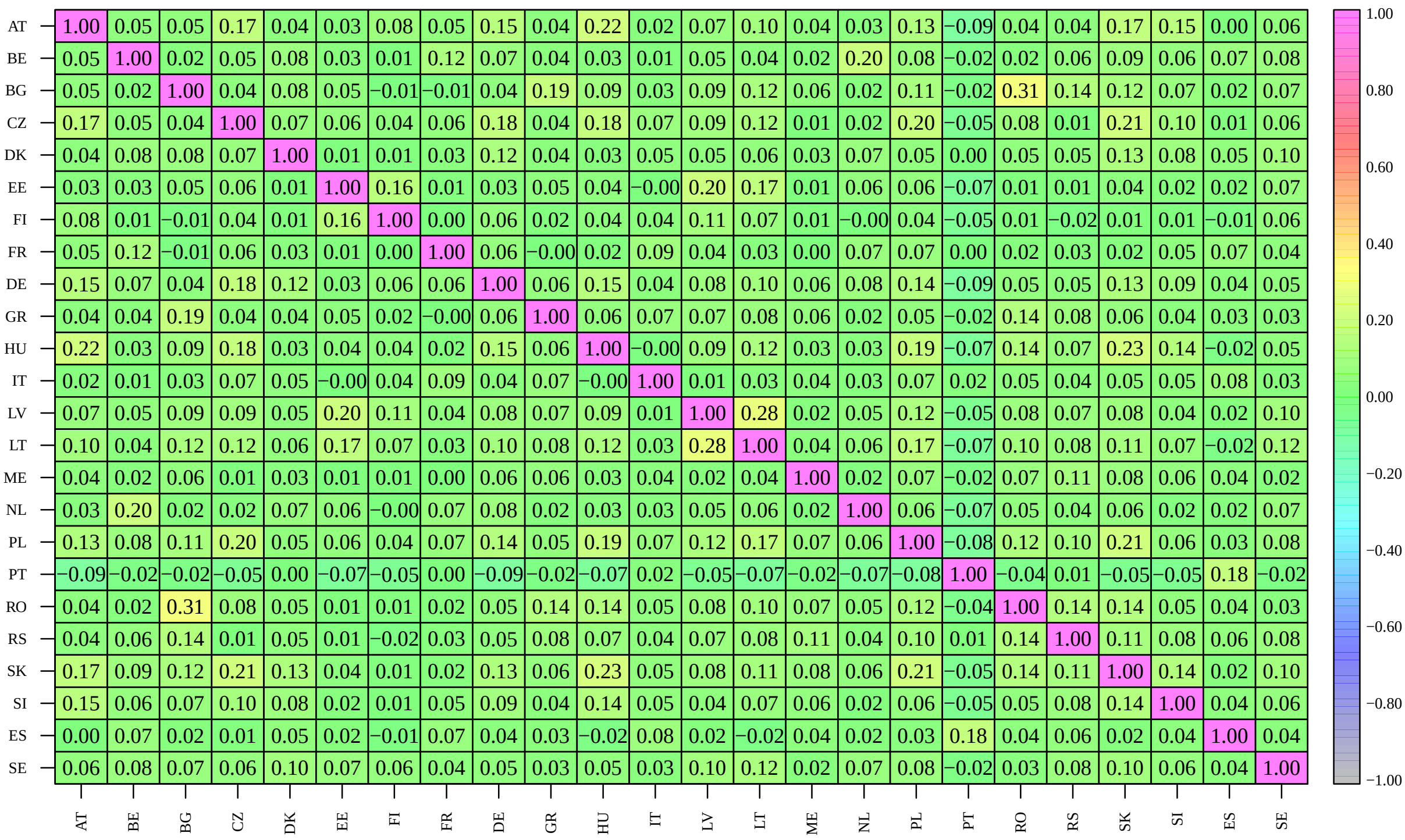

Figure 24: The in-sample correlation matrix $\widehat{\text{Corr}}(\boldsymbol{\epsilon}_t^{\text{AR}})$ for $t \bmod 24 = 8$ from February 6th, 2019 to February 1st, 2023 and the VECM model for the socio-economic state.

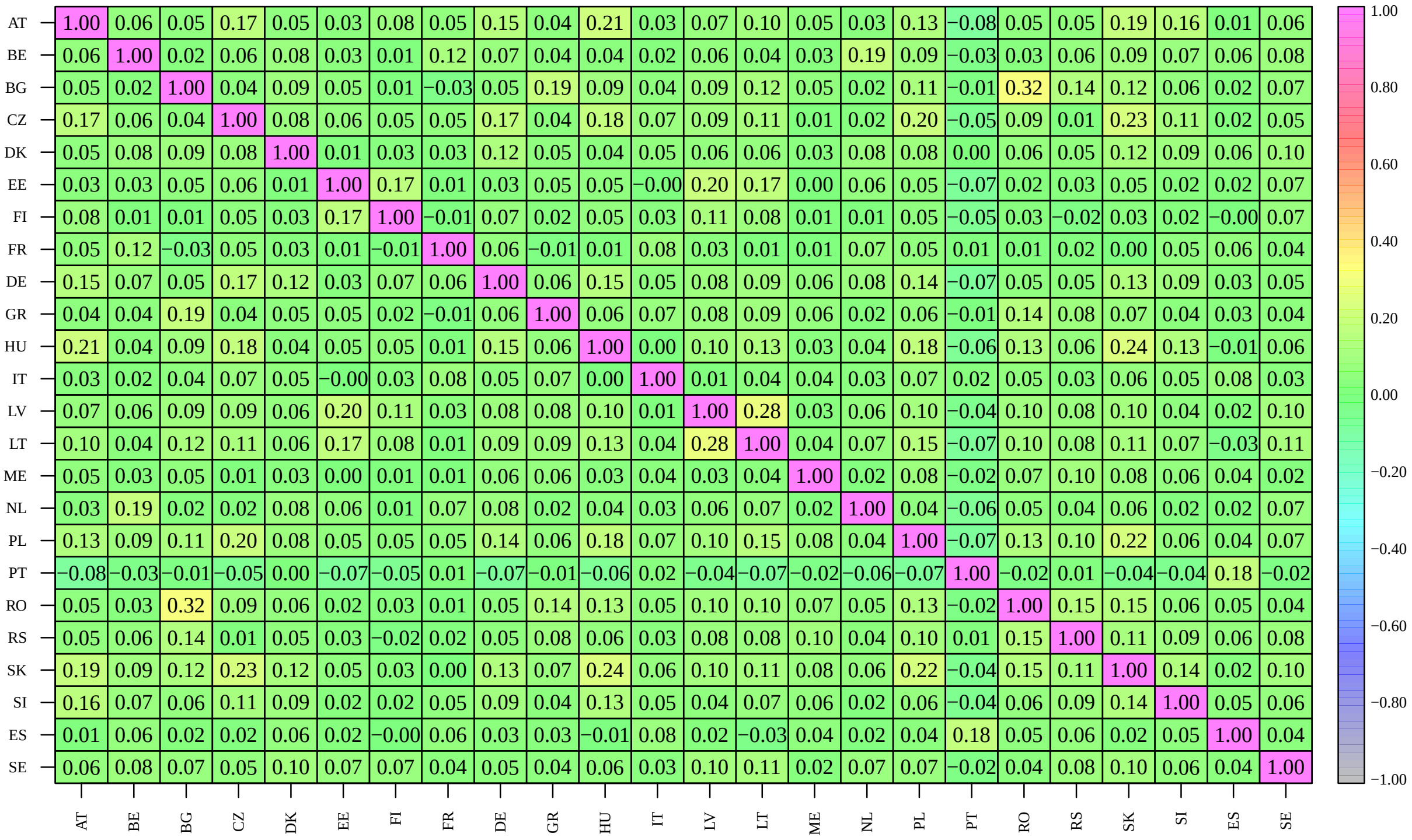

Figure 25: The in-sample correlation matrix $\widehat{\mathrm{Corr}}(\boldsymbol{\epsilon}_t^{\mathrm{AR}})$ for $t$ mod $24 = 8$ from February 6th, 2019 to February 1st, 2023 and the VETS model for the socio-economic and political state.

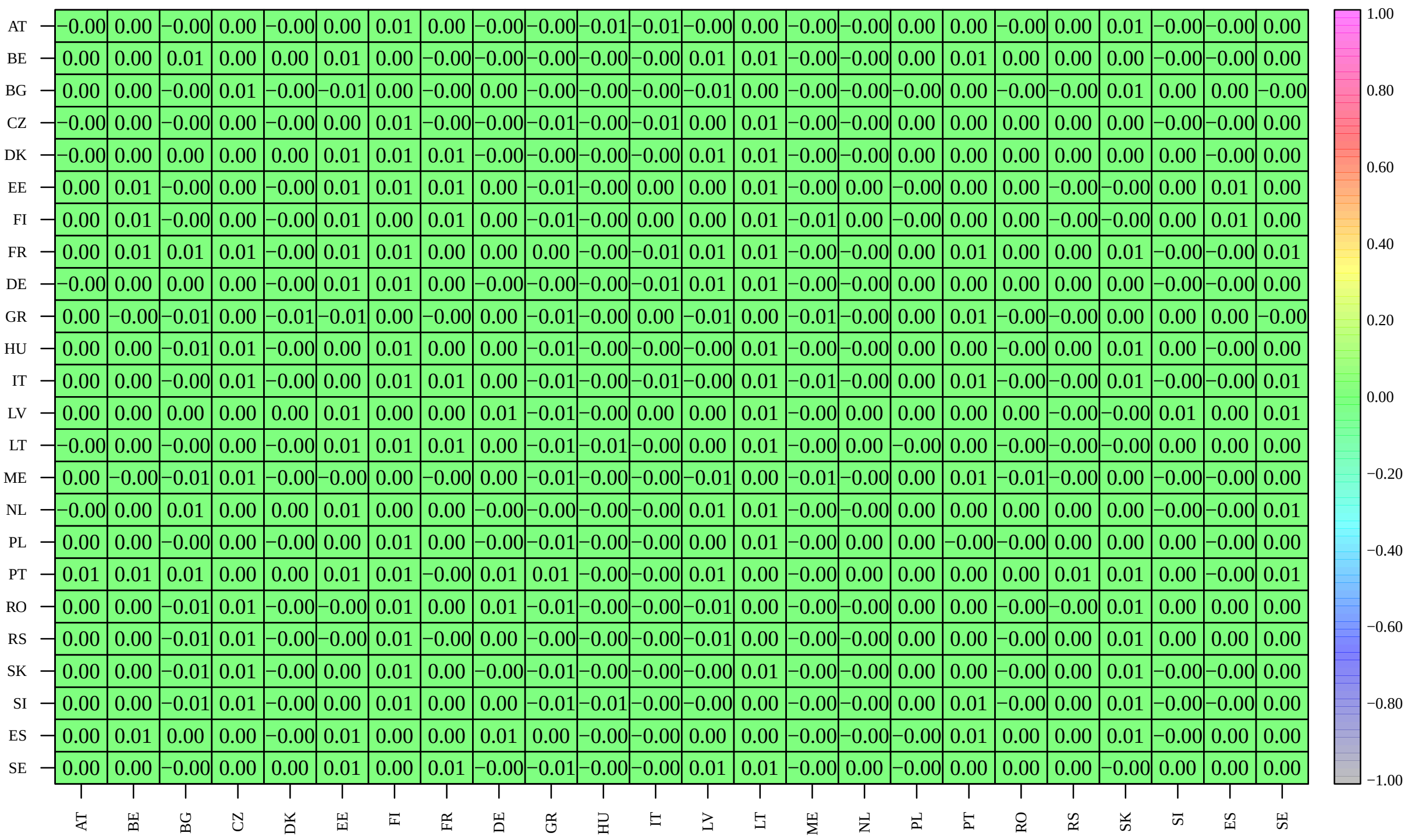

Figure 26: The in-sample correlation matrix $\widehat{\mathrm{Corr}}(\boldsymbol{\epsilon}_{t=1,\dots,T}^{\mathrm{Temp}}, \boldsymbol{\epsilon}_{t=1,\dots,T}^{\mathrm{VECM}})$ for $t$ from February 6th, 2019 to February 1st, 2023, $\alpha = 1/24$.

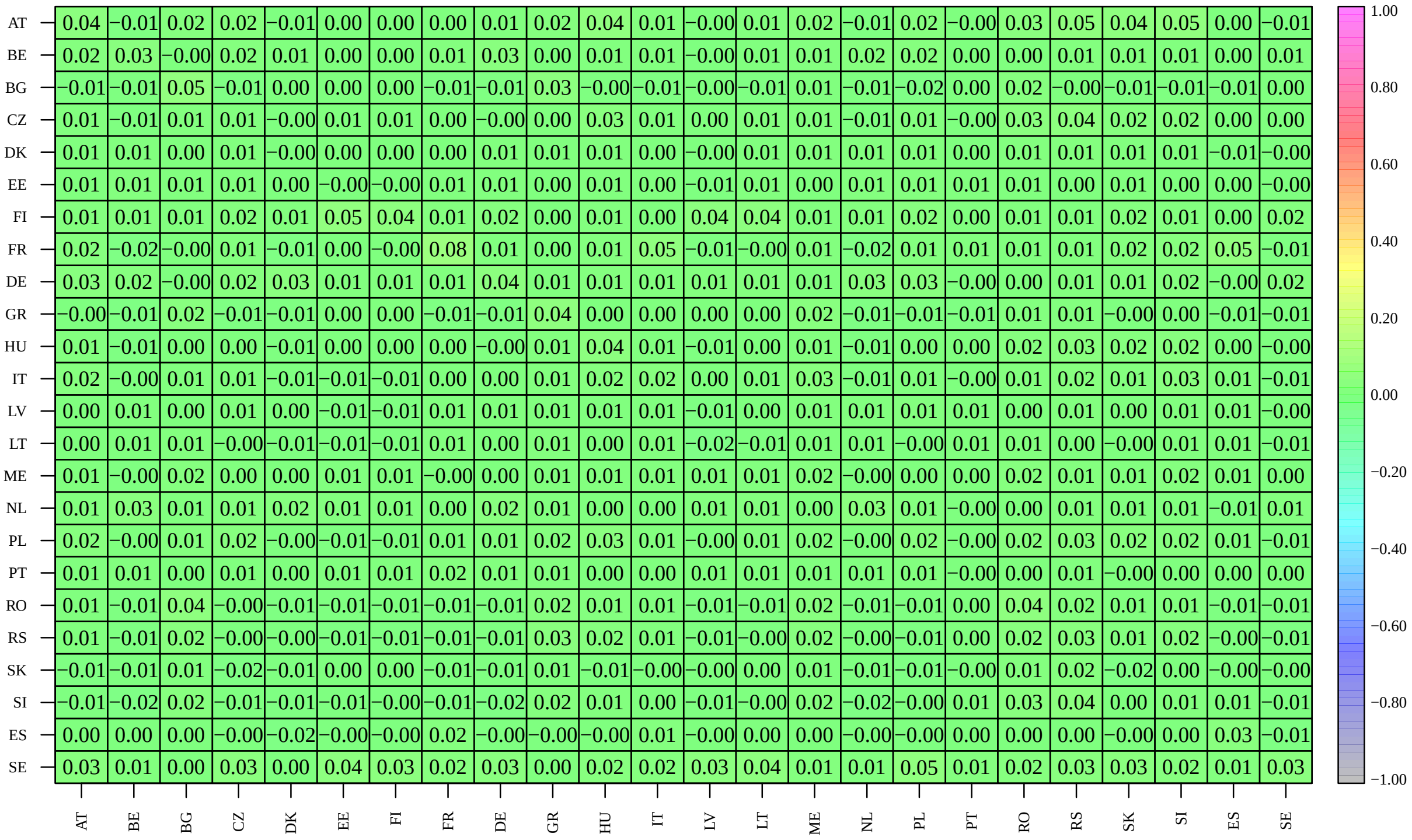

Figure 27: The in-sample correlation matrix $\widehat{\mathrm{Corr}}(\boldsymbol{\epsilon}^{\mathrm{Temp}}_{t=1,\dots,T}, \boldsymbol{\epsilon}^{\mathrm{AR}}_{t=1,\dots,T})$ for $t$ from February 6th, 2019 to February 1st, 2023, $\alpha = 1/24$ and VECM model for the socio-economic and political state.

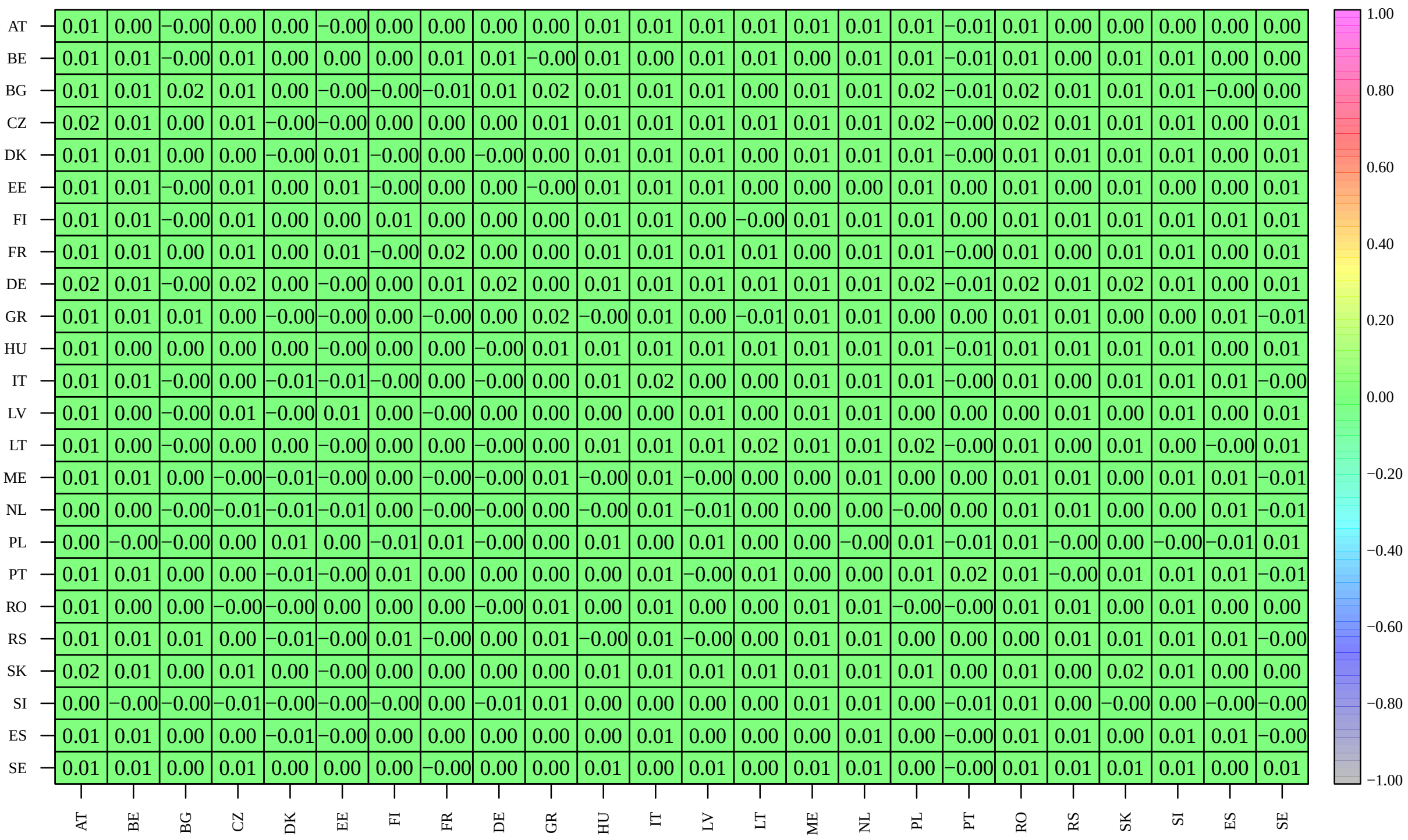

Figure 28: The in-sample correlation matrix $\widehat{\mathrm{Corr}}(\boldsymbol{\epsilon}^{\mathrm{VECM}}_{t=1,\dots,T}, \boldsymbol{\epsilon}^{\mathrm{AR}}_{t=1,\dots,T})$ for $t$ from February 6th, 2019 to February 1st, 2023 and VECM model for the socio-economic and political state.

## B. Figures

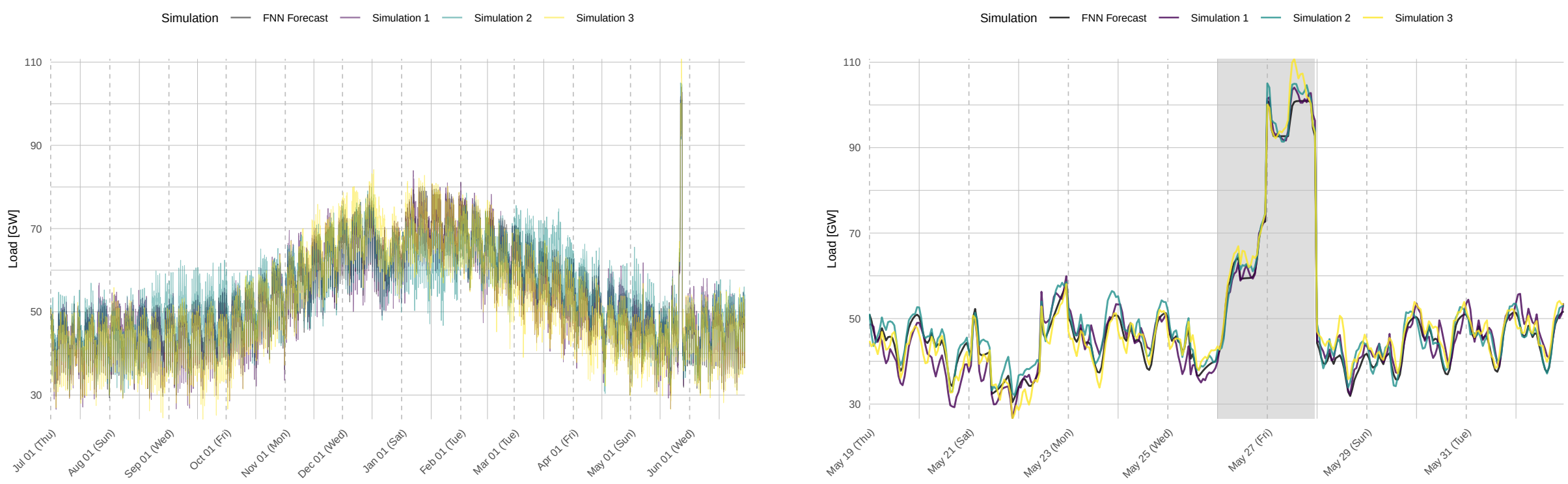

Figure 29: Spikes in FNN forecasted and simulated load around Abolition De L'esclavage (27th May, Guadeloupe) for in-sample data from France, July 6th, 2017 to July 1st, 2021.

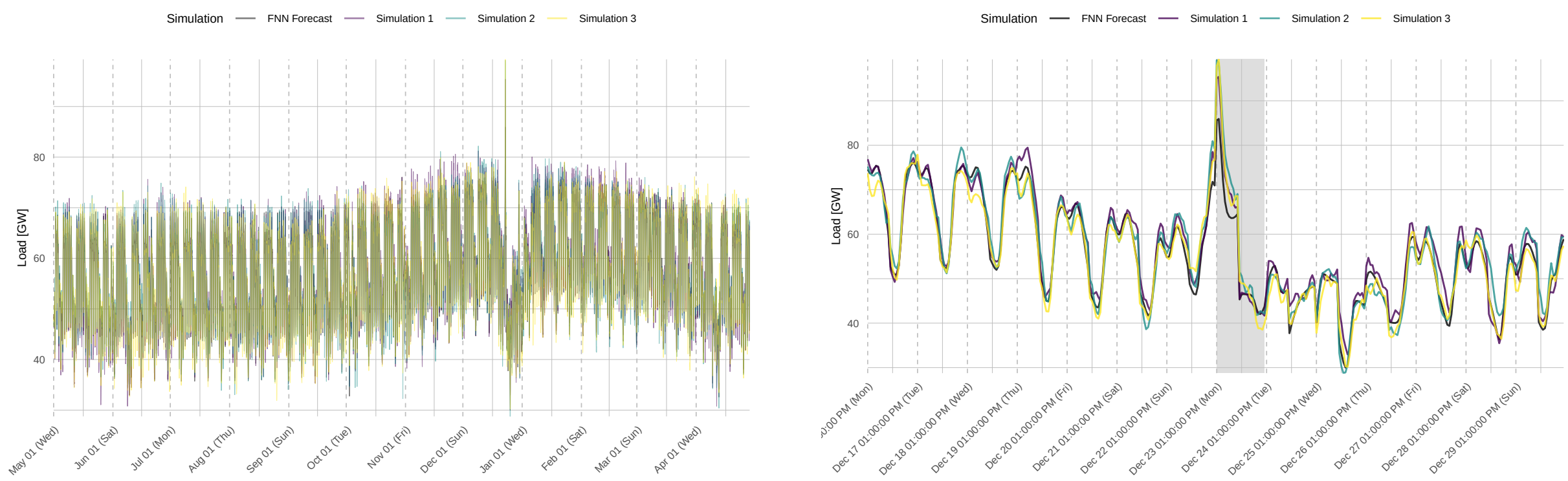

Figure 30: Spikes in FNN forecasted and simulated load around Christmas Eve (24th Dec.) for in-sample data from Germany, April 6th, 2015 to May 1st, 2019.

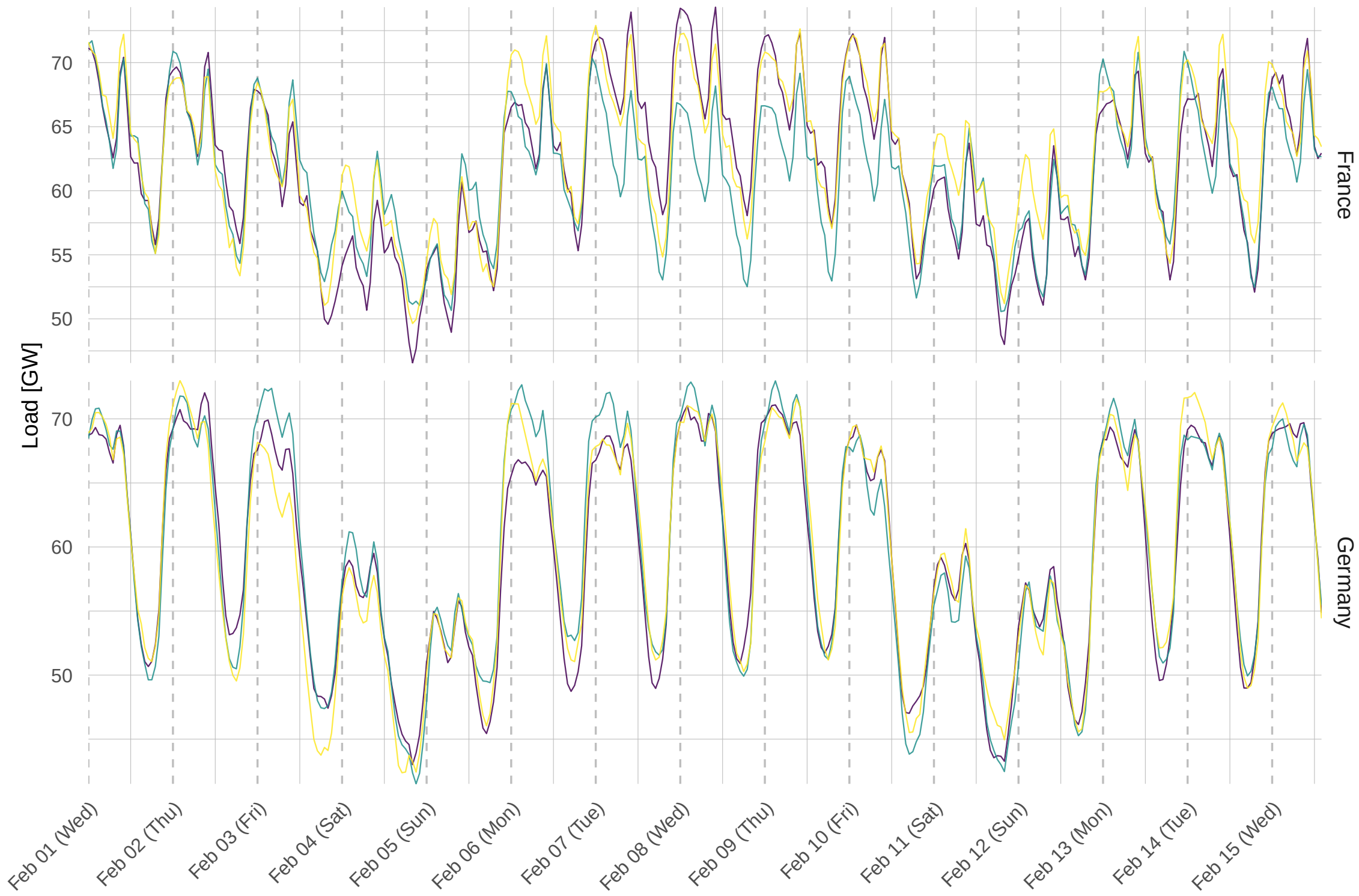

Figure 31: Three sampled load trajectories for the VAR model (forecasting experiment $N = 50$) in France and Germany forecasting the first two weeks of the one-year forecasting horizon (February 1st, 2023 to February 15th, 2023).

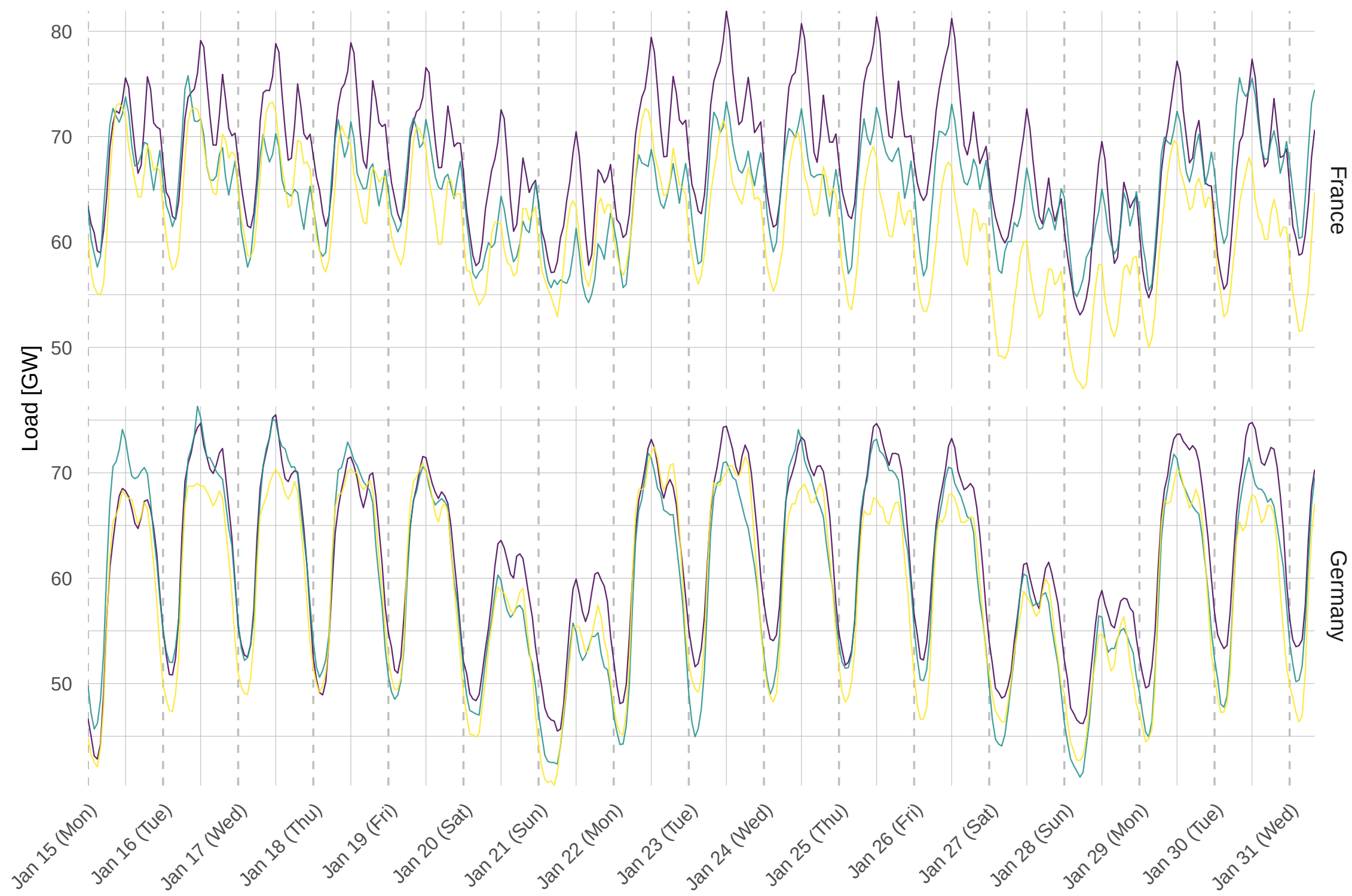

Figure 32: Three sampled load trajectories from the VAR model (forecasting experiment $N = 50$) in France and Germany forecasting the last two weeks of the one-year forecasting horizon (January 15th, 2024 to January 31st, 2024).

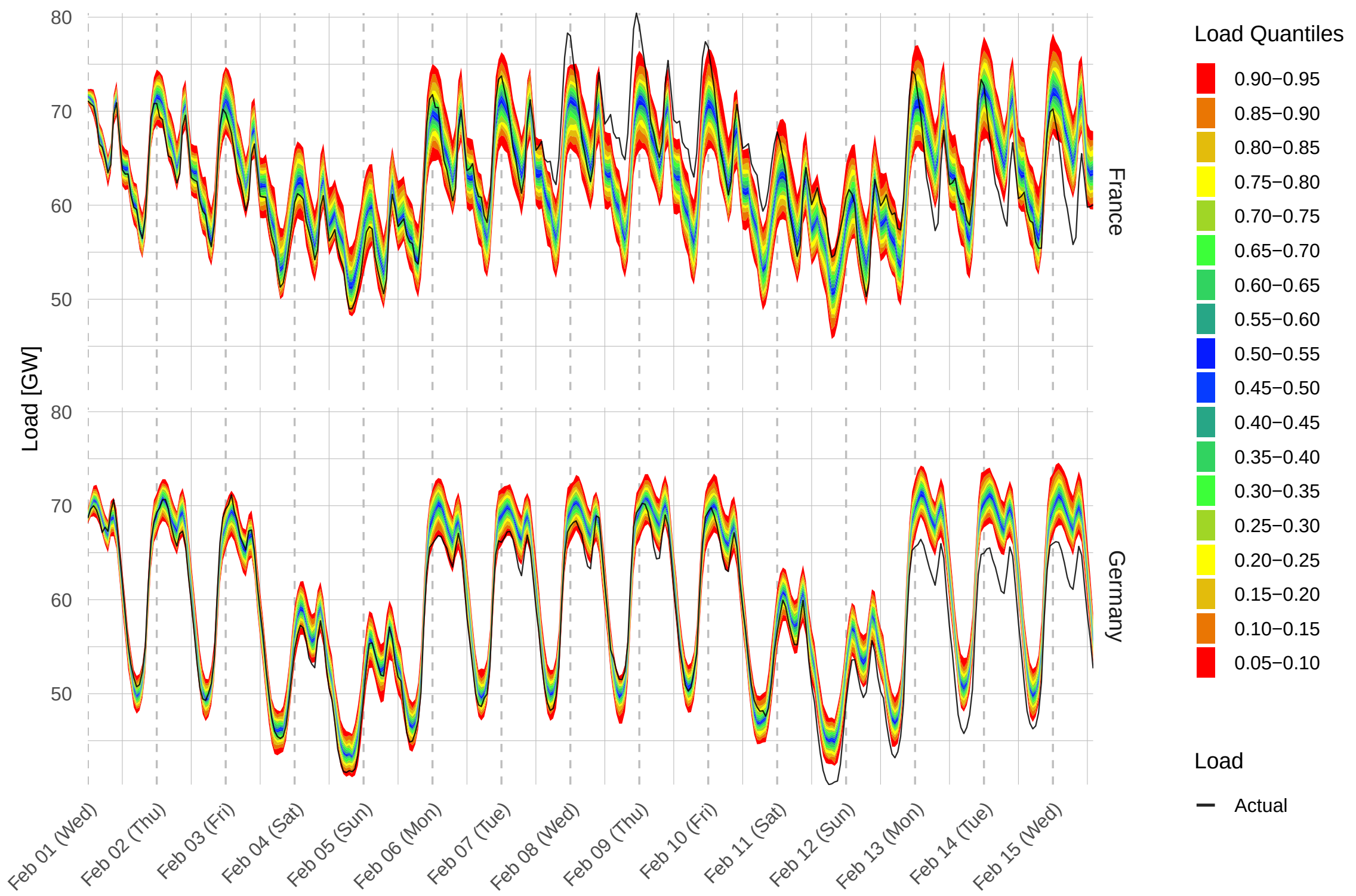

Figure 33: Quantiles of the VAR model (forecasting experiment $N = 50$) in France and Germany forecasting the first two weeks of the one-year forecasting horizon (February 1st, 2023 to February 15th, 2023).

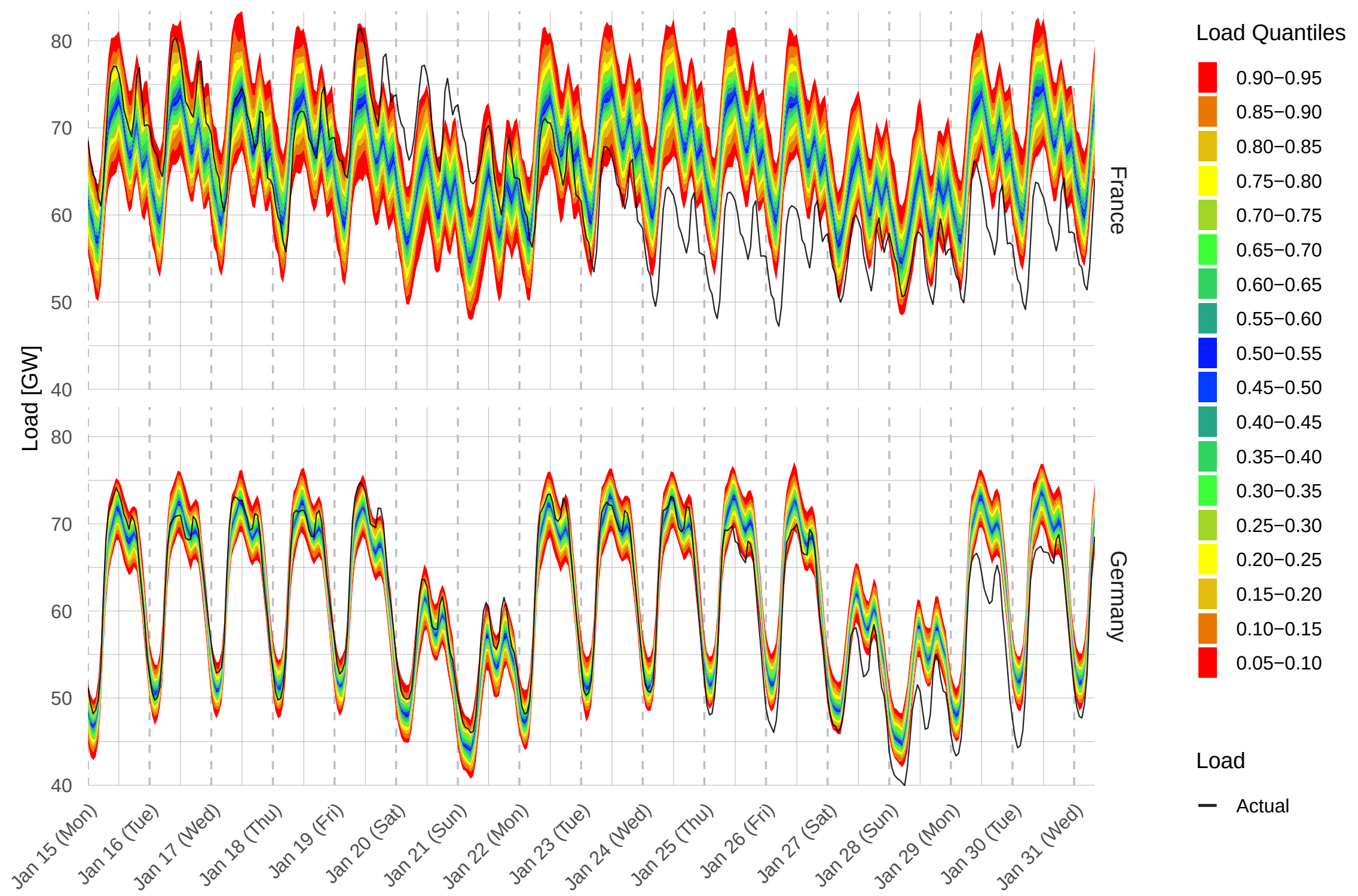

Figure 34: Quantiles of the VAR model (forecasting experiment $N = 50$) in France and Germany forecasting the last two weeks of the one-year forecasting horizon (January 15th, 2024 to January 31st, 2024).

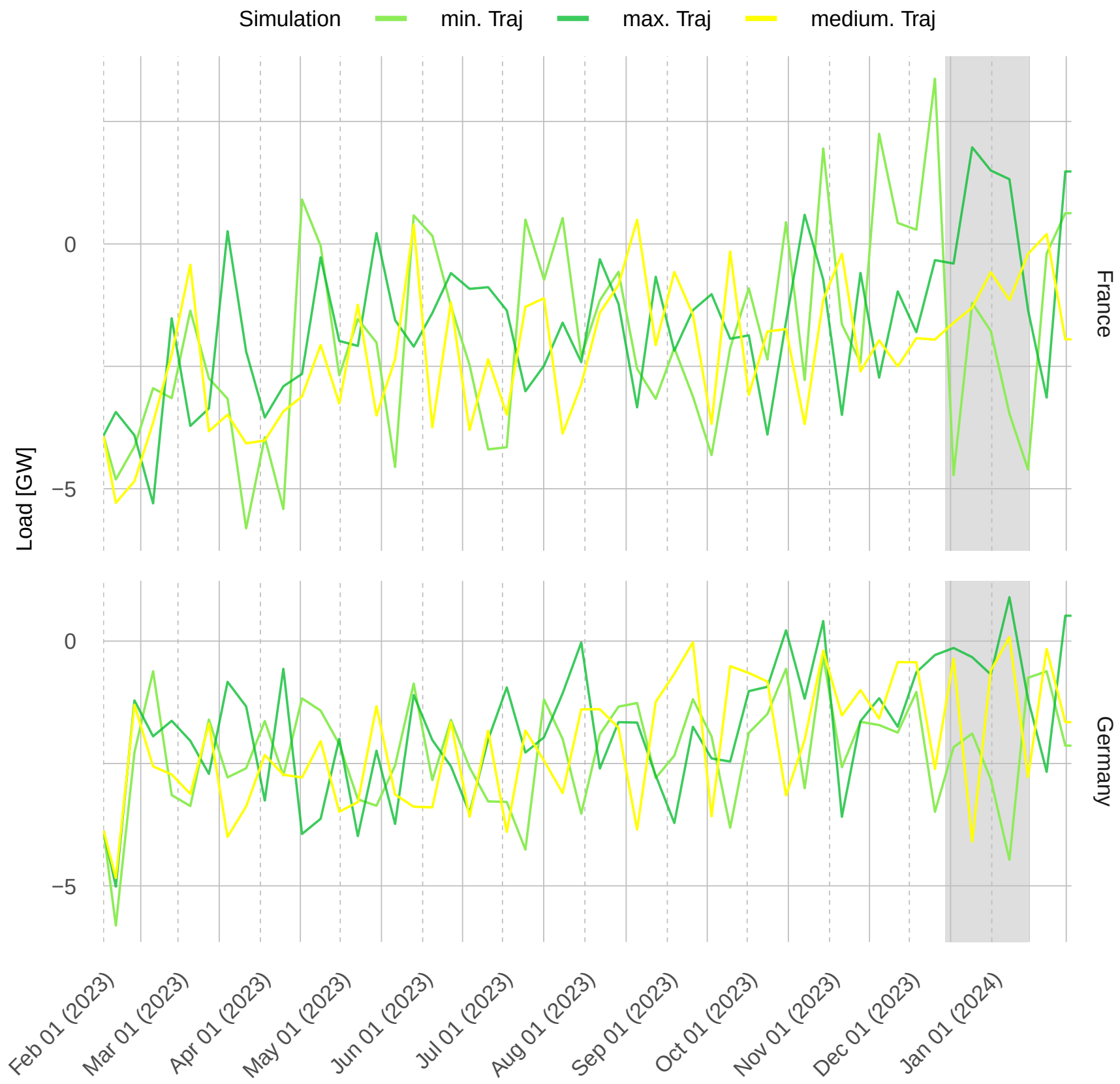

Figure 35: Minimum and maximum socio-economic state trajectories for VAR, medium trajectory in France and Germany for the one-year forecasting horizon of forecasting experiment $N = 50$.

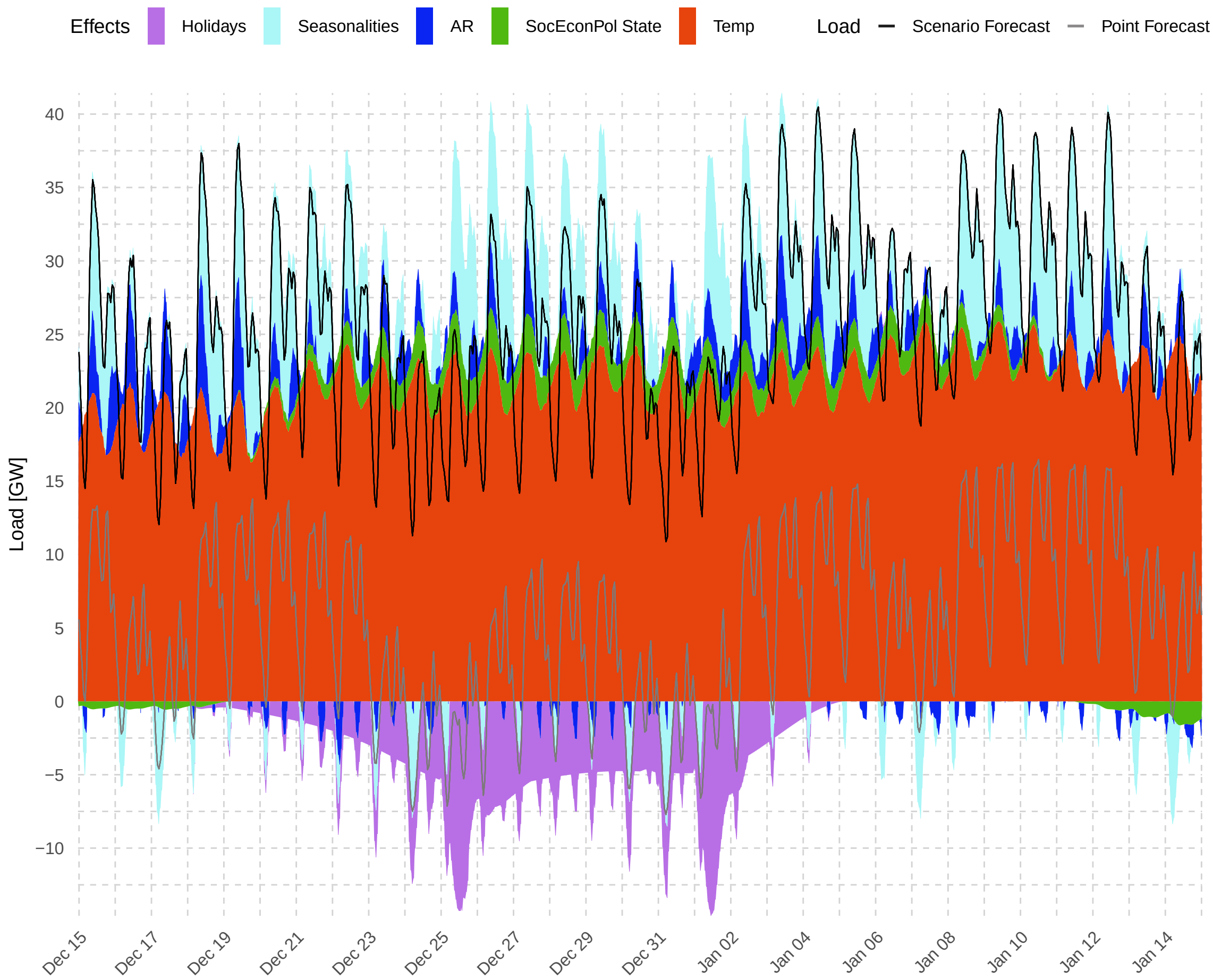

Figure 36: Forecasted load scenario (see (6) - (8)) decomposed in its modeling components for min. temperature, max. socio-economic state and max. autoregressive trajectories along with point forecasted load reduced by the estimated intercept for the VAR model (forecasting experiment $N = 50$) in France from December 15th, 2023 to January 14th, 2024.

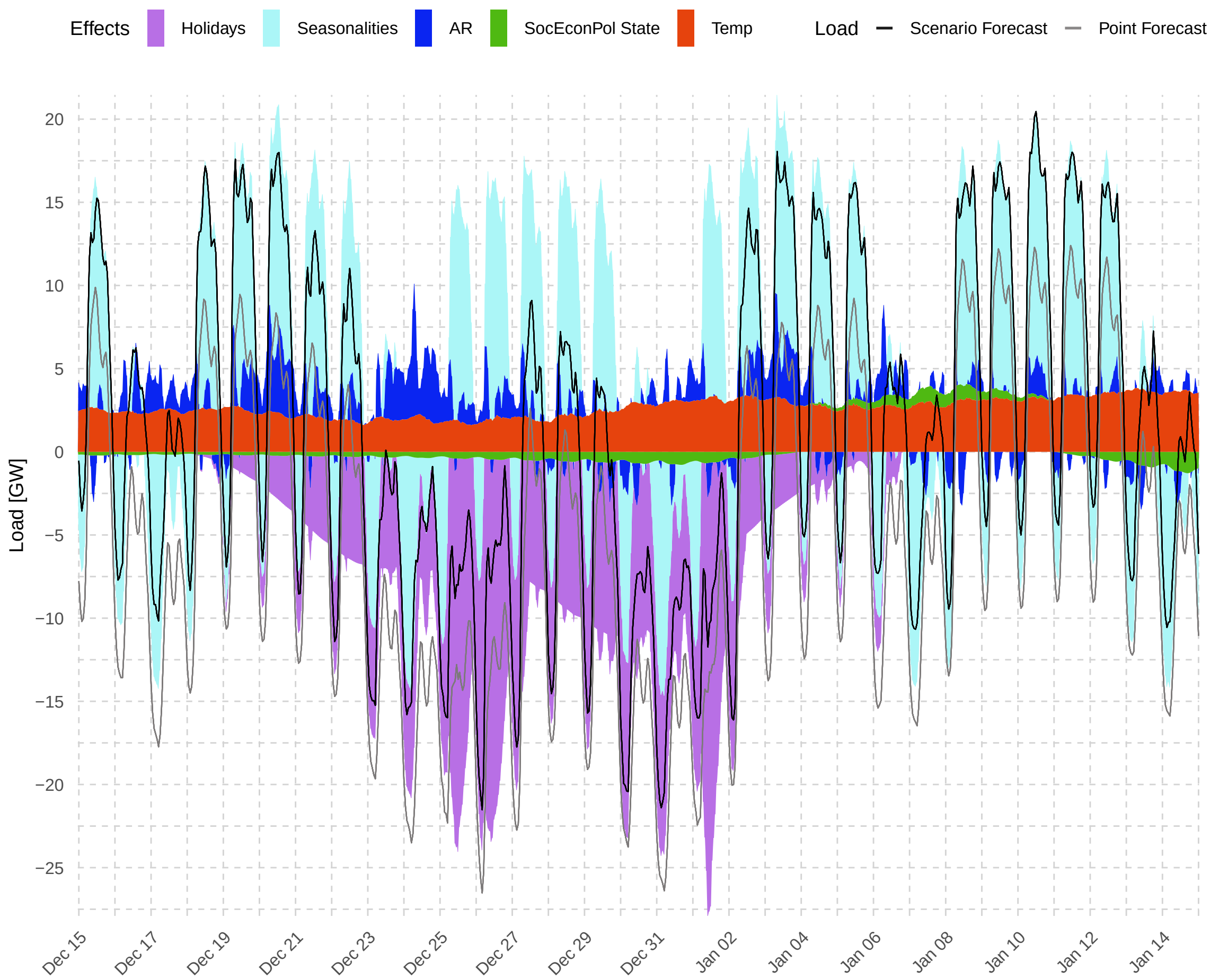

Figure 37: Forecasted load scenario (see (6) - (8)) decomposed in its modeling components for min. temperature, max. socio-economic state and max. autoregressive trajectories along with point forecasted load reduced by the estimated intercept for the VAR model (forecasting experiment $N = 50$) in Germany from December 15th, 2023 to January 14th, 2024.